\documentclass[11pt,letterpaper]{article}
    
\usepackage{fullpage}
    \usepackage{graphicx} 
    \usepackage{todonotes}
    \usepackage[normalem]{ulem}
    \usepackage{enumitem}

    \usepackage{wrapfig}

    \usepackage{amsthm}
    \usepackage{amsmath}
    \usepackage{amsfonts}
    \usepackage{amssymb}
    \usepackage{xcolor}
    \usepackage{fullpage}
    \usepackage{xspace}
    \usepackage{mdframed}
    \usepackage{hyperref}
    \usepackage{thm-restate}
    \usepackage{lineno}

    \usepackage{array}
    \usepackage{caption}

\newcolumntype{L}[1]{>{\raggedright\arraybackslash}p{#1}}
    
    \usepackage{authblk}
    \usepackage{comment}
    \usepackage{float}

    \newcommand{\reach}{\textnormal{\texttt{Reach}}}
    \newcommand{\cutcover}{\textnormal{\texttt{CutCover}}}

    \newcommand{\opt}{\texttt{opt}}

    \newcommand{\cI}{{\mathcal{I}}\xspace}

    \newcommand{\cT}{{\mathcal{T}}\xspace}

    \newcommand{\flow}{\lambda}

    \newcommand{\mwc}{\textnormal{mwc}\xspace}

    \newcommand{\bigslim}{{big-slim}}
    \newcommand{\slim}{{slim}}
    \newcommand{\bg}{{big}}
    
    \newcommand{\Bigslim}{{Big-slim}}
    \newcommand{\lslf}{large-size-large-flow}
    \newcommand{\Lslf}{Large-size-large-flow}

    \newcommand{\bch}{big-component-hitter}
    \newcommand{\Bch}{Big-component-hitter}

    \newcommand{\defparproblem}[4]{
        \vspace{3mm}
        \noindent\fbox{
            \begin{minipage}{.95\textwidth}
                \begin{tabular*}{\textwidth}{@{\extracolsep{\fill}}lr} \textsc{#1}  & {\bf{Parameter:}} #3 \\ \end{tabular*}
                {\bf{Input:}} #2  \\
                {\bf{Question:}} #4
            \end{minipage}
        }
        \vspace{2mm}
    }

    \newtheorem{lemma}{Lemma}[section]
    
    \newtheorem{claim}{Claim}
    \newtheorem{proposition}{Proposition}[section]

    \newtheorem{definition}{Definition}[section]
    
    \providecommand{\bound}{\textnormal{\texttt{bound}}}

    \title{
    Quasi-polynomial-sized vertex multicut-mimicking networks and kernelization for Vertex Multiway Cut}
    
    \author[1]{Eduard Eiben
    }
    
    \author[2]{Roohani Sharma\thanks{Supported by the Young Scientist Fellowship of the Institute for Basic Science (IBS-R029-Y8).}}
    
    \author[1]{Magnus Wahlstr\"om
    }
    
    \affil[1]{Royal Holloway, London, United Kingdom\\
    
    \texttt{eduard.eiben@rhul.ac.uk, magnus.wahlstrom@rhul.ac.uk}}
    
    \affil[2]{DIMAG, Institute for Basic Science, South Korea\\
    
    \texttt{roohani@ibs.re.kr}}
    \date{}
    
\begin{document}
    \maketitle
    \thispagestyle{empty}

\begin{abstract}

We give deterministic polynomial-time constructions of {\em vertex multicut-mimicking networks} of quasi-polynomial size.
Given a graph $G$, a terminal set $T$, and a positive integer
$c$, our algorithm constructs a graph with
$|T|^7 c^{O(\log c)}$ vertices that preserves every minimum vertex multiway cut value of at most $c$, for every partition of the terminals.
Taking $c=|T|$ gives an exact vertex multicut-mimicking
network with $|T|^{O(\log |T|)}$ vertices, resolving
an open question of Wahlstr\"om [ACM Transactions on Algorithms 2022], who showed a bound of $\kappa^{O(\log^3 \kappa)}$ for the edge multicut-mimicking variant, where $\kappa:= \sum_{t \in T} |N_G(t)|$ is the capacity of the terminal set,
which was later improved to $\kappa^{O(\log^2 \kappa)}$ by Anand, Lee, Li, and
Saranurak [STOC 2024].

As applications of our vertex multicut-mimicking network, 
we obtain deterministic kernels of
size $k^{O(\log k)}$ for
\textsc{Vertex Multiway Cut},
$p^{O(1)}k^{O(\log k)}$ for
\textsc{Vertex Multicut} with $p$ requests and
$|\Gamma|^{O(1)}k^{O(\log k)}$ for
\textsc{Group Feedback Vertex Set}, where
 $k$ is the deletion budget and $\Gamma$ is a finite group.
Our first kernelization result, in particular, resolves an open question by Wahlstr\"om [ACM Transactions on Algorithms 2022],
which was reiterated in a recent survey on kernelization by Jansen [Computer Science Review 2026].
Also, to the best of our knowledge, our kernel for {\sc Vertex Multicut} is the first one whose size is completely polynomial in $p$ and (quasi-polynomial) super-polynomial only in $k$.

Finally, again using our vertex multicut-mimicking network for graphs, we obtain a hyperedge multicut-mimicking network for hypergraphs, which preserves hypergraph edge-multicuts of size up to $c$,
of size
$|T|^{O(1)} c^{O(\log c)}$.
Our result removes the dependence on the rank of the hypergraph from the exponent of the size bound, that existed in prior work. 
\end{abstract}

\newpage
\setcounter{page}{1}
\pagenumbering{arabic}\section{Introduction}\label{sec:intro}

The computation of minimum cuts is fundamental to several algorithms for connectivity-based,
graph partitioning and network design problems.
When many minimum cut queries concern the same graph,
compact
representations can encode their answers without storing
a separate solution for each query.
For example, a Gomory-Hu tree represents a minimum
$s$-$t$ cut and its value for every pair of vertices
$s,t\in V(G)$ using only $|V(G)|-1$ weighted
edges~\cite{GomoryHu61}.

Suppose now that the minimum cut queries involve only a designated set
$T\subseteq V(G)$ of terminals, but may ask to separate
two arbitrary subsets of $T$. 
That is, we are interested in compactly storing the value of a minimum $X$-$Y$ cut in a graph $G$, for every possible $X,Y \subseteq T$.
In this case the number of potential queries is exponential in $|T|$.
Storing one minimum cut value for every choice of $X,Y \subseteq T$ 
already gives a representation with exponential in $|T|$ many entries (independent of the size of the graph).
The interesting question here is whether the same information can be encoded
in a much smaller graph containing $T$, one whose size is
sub-exponential, or quasi-polynomial or polynomial in $|T|$.

Cut sparsifiers and mimicking networks seek such
representations preserving the relevant minimum cut values
approximately or exactly.
Given a graph $G$ and a designated set $T\subseteq V(G)$
of terminals, a {\em cut sparsifier} is a smaller graph containing
$T$ that {\em approximately} preserves minimum cut values between
any {\em two} subsets of $T$.
The balance here is in the trade-off between its size and the approximation quality in guarantees~\cite{DBLP:journals/siamcomp/BenczurK15,DBLP:journals/siamcomp/SpielmanT11,DBLP:conf/focs/ChenGHPS20,DBLP:conf/focs/Madry10,DBLP:journals/jacm/ChenKLPGS25,DBLP:conf/focs/SaranurakY22,HagerupKNR98JCSS,DBLP:conf/focs/Moitra09,DBLP:conf/stoc/LeightonM10,DBLP:conf/focs/MakarychevM10,DBLP:journals/siamcomp/EnglertGKRTT14,DBLP:conf/stoc/Chuzhoy12,DBLP:conf/soda/AndoniGK14,DBLP:conf/soda/KrauthgamerR13,DBLP:journals/siamdm/KrauthgamerR20,DBLP:journals/algorithmica/KarpovPZ19,ChalermsookDKLL21,DBLP:conf/soda/KrauthgamerM23,DBLP:conf/mfcs/Das0024}. 
When all minimum cut values (for two subsets of the terminal set) are preserved {\em exactly}, the replacement
is called a \emph{mimicking network}, a notion introduced
by Hagerup, Katajainen, Nishimura, and
Ragde~\cite{HagerupKNR98JCSS} and widely studied thereafter~\cite{HagerupKNR98JCSS,DBLP:conf/soda/KrauthgamerR13,DBLP:journals/algorithmica/KarpovPZ19,DBLP:journals/siamdm/KrauthgamerR20,ChalermsookDKLL21,DBLP:conf/focs/SaranurakY22,DBLP:conf/mfcs/Das0024}.

The present work concerns mimicking networks for a more general
family of cut queries.
Let $G$ be a graph and
$T\subseteq V(G)$ be a set of $k$ terminals.
We refer to $(G,T)$ as a terminal network.
For a partition $\cT=\{T_1,\ldots,T_t\}$ of $T$,
a \emph{vertex $\cT$-multiway cut ($\cT$-\mwc)}
in $G$ is a set of vertices $X\subseteq V(G)$
such that in $G-X$, each connected component contains
terminals from at most one part of $\cT$.
A vertex \emph{multicut-mimicking network} of $(G,T)$
is a graph $G'$ with $T\subseteq V(G')$ such that 
for
every $T_0 \subseteq T$ and for every 
partition $\cT$ of $T\setminus T_0$,
the size of a minimum 
$\cT$-\mwc\ is identical in $G-T_0$ and
in $G'-T_0$.
Note that when $t=2$, the resulting network is a mimicking network, thus multicut-mimicking networks are a generalization. 
In our setup, for the vertex deletion case, we allow
the multiway cut solution to intersect the terminals $T$,
and we justify this assumption towards the end of
this section\footnote{
Note that the ability to choose a partition of $T$
allows to emulate undeletable terminals by duplicating
them, up to some cost bound $k$.
}.
The question we are investigating here is the following.

\begin{center}
{\em What is the minimum size $|V(G')|$
for a vertex multicut-mimicking network, as a function
of $k$?}\footnote{
The name \emph{multicut-mimicking} (rather than
\emph{multiway cut-mimicking}) comes from the observation
that a terminal network $(G',T)$ that preserves the size
of a $\cT$-\mwc\ for every partition $\cT$ of $T$ also
naturally preserves the minimum size of a multicut for
every set of \emph{cut requests}
$R\subseteq\binom{T}{2}$.
See Wahlström~\cite[Prop.~2.2]{DBLP:journals/talg/Wahlstrom22};
the same statement holds for vertex cuts.
}
{\em 
In particular, can we bound $|V(G')|$
(quasi-)polynomially in $k$?
}

{\em
Further, can one also find such mimickers efficiently (in polynomial-time)?
}
\end{center}
This question is also motivated by the notorious open question
of the existence of a polynomial kernel for
\textsc{Vertex Multiway Cut} (see later),
but is also interesting in its own right as a structural
question in graph theory about compressing information
about exponential behaviours in 
(quasi-)polynomial space:
is there a graph on $k^{O(1)}$ or even
$k^{O(\log k)}$ vertices that realizes the behaviour
of a minimum multiway cut for each of the
$2^{\Theta(k\log k)}$ partitions of $T$?

When we omit the prefix vertex in front of multicut-mimicking network we refer to the one where the separators are edge cuts instead of vertex cuts. We also note that, generally, dealing with vertex cuts is much more challenging than dealing with edge cuts, in undirected graphs, and this happens to be the case also in our work.
It is not difficult to observe that the existence and construction of vertex multicut-mimicking networks imply the same, with same bounds, for (edge) multicut-mimicking networks.

As a remark, one can also note
that in the case of vertex multicut-mimicking networks we are preserving information about objects ($\mathcal{T}$-\mwc) that are NP-hard to compute even for one query~\cite{DBLP:conf/icalp/GargVY94},
unlike mimicking networks, 
but still hoping for a construction that is both compact and 
computable in ``purely'' polynomial time, without assuming that the terminals or other parameters of the instance are constants.
Below we describe the landscape of relevant results and how our results fit there. We also refer the reader to Tables~\ref{tab:graph-mimicking-comparison},~\ref{tab:hypergraph-mimicking-comparison} and~\ref{tab:kernelization-comparison} for comparison between known results and our results.

\begin{table}[!htbp]
\centering
\setlength{\tabcolsep}{2.7pt}
\renewcommand{\arraystretch}{1.0}

\begin{tabular}{@{}
    p{0.08\linewidth}
    p{0.12\linewidth}
    p{0.21\linewidth}
    p{0.17\linewidth}
    p{0.38\linewidth}@{}}
\hline
\textbf{Cuts}
& \textbf{Terminal parts}
& \textbf{Approximation}
& \textbf{Construction}
& \textbf{Size} \\
\hline

Edge
& $2$
& $1$
& Deterministic
& $2^{2^k}$ vertices
  \cite{HagerupKNR98JCSS} \\

& & &
---
& $2^{\Omega(k)}$ vertices necessary
  \cite{DBLP:conf/soda/KrauthgamerR13} \\

& & &
\textcolor{blue}{Deterministic}
& $O(\kappa^3)$ edges
  \cite{DBLP:journals/jacm/KratschW20} \\

& & &
{Randomized}
& $O(kc^3\log^{3/2}(kc))$ edges${}^{\dagger}$
  \cite{DBLP:conf/esa/JiangHSZ22} \\

\cline{3-5}

& &
$O\!\left(\dfrac{\log k}{\log\log k}\right)$
& Deterministic
& $k$ vertices
  \cite{DBLP:conf/focs/MakarychevM10} \\

\hline

Vertex
& $2$
& $1$
& Deterministic
& $O(k^3)$ vertices
  \cite{DBLP:journals/jacm/KratschW20,
        DBLP:journals/corr/abs-2609-29141} \\

&
$2$
& $1$
& Deterministic\newline
  $m^{1+o(1)}$\newline
  $\cdot 2^{O(c^2)}$ time
& $k\,2^{O(c^2)}$ edges${}^{\dagger}$
  \cite{DBLP:conf/focs/SaranurakY22} \\

\noalign{\hrule height 1.2pt}

Edge
& $s$
& $1$
& \textcolor{blue}{Deterministic}
& $O(\kappa^{s+1})$ edges
  \cite{DBLP:journals/jacm/KratschW20} \\

&
Unbounded
& $2$
& \textcolor{blue}{Deterministic}
& $O(\kappa^3)$ edges
  \cite{DBLP:journals/talg/Wahlstrom22} \\

&
Unbounded
& $1$
& {Randomized}
& $\kappa^{O(\log^2\kappa)}$ edges
  \cite{DBLP:journals/talg/Wahlstrom22,
        DBLP:conf/stoc/0001L0S24} \\

\hline

Vertex
& $s$
& $1$
& \textcolor{blue}{Deterministic}
& $O(k^{s+1})$ vertices
  \cite{DBLP:journals/jacm/KratschW20} \\

&
Unbounded
& $2$
& Randomized
& $O(k^3)$ vertices
  \cite{DBLP:conf/soda/Wahlstrom24} \\

&
\textcolor{red}{Unbounded}
& \textcolor{red}{$1$}
& \textcolor{red}{Deterministic}
& \textcolor{red}{
  $k^7 c^{O(\log c)}$ vertices${}^{\dagger}$
  (Theorem~\ref{thm:budgeted-vertex-mimicking})\newline
    $k^{O(\log k)}$ vertices
  (Corollary~\ref{thm:full-vertex-mimicking})} \\

\hline
\end{tabular}

\caption{
Graph sparsifiers and mimicking networks.
Here $k:=|T|$, and
$\kappa:=\sum_{t\in T}\deg_G(t)$ denotes the terminal
capacity.
The rows indexed by $s$ preserve multiway cuts of
terminal partitions into at most $s$ parts, for fixed $s$.
The construction column describes whether the algorithm is randomized or deterministic, and unless mentioned otherwise the time is purely polynomial.
A dagger denotes connectivity-$c$ preservation, with cut
values truncated at $c$.
The red row denotes our result.
Blue construction labels denote results originally obtained
by randomized algorithms whose constructions admit
deterministic implementations using the cut-covering and
representative-family machinery of Koana and
Kumabe~\cite{DBLP:journals/corr/abs-2609-29141}.
}
\label{tab:graph-mimicking-comparison}

\end{table}

\paragraph{Mimicking networks and cut sparsifiers.}

For exact minimum edge-cut preservation, 
the size of a mimicking
network can be exponential in the number of terminals.
Hagerup, Katajainen, Nishimura, and
Ragde~\cite{HagerupKNR98JCSS}
showed that every edge-weighted graph with $k$ terminals
admits a mimicking network with at most $2^{2^k}$ vertices.
Krauthgamer and
Rika~\cite{DBLP:conf/soda/KrauthgamerR13}
proved that $2^{\Omega(k)}$ vertices are necessary
for some terminal networks.

Allowing to only preserve approximations of a minimum cut for every query, 
give a spectrum of different tradeoffs between the 
size of the mimicking network and its  quality. We do not go deep into surveying this literature here.
Moitra~\cite{DBLP:conf/focs/Moitra09} studied cut
sparsifiers whose vertex set consists only of the terminals.
Such sparsifiers of quality $O(\log k/\log\log k)$
exist and can be constructed in polynomial
time~\cite{DBLP:conf/focs/MakarychevM10,
DBLP:journals/siamcomp/EnglertGKRTT14}.

In a seminal work, Kratsch and Wahlstr\"om~\cite{DBLP:journals/jacm/KratschW20} showed that, for edge cuts, if the size of the mimicking network is measured in terms of the {\em capacity of the terminals}, that is $\kappa:= \sum_{t\in T} |N_G(t)|$, then one can bypass the exponential-in-$k$ lower bound, and
one can obtain mimicking networks of size $O(\kappa^3)$.
In fact, their result also works for vertex cuts: as long as the vertex cuts are allowed to delete terminals,
one can obtain vertex mimicking network of size $O(k^3)$ with $k$ terminals.
This result is known as the {\em cut-covering lemma} in the literature and has had a huge impact on the landscape of kernelization and parameterized complexity.
The construction of the cut-covering lemma of Kratsch and Wahlstr\"om~\cite{DBLP:journals/jacm/KratschW20} was randomized and has very recently been derandomized by Koana and Kumabe~\cite{DBLP:journals/corr/abs-2609-29141}, preserving the same asymptotic size bounds.

Chalermsook, Das, Kook, Laekhanukit, Liu, Peng,
Sellke, and Vaz~\cite{ChalermsookDKLL21}
introduced connectivity-$c$ mimicking networks,
which preserve minimum edge-cut values between
disjoint terminal subsets up to a prescribed threshold.
They established a bound of $O(kc^4)$ edges and
gave applications to dynamic connectivity and
survivable network design.
Liu~\cite{Liu23aITCS}
improved the existence bound to $O(kc^3)$ edges.
He also gave a randomized polynomial-time construction
with $O(kc^3\log^{3/2}n\log\log n)$ edges,
where $n:=|V(G)|$~\cite[Theorem~1]{Liu23aITCS}.
Jiang, Huang, Saranurak, and
Zhang~\cite{DBLP:conf/esa/JiangHSZ22}
subsequently obtained a polynomial-time construction
with $O(kc^3\log^{3/2}(2kc))$ edges,
removing the dependence on $n$ from the output bound,
and extended these guarantees to hypergraphs.

\paragraph{Multicut-mimicking networks.}

The result of Kratsch and
Wahlstr\"om~\cite{DBLP:journals/jacm/KratschW20}
for mimicking networks also extend to the case of {\em multicut-mimicking} networks in the case where terminal partitions are restricted to ones that have {\em bounded-many parts}.
That is, Kratsch and Wahlstr\"om~\cite{DBLP:journals/jacm/KratschW20}
showed that, for every fixed integer $s$, a terminal
network with $k:=|T|$ terminals admits a multicut-mimicking network with
$O(k^{s+1})$ (resp. $O(\kappa^{s+1})$) vertices which preserves minimum vertex (resp. edge) multiway cut
values for all partitions of $T$ into at most $s$ parts.
This gives a polynomial bound on the size
for every fixed number of parts, with an exponent depending on that number.
This is also constructible in randomized polynomial time and using the recent result of Koana and Kumabe~\cite{DBLP:journals/corr/abs-2609-29141} can be derandomized.

Later, just for edge multiway cuts, Wahlstr\"om~\cite{DBLP:journals/talg/Wahlstrom22}
proved that every undirected terminal network $(G,T)$
admits an exact multicut-mimicking network with
$\kappa^{O(\log\kappa)}$ edges, where
$\kappa:=\sum_{t\in T}\deg_G(t)$ is the total terminal
capacity.
The network preserves minimum multiway cut values for {\em every partition} of $T$, without restricting the
number of parts.
He also gave a randomized polynomial-time construction for such multicut-mimicking networks, of size $\kappa^{O(\log^3\kappa)}$, combining
representative-set methods with recursive decomposition along sparse cuts.
Using improved approximation algorithms for
\textsc{Small Set Expansion}, Anand, Lee, Li, and
Saranurak~\cite{DBLP:conf/stoc/0001L0S24}
improved the constructive bound to
$\kappa^{O(\log^2\kappa)}$.

Unlike the $O(\kappa^{s+1})$ bound, this quasi-polynomial bound does not extend to {\em vertex} multicut-mimicking networks.
In fact,
Wahlstr\"om~\cite{DBLP:journals/talg/Wahlstrom22}
explicitly asked whether {\em quasi-polynomial-sized
vertex multicut-mimicking networks} also exist, and if so, if they can be
constructed efficiently (the number of parts of a terminal partition is unrestricted).
In this work we answer this question positively, and in fact give a better size bound than that of Wahlstr\"om's for edge multicut-mimicking network~\cite{DBLP:journals/talg/Wahlstrom22}.

\paragraph{Our result: Multicut-mimicking networks of quasi-polynomial size for vertex cuts.}

Despite several advances in the field of cut sparsifiers and mimicking networks, the question of designing
sub-exponential multicut-mimicking networks for the vertex
deletion setting remained open.
In this paper we show the existence of a
vertex multicut-mimicking network of size $k^{O(\log k)}$
where $k$ is the number of terminals.
We thereby resolve Wahlstr\"om's open
question~\cite{DBLP:journals/talg/Wahlstrom22} and also strictly improve the size bound of the constructive (edge) multicut-mimicking networks of Wahlstr\"om~\cite{DBLP:journals/talg/Wahlstrom22} and
Anand, Lee, Li and Saranurak~\cite{DBLP:conf/stoc/0001L0S24} (by shaving off log factor in the exponent in the latter).
Moreover, we show that such a vertex multicut-mimicking network can also
be found in {\em deterministic} polynomial-time, whereas the ones in
\cite{DBLP:journals/talg/Wahlstrom22,DBLP:conf/stoc/0001L0S24} rely on randomized algorithms. 
Our result answers an open problem posed by Jansen in his recent survey on Kernelization~\cite{DBLP:journals/csr/Jansen26}.
We also bypass the need for a vertex analogue of the \textsc{Small Set Expansion}-approximation algorithms which were used in the constructivity of the quasi-polynomial bound for edge multicut-mimicking networks of Wahlstr\"om~\cite{DBLP:journals/talg/Wahlstrom22}. This is essential since
an analogous approximation algorithm for \textsc{Small Set Vertex Expansion} is excluded under current conjectures in the area~\cite{DBLP:conf/stoc/0001L0S24}.

We, in fact, prove an even more general result.
In the spirit of the connectivity-$c$ mimicking networks introduced by 
Chalermsook, Das, Kook, Laekhanukit, Liu, Peng, Sellke and Vaz~\cite{ChalermsookDKLL21},
and later studied in~\cite{Liu23aITCS,DBLP:conf/focs/SaranurakY22,DBLP:conf/esa/JiangHSZ22,DBLP:conf/isaac/Cho024},
we define {\em connectivity-$c$ multicut-mimicking networks} and 
obtain the described results for them.

  For an integer $c \geq 1$,
    a \emph{connectivity-$c$ vertex multicut-mimicking network} of a terminal network $(G,T)$ is a pair $(G',T)$ such that for any $T_0 \subseteq T$ and $\mathcal{T}$ which is a partition of $T \setminus T_0$,

    $$\min\{ \opt_{G-T_0}(\mathcal{T} ), c+1  \} = \min \{\opt_{G'-T_0}(\mathcal{T}) , c+1\},$$
    
    where $\opt_{H}(\mathcal{T})$ denotes the size of a minimum (vertex) $\mathcal{T}$-\mwc\ in the graph $H$.
    Note that {connectivity-$|T|$ vertex multicut-mimicking network} is a vertex multicut-mimicking network.
    Our main result is Theorem~\ref{thm:budgeted-vertex-mimicking}.
    
    \begin{restatable}[Bounded connectivity vertex multicut-mimicking network]{theorem}{thmbudgetedvertex}
    \label{thm:budgeted-vertex-mimicking}
    Given a terminal network $(G,T)$
    and a positive integer $c$,
    there is a deterministic
    polynomial-time algorithm 
    that outputs 
    a connectivity-$c$ vertex multicut-mimicking network $(G',T)$ of $(G,T)$,
    where $|V(G')| = |T|^7 \cdot c^{O(\log c)}$.
    \end{restatable}

    \begin{restatable}[Vertex multicut-mimicking network]{corollary}{thmfullvertex}
    \label{thm:full-vertex-mimicking}
    Let $(G,T)$ be a terminal network with $|T|=k \geq 2$.
    There is a deterministic polynomial-time algorithm that outputs
    a vertex multicut-mimicking network for $(G,T)$
    of size $k^{O(\log k)}$.
    \end{restatable}

There are several advantages of studying such a ``fine-grained'' bound (depending on $k$ and $c$) on the size of a mimicker.
As we will later see, in our applications to kernelization, 
a lot of times the number of terminals may be much larger than the size of the sought cut, and in those cases the bound of Theorem~\ref{thm:budgeted-vertex-mimicking} tells that the size of our mimicker is cheap (polynomial) in terms of the number of terminals and slightly-expensive (quasi-polynomial) only in the size of the separator.

We also remark that the polynomial factor in Theorem~\ref{thm:budgeted-vertex-mimicking} is not tight, and in fact it can be brought down to $|T|^3$. But we keep it this way, in order to avoid adding extra arguments to the proof.

\paragraph*{Bounded vs unbounded capacity.}
In all the above results concerning vertex (multiway) cuts, it is crucial that we consider
terminals to be deletable, or more generally, that
the parameter $k$ corresponds to the total capacity
of the terminal set.
If $T$ is undeletable (or, in the edge case,
if we use the parameter $|T|$ instead of the capacity
of $T$), then the size of a mimicking network may be
forced to be exponential, and the best known upper bound is
double-exponential, even for the case of
edge cuts~\cite{HagerupKNR98JCSS,KhanR14Mimic}.
This stark difference between a double-exponential
upper bound for the size of the undeletable terminal set parameter 
and a
cubic bound for the capacity, or size of the deletable terminal set, parameter is intriguing.
One major open question is whether this difference
disappears for approximate cut sparsifiers.
It is a long-standing open question in sparsification
whether a $(1+\varepsilon)$-approximate cut sparsifier
of size polynomial in the terminal set 
exists for edge cuts,
going back to
Moitra~\cite{DBLP:journals/siamcomp/Moitra13}
(conference version 2009~\cite{DBLP:conf/focs/Moitra09});
we still do not know even if an $O(1)$-approximate
polynomial-size cut sparsifier for edge cuts exists
in the unbounded capacity setting.
This is especially intriguing in that the methods used
for the cubic bound (i.e.\ the cut-covering lemma and
related results) are completely different from the
methods generally used in the sparsification literature.
See more in the related works paragraph.

\paragraph{Our result: Generalization to hypergraphs.}
Hypergraph cut sparsification has been studied for both
approximate preservation of all minimum cuts and exact preservation
of terminal minimum cuts up to a prescribed threshold.
Chen, Khanna and Nagda~\cite{DBLP:conf/focs/ChenKN20}
showed that every $n$-vertex hypergraph admits a weighted
cut sparsifier with $\widetilde O(n/\varepsilon^2)$
hyperedges, preserving the value of every cut within
a factor of $1\pm\varepsilon$.
Khanna, Putterman and Sudan~\cite{DBLP:conf/focs/KhannaP024}
subsequently developed near-optimal linear sketches for recovering hypergraph cut sparsifiers.
These results reduce the number of hyperedges while retaining the original vertex set and allowing
approximation in the cut values.

In the context of the present work we are interested in preserving (multiway) cuts when the (multiway) cut queries are over a prescribed terminal set $T$.
In that context,
Jiang, Huang, Saranurak and
Zhang~\cite{DBLP:conf/esa/JiangHSZ22}
introduced connectivity-$c$ mimicking hypergraph networks for
hyperedge cuts.
They showed that $O(|T|c^3)$ hyperedges suffice to preserve
minimum hyperedge cuts between any two terminal subsets
up to threshold $c$.
They also gave a polynomial-time construction with
$O(|T|c^3\log^{3/2}(2|T|c))$ hyperedges,
without requiring either the rank or $c$ to be fixed.
Cho and Oh~\cite{DBLP:conf/isaac/Cho024}
subsequently extended this framework to hypergraph multiway (edge) cuts,
preserving minimum multiway cut values for unrestricted terminal
partitions, with
$|T|c^{O(r\log c)}$ hyperedges, where $r$ is the maximum hyperedge size in the input hypergraph, referred to as the {\em rank} of the hypergraph. Their running time is polynomial for fixed $r$ and $c$.
Our result removes the rank dependence for unrestricted terminal partitions, bounds the total output encoding
length, and gives a deterministic ``purely'' polynomial-time
construction without fixing either parameter.

We show in Theorem~\ref{thm:hypergraph} that
given a hypergraph terminal network $(H,T)$ and a connectivity upper bound
$c$, one can construct a
connectivity-$c$ hyperedge multicut-mimicker of  size
$|T|^{O(1)} \cdot c^{O(\log c)}$.
The bound is independent of the rank of the hypergraph, that is, the maximum
hyperedge size.
 See Table~\ref{tab:hypergraph-mimicking-comparison} for a comparison of our result with previous ones, on hypergraphs.

For a hypergraph terminal network $(H,T)$ and a
partition $\mathcal T$ of $T$, a
$\mathcal T$-edge-\mwc\ is a set
$\mathcal X\subseteq E(H)$ whose deletion leaves
every connected component containing terminals
from at most one part of $\mathcal T$.
Write $\opt_H^{\mathrm{edge}}(\mathcal T)$ for its
minimum size.
A connectivity-$c$ hyperedge multicut-mimicking
network $(H',T)$ preserves
$\min\{\opt_H^{\mathrm{edge}}(\mathcal T),c+1\}$
for every partition $\mathcal T$ of $T$.

  \begin{restatable}[Edge multicut-mimicking hypergraph network]
    {theorem}{thmhypergraph}
    \label{thm:hypergraph}
    Given a hypergraph terminal network $(H,T)$ 
    and a positive integer $c$, there is a deterministic
    polynomial-time algorithm that outputs a hypergraph terminal
    network $(H',T)$ such that $(H',T)$ is a connectivity-$c$
    hyperedge multicut-mimicking network of $(H,T)$.
    The size of $H'$ is
    $|T|^{O(1)} \cdot c^{O(\log c)}$.
    \end{restatable}

Given that our result for graphs is for {\em vertex} multicut-mimicking, one is tempted to ask if an analogous result holds for hypergraphs and vertex multiway cuts.
That is, is there a vertex multicut-mimicking hypergraph whose size depends (quasi-polynomially) on $c$ and $|T|$?
The answer to this depends on what notion of vertex deletion we choose for hypergraphs. There are two standard definitions of vertex deletion in hypergraphs: weak vertex deletion and strong vertex deletion.
Under {\em weak vertex deletion}, deleting a set of vertices $X$ from the hypergraph results in a new hypergraph where the set $X$ is removed from every hyperedge.
Under {\em strong vertex deletion}, deleting a set of vertices $X$ from the hypergraph, results in a new hypergraph where each hyperedge that intersects with $X$ is deleted.

It is easy to observe that under the notion of weak vertex deletion, hypergraphs can be looked at as graphs on the same vertex set, and where for each hyperedge there is a clique on the corresponding vertices in the graph.
Thus, under weak vertex deletion,
from Theorem~\ref{thm:budgeted-vertex-mimicking} we get a connectivity-$c$ (weak) vertex multicut-mimicking hypergraph terminal network of size $|T|^{O(1)} \cdot c^{ O( \log c)}$.
But the situation is drastically different under the strong vertex deletion notion.

Under the strong vertex deletion notion, we observe that, 
deterministic polynomial-time construction of a strong
vertex mimicking (not even multicut-mimicking) hypergraph terminal network with {\em any} computable bound on its size depending only on $|T|$ and $c$  (not just quasi-polynomial) would imply
$\mathrm{FPT}=\mathrm{W}[2]$.
See Appendix~\ref{app:strong-vertex-deletion} for this observation.

\paragraph{Applications to kernelization.}
Efficient pre-processing can reduce the amount of data (equivalently the size of the instance) that an algorithm for an NP-hard problem must handle.
Kernelization provides a framework for proving guarantees on the size of the instance that remains after efficient pre-processing.

A {\em kernelization algorithm} is a polynomial time algorithm that reduces an instance of a parameterized, usually NP-hard, problem to an equivalent instance,
whose size is
bounded by a function of only the parameter.
A {\em polynomial kernel} requires this size bound to be a polynomial function of the parameter.

\textsc{(Vertex) Multiway Cut} and \textsc{(Vertex) Multicut} are two fundamental NP-hard generalizations of the classical
{\sc Minimum $s$--$t$ Separator} problem.
In \textsc{(Vertex) Multiway Cut}, the objective is to separate
every pair of distinct specified terminals.
In \textsc{(Vertex) Multicut}, only a specified collection of
$p$ terminal pairs must be separated.
Here are the formal definitions.

    \defparproblem{\textsc{Vertex Multiway Cut}}
    {An undirected graph $G$, a set $T\subseteq V(G)$ of terminals
    and a non-negative integer $k$}
    {$k$}
    {Does there exist a set $X\subseteq V(G)\setminus T$
    of size at most $k$ such that every connected component
    of $G-X$ contains at most one terminal?}

 \defparproblem{\textsc{Vertex Multicut}}
    {An undirected graph $G$, a set
    $\mathcal T\subseteq\binom{V(G)}2$ of size $p$
    and a non-negative integer $k$}
    {$k,p$}
    {Does there exist a set
    $X\subseteq V(G)\setminus
     \bigcup_{\{s,t\}\in\mathcal T}\{s,t\}$
    of size at most $k$ such that in $G-X$ there is no $s$-$t$ path for any $(s,t) \in \mathcal{T}$?
    }

The question of the existence of a polynomial kernel for {\sc (Vertex) Multiway Cut} is central to the area of kernelization~\cite{Book_kernelization_FLSZ19,DBLP:journals/csr/Jansen26}, as also reflected in a recent survey on Kernelization by Jansen~\cite{DBLP:journals/csr/Jansen26}.

In this work, we consider the known vertex-deletion versions of these problems, where the
solution contains at most $k$ vertices and terminals
are {\em undeletable}.
For \textsc{Vertex Multiway Cut}, the parameter is $k$;
for \textsc{Vertex Multicut}, we use $k+p$.
We note that if the terminals are deletable, using the cut-covering lemma and matroid-based tools developed by Kratsch and Wahlstr\"om~\cite{DBLP:journals/jacm/KratschW20}, {\sc Multiway Cut} already admits  a randomized polynomial kernel.
Thus, the difficulty for the {\sc Vertex Multiway Cut} problem 
lies when the terminals are undeletable.

Interestingly, we can make progress on the kernelization question for these problems that do not allow deleting terminals, by still using our vertex multicut-mimicking networks that allow deletable terminals. The trick is simple and uses a result by~\cite{DBLP:journals/disopt/Guillemot11a} that we will see later during the proofs.

Polynomial kernels for the undeletable terminals case and the edge-deletion case 
are known when the number of terminals and terminal pairs respectively is fixed.
Kratsch and Wahlstr\"om
\cite{DBLP:journals/jacm/KratschW20}
gave a randomized kernel with $O(k^{s+1})$ vertices for
\textsc{Vertex Multiway Cut} with $s$ terminals, and a
randomized kernel with
$O(k^{\lceil\sqrt{2p}\rceil+1})$ vertices for
\textsc{Vertex Multicut} with $p$ terminal pairs.
A recently announced result of Koana and
Kumabe~\cite{DBLP:journals/corr/abs-2609-29141}
implies
deterministic polynomial kernels in these
fixed-terminal settings.

For edge deletion versions only, Wahlstr\"om
\cite{DBLP:journals/talg/Wahlstrom22}
obtained quasi-polynomial kernels through multicut-mimicking networks.
Combined with the improved approximation of {\sc Small Set Expansion} problem by Anand, Lee, Li, and
Saranurak~\cite{DBLP:conf/stoc/0001L0S24},this gives
randomized kernels of size $k^{O(\log^2 k)}$ for
\textsc{Edge Multiway Cut} and
$(k+p)^{O(\log^2(k+p))}$ for \textsc{Edge Multicut}.
Analogous results for vertex-deletion variants were posed open and have remained open since, 
and we resolve them in this work.
In fact, we obtain better bounds than the ones in the edge-deletion case.
Since the vertex-deletion variants generalize the edge-deletion variants, 
our results also improve these edge-deletion variants bounds.

\begin{table}[!htbp]
\centering
\small
\setlength{\tabcolsep}{3pt}

\begin{tabular}{@{}
    L{0.15\linewidth}
    L{0.12\linewidth}
    L{0.17\linewidth}
    L{0.26\linewidth}
    L{0.42\linewidth}@{}}
\hline
\textbf{Cuts}
& \textbf{Terminal Parts}
& \textbf{Approximation}
& \textbf{Construction}
& \textbf{Size} \\
\hline

Hyper\-edge
& $2$
& $1$
& Randomized\newline  $O(p^{1+o(1)} +n (rc\log n)^{O(rc)} \log m)$ 
& $O(kc^3)$ hyperedges${}^{\dagger}$
  \cite{DBLP:conf/esa/JiangHSZ22} \\

& &
& Randomized 
& $O(kc^3\log^{3/2}(2kc))$ hyperedges${}^{\dagger}$
  \cite{DBLP:conf/esa/JiangHSZ22} \\

\noalign{\hrule height 1.2pt}

Hyper\-edge
& Unbounded
& $1$
& Deterministic\newline
 $p^{1+o(1)}+k(c^r\log n)^{\widetilde O(rc)}m$
& $kc^{O(r\log c)}$ hyperedges${}^{\dagger}$
  \cite{DBLP:conf/isaac/Cho024} \\

\textcolor{red}{Hyper\-edge}
& \textcolor{red}{Unbounded}
& \textcolor{red}{$1$}
& \textcolor{red}{Deterministic}
& \textcolor{red}{
  $k^{O(1)}c^{O(\log c)}$
  encoding length${}^{\dagger}$\newline
  (Theorem~\ref{thm:hypergraph})} \\

\hline

\textcolor{red}{Weak vertex}
& \textcolor{red}{Unbounded}
& \textcolor{red}{$1$}
& \textcolor{red}{Deterministic}
& \textcolor{red}{
  $k^{O(1)}c^{O(\log c)}$
  encoding length${}^{\dagger}$\newline
  (Theorem~\ref{thm:budgeted-vertex-mimicking},
  via clique hyperedges)} \\

\textcolor{red}{Strong vertex}
& \textcolor{red}{$2$}
& \textcolor{red}{$1$}
& \textcolor{red}{--}
& \textcolor{red}{
  No computable vertex bound\newline
  (Appendix~\ref{app:strong-vertex-deletion})} \\

\hline
\end{tabular}

\caption{
Hypergraph sparsifiers and mimicking networks.
Here $k:=|T|$, $n:=|V(H)|$, $m:=|E(H)|$,
$r$ is the rank, and $p:=\sum_{e\in E(H)}|e|$.
The construction column describes the whether the algorithm is randomized or deterministic, and unless mentioned otherwise the time is purely polynomial.
A dagger denotes preservation after truncation at $c+1$,
normalizing threshold conventions across references.
Red denotes this work.
For strong deletion, unless $\mathrm{FPT}=\mathrm{W}[2]$,
no deterministic polynomial-time construction admits
any computable vertex bound depending only on $k,c$.
This obstruction concerns unrestricted rank and already
holds for one binary partition, with paid terminal
deletion and no free terminal deletions.
}
\label{tab:hypergraph-mimicking-comparison}
\end{table}

\begin{table}[!htbp]
\centering
\small
\setlength{\tabcolsep}{3pt}

\begin{tabular}{@{}
    L{0.12\linewidth}
    L{0.15\linewidth}
    L{0.13\linewidth}
    L{0.14\linewidth}
    L{0.40\linewidth}@{}}
\hline
\textbf{Deletion}
& \textbf{Problem}
& \textbf{Parameter}
& \textbf{Construction}
& \textbf{Size} \\
\hline

Edge
& \textsc{Multi\-way Cut}
& $k$
& Randomized
& $k^{O(\log^2 k)}$ size
  \cite{DBLP:journals/talg/Wahlstrom22,
        DBLP:conf/stoc/0001L0S24} \\

Vertex
&
& $k$, fixed $t$
& \textcolor{blue}{Deterministic}
& $O(k^{t+1})$ vertices
  \cite{DBLP:journals/jacm/KratschW20,
        DBLP:journals/corr/abs-2609-29141} \\

&
& \textcolor{red}{$k$}
& \textcolor{red}{Deterministic}
& \textcolor{red}{
  $k^{O(\log k)}$ size\newline
  (Theorem~\ref{thm:mwc-kernel})} \\

\noalign{\hrule height 1.2pt}

Edge
& \textsc{Multi\-cut}
& $k,p$
& Randomized
& $(k+p)^{O(\log^2(k+p))}$ size
  \cite{DBLP:journals/talg/Wahlstrom22,
        DBLP:conf/stoc/0001L0S24} \\

Vertex
&
& $k$, fixed $p$
& \textcolor{blue}{Deterministic}
& $O(k^{\lceil\sqrt{2p}\rceil+1})$ vertices
  \cite{DBLP:journals/jacm/KratschW20,
        DBLP:journals/corr/abs-2609-29141} \\

&
& \textcolor{red}{$k,p$}
& \textcolor{red}{Deterministic}
& \textcolor{red}{
  $p^{O(1)}k^{O(\log k)}$ size\newline
  (Theorem~\ref{thm:mc-kernel})} \\

\noalign{\hrule height 1.2pt}

Edge
& \textsc{Group Feedback Set}
& $k,|\Gamma|$
& Randomized
& $|\Gamma|^{O(1)}k^{O(\log^2 k)}$ size
  \cite{DBLP:journals/talg/Wahlstrom22,
        DBLP:conf/stoc/0001L0S24} \\

Vertex
&
& $k$, fixed $\Gamma$
& \textcolor{blue}{Deterministic}
& $O((k\log k\log\log k)^{|\Gamma|+1})$ vertices
  \cite{DBLP:journals/jacm/KratschW20,
        DBLP:journals/corr/abs-2609-29141} \\

&
& \textcolor{red}{$k,|\Gamma|$}
& \textcolor{red}{Deterministic}
& \textcolor{red}{
  $|\Gamma|^{O(1)}k^{O(\log k)}$
  encoding length\newline
  (Theorem~\ref{thm:kernel-group-fvs})} \\

\hline
\end{tabular}

\caption{
Kernelization results for edge and vertex deletion.
Here $k$ is the deletion budget,
$t$ is the number of Multiway Cut terminals and
$p$ is the number of Multicut requests.
Terminals are undeletable in the vertex cut problems.
All constructions take polynomial time, with the
quantities explicitly marked fixed held constant.
In the uniform group-feedback rows, $\Gamma$ is
supplied by its multiplication table.
Blue denotes deterministic consequences of the
representative-family machinery of Koana and
Kumabe~\cite{DBLP:journals/corr/abs-2609-29141}.
Red denotes this work.
}
\label{tab:kernelization-comparison}
\end{table}

Our vertex multicut-mimicking network of Corollary~\ref{thm:full-vertex-mimicking} gives a deterministic
kernel for \textsc{Vertex Multiway Cut} of size $k^{O(\log k)}$
(Theorem~\ref{thm:mwc-kernel}).
This answers the quasi-polynomial-sized kernel question 
posed by Wahlstr\"om~\cite{DBLP:journals/talg/Wahlstrom22}
and repeated in Jansen's survey~\cite{DBLP:journals/csr/Jansen26}.
The connection here is that an instance of
\textsc{Vertex Multiway Cut} can be reduced to a
multiway-cut query on a network with $O(k)$ deletable
terminals~\cite{DBLP:journals/jacm/KratschW20}.
Replacing this network by our mimicking network therefore
bounds the entire reduced instance by a quasi-polynomial function of  $k$.

  \begin{restatable}[Vertex Multiway Cut]{theorem}{thmkernelmwc}
\label{thm:mwc-kernel}
\textnormal{\textsc{Vertex Multiway Cut}} admits a deterministic
kernel of size
$k^{O(\log k)}$.
\end{restatable}

The connectivity-$c$ version presented in Theorem~\ref{thm:budgeted-vertex-mimicking} gives a more refined
consequence for \textsc{Vertex Multicut}:
a deterministic kernel of size
$p^{O(1)} \cdot k^{O(\log k)}$
(Theorem~\ref{thm:mc-kernel}).
The dependence on the number of terminal pairs is polynomial in the kernel size, while the quasi-polynomial factor depends only on the deletion budget.
    
    \begin{restatable}[Vertex Multicut]{theorem}{thmkernelmc}
    \label{thm:mc-kernel}
    \textnormal{\textsc{Vertex Multicut}} admits a deterministic
    kernel of size
    $p^{O(1)} \cdot k^{O(\log k)}$.
    \end{restatable}

Finally, our construction also applies to the
\textsc{Group Feedback Vertex Set} problem. The problem is defined below.

\defparproblem{\textsc{Group Feedback Vertex Set}}
{A finite group $(\Gamma,\otimes)$ specified by its
multiplication table, a directed graph $D$, an arc labelling
$\phi:E(D)\to\Gamma$, and a non-negative integer $k$}
{$k+|\Gamma|$}
{Does there exist $X\subseteq V(D)$ of size at most $k$
such that $D-X$ has a consistent labelling, that is,
a function $\pi:V(D)\setminus X\to\Gamma$ satisfying
$\pi(u)\otimes\phi((u,v))=\pi(v)$
for every $(u,v)\in E(D-X)$?}

    \begin{restatable}[Group FVS]{theorem}{thmgroupfvs}
    \label{thm:kernel-group-fvs}
    \textnormal{\textsc{Group Feedback Vertex Set}}
    admits a deterministic kernel of size
    $|\Gamma|^{O(1)} \cdot k^{O(\log k)}$.
    \end{restatable}

{\sc Group Feedback Vertex Set} had a kernel of size $k^{O(|\Gamma| +1) }$
\cite{DBLP:journals/jacm/KratschW20,
      DBLP:journals/corr/abs-2609-29141}.
Our bound instead separates a polynomial dependence on
the group order from a quasi-polynomial dependence on
the deletion budget.
As a final remark, {\sc Group Feedback Vertex Set} generalizes many interesting problems 
like {\sc Odd Cycle Transversal} and {\sc Signed Graph Balancing} 
(see~\cite{DBLP:journals/algorithmica/CyganPP16}).

\paragraph*{Related work.}

Curiously, a polynomial mimicking network-style
construction is known for the \emph{path-packing dual}
of the multiway cut problem.
Let $(G,T)$ be a terminal network and $\cT$ a partition
of $T$.
Let the \emph{$\cT$-path packing number} in $G$ be the
maximum number of pairwise vertex-disjoint paths whose
endpoints are in different parts of $\cT$ and which are
internally disjoint from $T$.
Note that this is the natural dual to a $\cT$-mwc for
the setting of deletable terminals.
Wahlström~\cite{DBLP:conf/soda/Wahlstrom24} showed that
in randomized polynomial time, you can construct a
terminal network $(G',T)$ such that $G$ and $G'$ have
identical $\cT$-path packing numbers for every partition
of $T$, and $|V(G')|=O(|T|^3)$.
Let us for now refer to $(G',T)$ as a
\emph{Mader-mimicking network} for $(G,T)$.\footnote{
The definition of a Mader-mimicking network
in~\cite{DBLP:conf/soda/Wahlstrom24} is slightly different,
but~\cite{DBLP:conf/soda/Wahlstrom24} also shows the
above result.
}
Such a network does not preserve the sizes of
$\cT$-mwcs precisely, but was shown to imply a
\emph{2-approximate} multicut-mimicking network with
$O(|T|^3)$ vertices~\cite{DBLP:conf/soda/Wahlstrom24},
that is, the size of a multiway cut for every partition
of $T$ is preserved up to a factor of 2.

Mimicking networks are also known as
\emph{vertex sparsifiers with Steiner nodes},
especially in a context where the \emph{quality}
of the sparsification is traded against a smaller
replacement graph $G'$.
Moitra~\cite{DBLP:journals/siamcomp/Moitra13,
DBLP:conf/stoc/LeightonM10} studied cut sparsifiers
with no Steiner nodes, i.e.\ in our setting,
replacement terminal networks $(G',T)$ with $V(G')=T$.
The attainable quality is between
$O(\log k/\log\log k)$ and $\Omega(\sqrt{\log k}/\log\log k)$~\cite{
DBLP:conf/focs/MakarychevM10,
DBLP:journals/siamcomp/EnglertGKRTT14}.
A more ambitious goal is \emph{flow sparsifiers},
where we want to preserve the value of any multicommodity
flow over the terminals~\cite{DBLP:conf/stoc/LeightonM10,
DBLP:conf/soda/AndoniGK14}.
Recently, it was shown that quality-$(1+\varepsilon)$
flow sparsifiers require unbounded size~\cite{
DBLP:conf/soda/KrauthgamerM23,
DBLP:conf/soda/ChenT24a,
DBLP:conf/stoc/ChenTY26}.
See also Chuzhoy~\cite{DBLP:conf/stoc/Chuzhoy12}
for quality-$O(1)$ flow sparsifiers of size
$k^{O(\log\log k)}$ where $k$ is the terminal capacity.
Mimicking network and sparsification questions have also
been studied for hypergraphs~\cite{
DBLP:conf/focs/ChenKN20,
DBLP:conf/focs/KhannaP024,
DBLP:conf/esa/JiangHSZ22,
DBLP:conf/isaac/Cho024}.

In a different sense of sparsification -- generalising
edge sparsifiers rather than vertex sparsifiers --
the task of sparsifying a CSP was connected in a
breakthrough result to the notion of
\emph{non-redundancy}~\cite{DBLP:conf/stoc/BrakensiekG25},
leading to a range of improvements to CSP sparsification,
code sparsification, and related areas,
cf.~\cite{DBLP:conf/icalp/BrakensiekGP26,
DBLP:conf/soda/KhannaPS24,
DBLP:conf/stoc/KhannaP025,
DBLP:conf/icalp/KhannaP025}.

\section{Technical overview}
    \label{sec:technical-overview}
In this section, we only focus on giving the proof overview of the (connectivity-$c$) vertex multicut-mimicking result, that is of Corollary~\ref{thm:full-vertex-mimicking} directly.

Before moving on to the details,
we would like to emphasize that the jump from the edge multicut-mimicking to vertex multicut-mimicking is not merely a technicality.
As flagged by
Wahlström~\cite{DBLP:journals/talg/Wahlstrom22},
the vertex-deletion case is significantly more complex
than the edge-deletion case.
See below for a more precise discussion,
but there is a sense in which multiway edge cuts are
naturally sparse, in that deleting $k$ edges from a graph
can only affect $2k$ vertices and so with a large number
of components, each component is incident to only a
smaller number of deleted edges.
For the vertex-deletion case, this is not true - it can
easily happen for a $\cT$-mwc $X$ that every component
of $G-X$ sees all of $X$.
Another way to illustrate the challenge is via the
path-packing dual discussed above:
for \textsc{Edge Multiway Cut}, a 2-approximate
path-packing dual follows from the union of
\emph{isolating min-cuts}, and is hence a property of
standard 2-way max-flows, while for
\textsc{Vertex Multiway Cut} the 2-approximate
path-packing dual has the more complex structure of
``Mader paths''~\cite{SchrijverBook,DBLP:conf/soda/Wahlstrom24}.

We now proceed to the technical overview.
We first recall the methods behind
    the quasi-polynomial kernel for the edge-deletion case~\cite{DBLP:journals/talg/Wahlstrom22}.
    Let $(G,T)$ be a terminal network and let $k$ be the capacity of $T$.
    A \emph{query} of $(G,T)$ is a pair $(T_0, \cT)$ where
    $T_0 \subseteq T$ is a set of deleted terminals and $\cT$ a partition
    of $T \setminus T_0$. A \emph{solution} to $(T_0, \cT)$ is a
    minimum $\cT$-\mwc\ in $G-T_0$. Thus the goal in producing a
    multicut-mimicking network for $(G,T)$ is to construct a
    network $(G',T)$ which preserves the size of a solution for every
    query $(T_0, \cT)$ over $(G,T)$. In~\cite{DBLP:journals/talg/Wahlstrom22},
    the solution is an edge cut; in this paper, it will be a vertex cut. 
    (The use of a deletion set $T_0$ was not considered in~\cite{DBLP:journals/talg/Wahlstrom22},
    but becomes important in the vertex-deletion case.)
    Say that an edge $x=e$ (for edge cuts), respectively a vertex $x=v$ (for
    vertex cuts), is \emph{essential} in $(G,T)$ if there is
    a query $(T_0, \cT)$ for which every minimum solution deletes $x$. 
    
    The kernel of Wahlström~\cite{DBLP:journals/talg/Wahlstrom22}, like
    our current result, is based around a recursive replacement scheme.
    Let $f(k)=k^{\log^{O(1)} k}$ be the size bound on the
    multicut-mimicking network produced in~\cite{DBLP:journals/talg/Wahlstrom22}. 
    Then progress is made in one of two ways. 
    \begin{enumerate}
    \item If we find a sufficiently sparse cut $C=\delta(S)$,
      such that $|S| > f(|T \cap S|+|C|)$, then we can treat
      $S$ as a recursive instance $(G_S, T_S)$ of edge multicut sparsification,
      with terminals $T_S=(S \cap T) \cup N(S)$, and replace $G[S]$ 
      in $G$ by a mimicking network for $(G_S, T_S)$.
    \item If no such cut is possible, then we deploy a matroid-based
      marking scheme that marks at most $f(k)$ edges, with a guarantee 
      that every essential edge is marked.
    \end{enumerate}
    Once the computation reaches the end of the marking stage, we can thus
    select any unmarked edge $e$, replace $(G, T)$ by $(G/e, T)$,
    where $G/e$ is the result of contracting $e$ in $G$,
    and return a multicut-mimicking network for $(G/e, T)$.
    
    For edge cuts, the observation that underlies the result is the
    following. Let $Z$ be a solution to a query $(T_0, \cT)$ and order the
    connected components of $G-Z$ as $V_1, \ldots, V_r$ in non-increasing
    order of cardinality of $B_i=V_i \cap (T \cup V(Z))$. Then for each
    $i=1, \ldots, r$, $V_i$ with boundary $B_i$ is a candidate for
    recursive replacement. Thus, if we put aside the difficulty of finding
    all sparse cuts and settle for a non-constructive existence result,
    we may assume that $|V_i| \leq f(|B_i|)$ for every $i=1, \ldots, r$.
    Furthermore, $\sum_i |B_i| \leq |T|+|V(Z)| \leq 3k$,
    and in particular $|B_i| \leq 3k/i$ by the ordering of $V_i$.
    With $f(k)=k^{\Omega(\log k)}$ we get 
    $f(k/c) \leq f(k)/c^{\Omega(\log k)}=f(k)/k^{\Omega(\log c)}$,
    so the size $|V_i|$ is discounted by a factor polynomial in $k$.
    Thus the cut $G-Z$ has a \emph{small tail} in that there is a constant
    $s$ such that the total number of vertices in components $V_s, \ldots, V_r$
    is a $1/\text{poly}(k)$-fraction of all vertices of $G$. 
    
    This small-tail guarantee enables the use of a matroid-based marking
    scheme based on \emph{representative sets}, following previous
    kernelization results for \textsc{Multiway Cut}~\cite{DBLP:journals/jacm/KratschW20}
    (see also~\cite{Book_kernelization_FLSZ19}).
    Specifically, if the size of the tail is $n_s=|\bigcup_{i=s}^r V_i|$, 
    then using these methods we can mark $O(k^s n_s)$ edges with a
    guarantee that all essential edges are marked. The small-tail property
    above implies that there is a constant $s$ such that this is at most
    $f(k)$. 
    
    To make this constructive, Wahlström~\cite{DBLP:journals/talg/Wahlstrom22} showed 
    that for the edge-deletion case, sparse cuts according to item 1 above
    can be found using an approximation algorithm for \textsc{Small Set Expansion} (SSE).
    The quality of the approximation algorithm influences the size bound
    $f(k)$ of the mimicking network; for our purposes, the best result is
    a recent $O(\log k)$-approximation~\cite{DBLP:conf/stoc/0001L0S24}, 
    which gives the currently tightest bound of $f(k)=k^{O(\log^2 k)}$
    for a polynomial-time constructive algorithm.
    
    Compared to the above, the vertex-deletion case encounters the
    following additional difficulties.
    \begin{enumerate}
    \item The approximation guarantees for \textsc{Small Set Vertex Expansion}
      are far inferior to those of SSE, to the point that the
      approximation guarantee would not be usable in a reduction scheme,
      and there is evidence that this is a fundamental obstacle~\cite{DBLP:conf/stoc/0001L0S24}.
    \item The small-tail guarantee fails for vertex separations,
      since a set $Z$ of $k$ vertices can contribute to the boundary of every
      connected component of $G-Z$.
    \end{enumerate}
    In addition to overcoming these obstacles, our paper makes the following additional
    improvements.
    \begin{enumerate}[resume]
    \item Our size bound in terms of $k=|T|$ for the size of the
      multicut-mimicking network is $k^{O(\log k)}$,  
      improving on the previous $f(k)=k^{O(\log^2 k)}$.
    \item Our algorithm is deterministic rather than randomized, using
      recent results of Koana and Kumabe~\cite{DBLP:journals/corr/abs-2609-29141},
      with additional efforts made to ensure that our algorithm is fully deterministic.
      
    \item In addition, we provide a \emph{connectivity-$c$ multicut-mimicking network}, 
    where we take two parameters, $k=|T|$ and a second parameter $c$, and only
      guarantee to retain the solution size for queries with a solution of
      size at most $c$. For this case, we provide a bound on the mimicking
      network of $k^{O(1)} \cdot c^{O(\log c)}$, so the superpolynomial
      behavior is confined to the second parameter $c$.
    \end{enumerate}
    
    Of these, the main obstacle is the second one, as
    multicuts $G-Z$ with ``large tails'' provide
    significant additional challenges over the small-tail case.
    However, the absence of an approximation algorithm for sparse separations 
    requires additional care as well.

    \subsection{Algorithm overview}
    
    Let us put aside the bounded connectivity condition and focus on the
    construction of a vertex multicut-mimicking network. 
    We define a function $\bound(k)=k^{O(\log k)}$ such that for 
    every terminal network $(G,T)$, our algorithm produces a vertex
    multicut-mimicking network $(G',T)$ of $(G,T)$ with
    $|V(G')| \leq \bound(|T|)$; the precise value of $\bound(k)$ is given
    in Section~\ref{sec:outline}.
    Let $(G,T)$ be a terminal network and let $n=|V(G)|$ and $k=|T|$. 
    Our algorithm is inspired by the broad strategy of~\cite{DBLP:journals/talg/Wahlstrom22},
    but uses several new ideas  to handle the obstructions named above. 
    In outline, our algorithm has the following structure.
    
    \begin{enumerate}
    \item The procedure can at any time find a region whose size can be
      recursively reduced, similar to sparse cuts in the edge-deletion
      case. If such a region is found, the procedure terminates and returns it.
      This is done without relying on good approximations for {\sc Small Set Vertex Expansion}, unlike in the edge-case.
    \item If no such region is found, we employ two separate marking schemes:
      \begin{enumerate}
      \item A marking scheme for solutions with a small tail,
        similar to the one used previously~\cite{DBLP:journals/talg/Wahlstrom22}.
      \item A second marking scheme for queries with no small-tail solution,
        whose correctness instead relies on the large-tail property of the solution.
      \end{enumerate}
    \end{enumerate}
    
    Thus, instead of using an approximation algorithm to find sparse separations
    as~\cite{DBLP:journals/talg/Wahlstrom22}, the algorithm interleaves
    the detection of reducible regions with the rest of its operations,
    verifying connectivity properties as it goes along. 
    
    To be more precise, let $0\leq h<k$. A set $R \subseteq V(G)$ is an
    \emph{$h$-recursively useful region} if
    $R \cap T = \emptyset$, $N_G[R] \neq V(G)$, $|N_G(R)| \leq h$,
    and $|R| > 2 \cdot \bound(|N_G(R)|)$. 
    Also fix an upper bound $h\le\tfrac{14}{15}k$ on $h$.
    Then, whenever the algorithm detects an $h$-recursively useful region,
    it can recursively reduce its size without significant cost to the running time.
    Furthermore, let $(T_0,\cT)$ be a query and $X \subseteq V(G)$ a
    solution to $(T_0, \cT)$. Let the components of $G-(X\cup T_0)$ that contain
    terminals have vertex sets $V_1, \ldots, V_r$, ordered
    non-increasingly by cardinality. Then we say that $X$ has a
    \emph{small tail} if $\sum_{i=4}^r |V_i| \leq n/(12k^3)$. 
    We say that $(T_0,\cT)$ is a \emph{small-tail query} if there exists
    a solution with a small tail, and a \emph{no-small-tail query} otherwise.
    The marking procedures for small-tail queries and no-small-tail queries
    are given in Lemma~\ref{lem:smalltail} and Lemma~\ref{lem:nosmalltail}, respectively.
    We defer a discussion of the small-tail case to the main paper, given
    its similarity to previous work, and focus on the no-small-tail case.

    \subsection{The no-small-tail marking procedure}
    
    The marking procedure for the no-small-tail case is made up of several
    steps. (The names below are mnemonics for this overview only, and are
    not used in the main text.)
    \begin{enumerate}
    \item A \emph{degree-reduction} step (Section~\ref{sec:bounding-degree}), ensuring that $\deg_G(v) \leq 2k+1$ for
      every vertex in $G$.
    \item A \emph{solution highlighting} phase (Lemma~\ref{lem:marking-a-large-fraction}), in which we mark a set
      $M_1$ of vertices such that for any large-tail solution $X$,
      $|X \setminus M_1| \leq (4/15)k$.
    \item A \emph{shattering} phase (Lemma~\ref{lem:second-cut-marking}), in which we mark a second set $M_2$
      of vertices with the property that all essential vertices outside $M_2$ for
      no-small-tail queries are found in connected components of
      $G-M_2$ of size at most $3 \cdot \bound(4k/5)$.
    \end{enumerate}
    Given the guarantee in the last step we may simply mark all vertices
    in small connected components of $G-M_2$. Since $\bound(t)$ is quasi-polynomial,
    $\bound(4k/5)=\bound(k)/k^{\Omega(1)}$, and the degree
    reduction implies that the number of components is at most $2k \cdot |M_2|$.
    
    Throughout, we use the cut-covering lemma a.k.a mimicking network for vertex cuts~\cite{DBLP:journals/jacm/KratschW20}
    as a processing tool (in a deterministic version derived from
    Koana and Kumabe~\cite{DBLP:journals/corr/abs-2609-29141}).
    Specifically, given a graph $G$ with a set of terminals $S$,
    $\cutcover(G,S)$ returns a set $Z \subseteq V(G)$ with $|Z|=O(|S|^3)$
    such that for every partition $S=A \cup B$, there is a minimum
    $(A,B)$-separation $C$ with $C \subseteq Z$. The terminals are
    deletable, i.e.\ $C$ may overlap $A \cup B$.
    
    In addition we have the following procedure (Lemma~\ref{lem:exposed-region}).
    Let $M \supseteq T$ be a marked set of vertices,
    e.g.~$M=\cutcover(G,S)$ for some $S$,
    and let $D$ be a connected component of $G-M$ with $|D| > 3 \cdot \bound(h)$
    for some $h \leq k$. Let $C$ be a minimum $(D,T)$-separator in $G$.
    Then either $|C| > h$, and there is a large $(D,T)$-flow in $G$,
    or $C$ carves out an $h$-recursively useful region with $D$. 
    We say that $M$ is an \emph{$h$-large-size-large-flow modulator}
    if only the former case applies, i.e.\ every ``large'' component
    of $G-M$ has ``large'' flow to $T$.

    We defer the degree-reduction step to Section~\ref{sec:bounding-degree}
    and focus on the two main phases.
    
\paragraph*{Solution highlighting.}
Let $(T_0, \cT)$ be a no-small-tail query and let $X \subseteq V(G-T_0)$
be a solution. Let $V_1, \ldots, V_4$ be the largest four $T$-components of $G-(X \cup T_0)$.
Since $X$ has a large tail and at most $k$ components of $G-(X \cup T_0)$ contain terminals, $|V_i| \geq n/(12k^4)$ for each $i \in [4]$.
Our goal is to sample a set of vertices $S \subseteq V(G)$,
$|S|=k^{O(1)}$, such that for each $i \in [4]$,
$\lambda_G(V_i \cap (S \cup T), T) \geq \tfrac {14} {15} k$. 
We then compute $M_1=\cutcover(G, S \cup T)$ and argue that
$|X \setminus M_1| \leq \tfrac 4 {15} k$. 

The details of the latter argument are somewhat involved, but in overview,
for $i \in [4]$ let $S_i=V_i \cap (S \cup T)$,
let $A=S_1 \cup S_2$ and $B=S_3 \cup S_4$, and 
let $C \subseteq M_1$ be a minimum $(A,B)$-separator. 
Then $A$ and $B$ separately have two large flows into $X$ by the flow
assumption, and a careful counting argument (also accounting for the
possible behaviors of $|T \cap V_i|$, $i \in [4]$)
yields that every possible minimum $(A,B)$-separator $C$
must have $|X \setminus C| \leq 4k/15$. 

To find $S$, we argue that it essentially suffices to 
sample a large enough set of random vertices, followed by an
application of Lemma~\ref{lem:exposed-region} to check for recursively
useful regions. Indeed, intuitively, a large enough sample should hit
a large connected component $V_i$ in $G-X$ a large number of times,
and then either $\lambda_G(S \cap V_i, T) > h$ or
Lemma~\ref{lem:exposed-region} finds a sparse separation to recurse
on. (The actual argument, which uses a deterministic procedure, is
slightly different but uses the same idea.) 

\paragraph*{Shattering.}
We are now in the case that $|X \setminus M_1| \leq \tfrac{4}{15} k$,
Here on, we make use of an important recurring principle. Our
goal is to compute a set $M_2$ with the following property.
\begin{quote}
  For any $h \leq 4k/15$ and any connected component $D$ of $G-M_2$ 
  with $|V(D)| > 3 \cdot \bound(3h)$, if a vertex $x \in V(D)$ is essential
  for a no-small-tail query $(T_0,\cT)$ with solution $X$,
  then $|X \setminus M_1| > h$. 
\end{quote}
This is achieved in two steps. First, since $M_1$ is an
$h$-large-size-large-flow modulator, for any such component we have
$\lambda_G(V(D), T) > 3h$. To complete the property, we show 
Lemma~\ref{lem:relative-lslf}: Let $M_2=\cutcover(G, N_G[M_1])$
and let $x \in V(D)$ be essential for a no-small-tail query $(T_0,\cT)$
with solution $X$. Then $\lambda_G(V(D), T) \leq 3|X \setminus M_1|$. 
(In fact, this holds for any set $A \supseteq T$ in place of $M_1$.)

Given this property, it is enough to mark all vertices in connected
components in $G-M_2$ of size at most $3\bound(4k/5)$,
and given the degree bound there are at most $O(k|M_2|)$ such components.

\paragraph*{Bounded connectivity.}

Finally, for the bounded connectivity case with connectivity bound
$c \leq k/4$, we essentially reuse the above principle. Given that
$|X \setminus M_1| \leq c$, it suffices to mark components of size
$3\bound(3c) = c^{O(\log c)}$. (In fact, to reduce the dependency on
$k$, it suffices to invoke Lemma~\ref{lem:relative-lslf} using
the set $A=T$ instead of $A=M_1$.)

    \paragraph*{Organization of the paper.}
    Section~\ref{sec:prelims} contains preliminaries. Section~\ref{sec:bounding-degree}
    gives the procedure for bounding the degree of the graph. Section~\ref{sec:outline}
    gives the technical outline and supporting lemmas, Section~\ref{sec:no-small-tail-case}
    covers the no-small-tail case and Section~\ref{sec:small-tail-case} covers the small-tail case.
    Sections~\ref{sec:kernelization} and~\ref{sec:hypergraphs} give additional applications to kernelization
    and hypergraph networks, respectively, and Section~\ref{sec:conclusion} concludes the paper.
   
    We would like to point out again that the novelty of this work lies in the ideas in Section~\ref{sec:no-small-tail-case} which uses the setup and some auxiliary lemmas from Section~\ref{sec:outline}. Interested reader can skip Section~\ref{sec:bounding-degree} and assume that the graph has degree bounded by $O(k)$.
    
    \section{Preliminaries}\label{sec:prelims}\label{sec:cut-covering}\label{sec:two-approx}
    
   For every non-negative integer $n$, let
$[n]:=\{1,\ldots,n\}$.
All undirected graphs considered in this work are simple.
For a graph $G$ and $A\subseteq V(G)$, let $ N_G(A):=
 \{v\in V(G)\setminus A:
       uv\in E(G)\text{ for some }u\in A\}$ and $N_G[A]:=A\cup N_G(A)$.
We write $N_G(v):=N_G(\{v\})$.
An $A$--$B$ path has one endpoint in $A$ and the other
in $B$; paths of length zero are allowed.
A $v$--$Q$ path means a $\{v\}$--$Q$ path.
    
    \begin{definition}[$(A,B)$-separator]
        Let $G$ be a graph. For any $A,B \subseteq V(G)$, an {\em $(A,B)$-separator} in $G$ is a set $X \subseteq V(G)$ such that $G-X$ has no path from any vertex of $A \setminus X$ to any vertex of $B \setminus X$. Note that $X$ is allowed to intersect $A \cup B$.
    \end{definition}
    
    Let $\flow_G(A,B)$ denote the size of a minimum $(A,B)$-separator in $G$.
    For any $A \subseteq V(G)$, let $\reach_G(A)$ denote the set of vertices reachable from $A$ in $G$.

    \begin{definition}[$\mathcal{T}$-\mwc]
        Given a graph $G$, a set of terminals $T \subseteq V(G)$ and a partition $\mathcal{T}=\{T_1, \ldots, T_r\}$ of $T$,
        a {\em $\mathcal{T}$-\mwc} in $G$
        is a set $X \subseteq V(G)$ (note that $X$ is allowed to intersect $T$) 
        such that for any $i, j\in [r], 
        i \neq j$, 
        there is no path from any vertex of $T_i$ to any vertex of $T_j$ in $G-X$.
    \end{definition}
    For any $T_0\subseteq T$ and partition $\mathcal T$ of $T\setminus T_0$, a minimum $\mathcal T$-\mwc\ in $G-T_0$ has size at most $|T\setminus T_0|$.
    
    \begin{definition}[Vertex multicut-mimicking network]
        Given a graph $G$ and set $T \subseteq V(G)$ of terminals, 
        a {\em vertex multicut-mimicking network} for $(G,T)$ is a pair 
        $(G',T)$ where $G'$ is a graph, $T \subseteq V(G')$ and,
        for any $T_0 \subseteq T$, 
        and any partition $\mathcal{T}$ of $T \setminus T_0$, 
        the size of a minimum $\mathcal{T}$-\mwc\ in $G - T_0$ 
        is equal to the size of a minimum $\mathcal{T}$-\mwc\ in $G' -T_0$. 
    \end{definition}
    
    \begin{definition}[Terminal network]
    A terminal network is a pair $(G,T)$, where $G$ is an undirected
    graph and $T\subseteq V(G)$ is its set of terminals.
    \end{definition}
    
    \begin{definition}[Essential vertices]
        Let $(G,T)$ be a terminal network.
        Let $T_0 \subseteq T$ and let
        $\mathcal{T}=\{T_1,\ldots,T_r\}$ be a partition of $T\setminus T_0$.
        We say that a vertex $v \in V(G)$ is {\em essential for $(T_0,\mathcal{T})$} in $G$,
        if every minimum $\mathcal{T}$-\mwc\ in $G-T_0$ contains $v$.
         We say that a vertex $v \in V(G)$ is {\em essential for $(G,T)$} in $G$,
         if it is essential for some $(T_0,\mathcal{T})$ where $T_0 \subseteq T$ and $\mathcal{T}$ is a partition of $T \setminus T_0$.

    \end{definition}

    We use the cut-covering lemma of Kratsch and
    Wahlstr\"om~\cite{DBLP:journals/jacm/KratschW20}, in the
    deterministic form of Koana and
    Kumabe~\cite[Theorem~1.2]{DBLP:journals/corr/abs-2609-29141}.
    We need its guarantee also for overlapping terminal sets and after
    deleting a subset of the terminals for free.
    
    \begin{restatable}[Deterministic cut-covering]{lemma}{lemcutcover}
    \label{lem:cut-covering}
    Let $(G,T)$ be a terminal network.
    In deterministic polynomial time one can find
    $M\subseteq V(G)$ such that $T\subseteq M$, $|M|=O(|T|^3)$,
    and, for every $T_0\subseteq T$ and every two, possibly overlapping,
    sets $A,B\subseteq T\setminus T_0$, the set $M\setminus T_0$
    contains a minimum $(A,B)$-separator in $G-T_0$.
    \end{restatable}
    
    \begin{proof}
    If $T=\emptyset$, return $M:=\emptyset$.
    Otherwise, replace every edge of $G$ by two opposite arcs.
    For every $t\in T$, add a source-only copy $t^-$ with the arc
    $t^-t$ and a sink-only copy $t^+$ with the arc $tt^+$.
    Let $D$ be the resulting directed graph and put
    $T^-:=\{t^-:t\in T\}$ and $T^+:=\{t^+:t\in T\}$.
    Koana and
    Kumabe~\cite[Theorem~1.2]{DBLP:journals/corr/abs-2609-29141}
    gives a set $Z\subseteq V(D)$ of size
    $O((|T^-|+|T^+|)^3)=O(|T|^3)$ containing a minimum
    $(A^-,B^+)$-separator for every $A,B\subseteq T$.
    Here $A^-:=\{t^-:t\in A\}$ and $B^+:=\{t^+:t\in B\}$.
    Project each selected copy $t^-$ or $t^+$ to $t$, retain the
    selected original vertices, and add $T$. Call this set $M$.
    
    Every $(A,B)$-separator in $G$ is an $(A^-,B^+)$-separator
    in $D$. Conversely, project any separator of $A^-$ from $B^+$
    to original vertices. This gives an $(A,B)$-separator in $G$ without
    increasing its size: a surviving $A$-$B$ path would lift to a
    surviving $A^-$-$B^+$ path. Thus the minimum values agree,
    including when $A\cap B\ne\emptyset$, and $M$ contains a
    minimum $(A,B)$-separator for every $A,B\subseteq T$.
    
    Now fix $T_0\subseteq T$ and $A,B\subseteq T\setminus T_0$.
    Choose a minimum $(A\cup T_0,B\cup T_0)$-separator $F\subseteq M$
    in $G$. Every vertex of $T_0$ belongs to both endpoint sets,
    so $T_0\subseteq F$. Moreover, the minimum size of such a
    separator is $|T_0|$ plus the minimum size of an $(A,B)$-separator
    in $G-T_0$. Consequently, $F\setminus T_0\subseteq M\setminus T_0$
    is a minimum $(A,B)$-separator in $G-T_0$.
    \end{proof}
    
    We denote the algorithm of Lemma~\ref{lem:cut-covering} by $\cutcover(G,T)$.

    \section{Bounding the degree of the graph}\label{sec:bounding-degree}
    
    In this section,
    given a terminal network $(G,T)$,
    we design a polynomial-time procedure that
    bounds the maximum degree of $G$ by $2k$ while preserving a vertex multicut-mimicking network.
    In particular, we prove Lemma~\ref{lem:global-degree}.

    \begin{restatable}[Degree reduction]{lemma}{degreereduction}\label{lem:global-degree}   
    Let $(G,T)$ be a terminal network and let $k:=|T|$.
    In polynomial time, one can compute a vertex
    multicut-mimicking network $(G',T)$ for $(G,T)$ such that
    $V(G')=V(G)$ and the maximum degree of $G'$ is at most $2k$.
    \end{restatable}
    
    Since a vertex multicut-mimicking network for $(G,T)$ is also a connectivity-$c$ vertex multicut-mimicking network for $(G,T)$ for every $c$, the output of Lemma~\ref{lem:global-degree} 
    is also a connectivity-$c$ vertex multicut-mimicking network, for any choice of $c$.

    We prove Lemma~\ref{lem:global-degree} using Lemmas~\ref{lem:rr:add} and~\ref{lem:rr:delete}.
    
    \begin{lemma}[Adding edges]\label{lem:rr:add}
        Let $(G,T)$ be a terminal network,
        $u \in V(G)$
        and let $C_u$ be an arbitrary minimum $(N(u) \setminus T, T\setminus \{u\})$-separator in $G -\{u\}$.
        Say there exists $v \in C_u$ such that $uv \not \in E(G)$.
        Then $(G':=G\cup \{uv\},T)$ is a vertex multicut-mimicking network of $(G,T)$.
    \end{lemma}
    
    \begin{proof}
    Fix $T_0\subseteq T$ and a partition $\mathcal T$ of
    $T\setminus T_0$.
    Since $G'$ is a supergraph of $G$, every
    $\mathcal T$-\mwc\ in $G'-T_0$ is also a
    $\mathcal T$-\mwc\ in $G-T_0$.
    
    For the other direction, if $u\in T_0$, or if some minimum
    $\mathcal T$-\mwc\ in $G-T_0$ contains $u$, then such a
    $\mathcal{T}$-\mwc\ of $G-T_0$ is also a $\mathcal T$-\mwc\ in $G'-T_0$.
    
    Otherwise, choose a minimum $\mathcal T$-\mwc\ $X$ in
    $G-T_0$ that maximizes the size of the connected component containing
    $u$ in $G-(T_0\cup X)$.
    Let $V_u$ be the vertex set of this connected component, and put
    $G^{\texttt{side}}:=G-(T_0\cup X\cup V_u)$.
    If $C_u\cap V(G^{\texttt{side}})=\emptyset$, then
    $v\in T_0\cup X\cup V_u$.
    Thus adding $uv$ does not merge distinct components of
    $G-(T_0\cup X)$, and $X$ remains a
    $\mathcal T$-\mwc\ in $G'-T_0$.
    
    Suppose instead that 
    $C'_u:=C_u\cap V(G^{\texttt{side}})\ne\emptyset$.
    For ease of presentation, put
    $ A:=N_G(u)\setminus T$ and 
    $ B:=T\setminus\{u\}$, and
    define
    $ R:=\reach_{G-\{u\}-C_u}(A\setminus C_u)$
    and 
    $ X_u:=X\cap R$.
    Since $C_u$ is an $(A,B)$-separator in $G-\{u\}$,
    $  R\cap T=\emptyset$ and
    $ N_G(R)\subseteq C_u\cup\{u\}$.
    Also,
    $ N_G(V_u)\subseteq T_0\cup X$.
    
    \begin{claim}\label{claim:small-xu}
    We have $|X_u|<|C'_u|$.
    \end{claim}
    \begin{proof}
    Set
    $  X'':=(X\setminus X_u)\cup C'_u$.
    Since $C'_u$ is disjoint from $T_0\cup X$,
    $ X''\cap T_0=\emptyset$ and
    $ |X''|=|X|-|X_u|+|C'_u|$.
    
    We first show that $X''$ is a $\mathcal T$-\mwc\
    in $G-T_0$.
    Every vertex of $R$ is present in $G-(T_0\cup X'')$:
    the set $R$ is disjoint from $T_0$, from
    $X\setminus X_u$, and from $C'_u\subseteq C_u$.
    Moreover, every vertex of $R$ can reach a vertex of
    $A\setminus C_u$ within $R$.
    Since these vertices are neighbors of $u$, every vertex
    of $R$ belongs to the component containing $u$ in
    $G-(T_0\cup X'')$.
    
    The set $V_u$ also remains connected and contains $u$.
    Furthermore, the neighborhood inclusions above and
    $ C_u\setminus C'_u\subseteq T_0\cup X\cup V_u$
    give
    $  N_G(V_u\cup R)
        \subseteq T_0\cup(X\setminus X_u)\cup C'_u
        =T_0\cup X''$.
    Thus the component containing $u$ in
    $G-(T_0\cup X'')$ is exactly $V_u\cup R$.
    Since $R\cap T=\emptyset$, its terminals are precisely
    the terminals in $V_u$, which belong to at most one
    part of $\mathcal{T}$.
    
    Every other component of $G-(T_0\cup X'')$ avoids $R$,
    and hence avoids $X_u$.
    Since the remaining vertices of $X$ belong to $X''$,
    such a component is contained in a component of
    $G-(T_0\cup X)$.
    Its terminals therefore also belong to at most one
    part of $\mathcal T$.
    Consequently, $X''$ is a $\mathcal T$-\mwc\ in $G-T_0$.
    
    By the minimum cardinality of $X$,
    $ |X|\leq |X''|=|X|-|X_u|+|C'_u|$,
    so $|X_u|\leq |C'_u|$.
    Suppose equality holds.
    Then $X''$ is another minimum $\mathcal T$-\mwc\
    in $G-T_0$.
    Since $C'_u\ne\emptyset$, equality implies
    $X_u\ne \emptyset$.
    But $X_u\subseteq R\setminus V_u$, so the component
    $V_u\cup R$ containing $u$ strictly contains $V_u$.
    This contradicts the choice of $X$.
    Therefore $|X_u|<|C'_u|$.
    \end{proof}
    
    \begin{claim}\label{claim:custar-sep}
    The set
    $ C^*_u:=(C_u\setminus C'_u)\cup X_u$
    is an $(A,B)$-separator in $G-\{u\}$.
    \end{claim}
    
    \begin{proof}
    Suppose not, and let $P$ be a path from $a\in A$
    to a vertex of $B$ in $G-\{u\}-C^*_u$.
    Since $C_u$ is an $(A,B)$-separator in $G-\{u\}$,
    the path $P$ meets $C_u$.
    Let $c$ be its first vertex in $C_u$.
    As $P$ avoids $C_u\setminus C'_u$, we have $c\in C'_u$.
    
    Since $a\in N_G(u)\setminus T$ and
    $T_0\subseteq T$, we have $a\in V_u\cup X$.
    In particular, $a\ne c$, because
    $c\in V(G^{\texttt{side}})$.
    The prefix of $P$ before $c$ avoids $C_u$ and is
    therefore contained in $R$.
    
    If $a\in X$, then $a\in X\cap R=X_u$, contrary
    to the choice of $P$.
    Otherwise $a\in V_u$.
    Let $x$ be the first vertex of $P$ outside $V_u$.
    Then
    $ x\in N_G(V_u)\subseteq T_0\cup X$.
    Since $c\notin T_0\cup X$, the vertex $x$ occurs
    before $c$ and hence belongs to $R$.
    As $R\cap T_0=\emptyset$, we obtain
    $x\in X\cap R=X_u$, again a contradiction.
    \end{proof}
    
    Since $X_u\cap C_u=\emptyset$,
    Claim~\ref{claim:small-xu} gives
    $|C^*_u|=|C_u|-|C'_u|+|X_u|<|C_u|$.
    Together with Claim~\ref{claim:custar-sep}, this
    contradicts the minimum cardinality of $C_u$ as an
    $(A,B)$-separator in $G-\{u\}$.
    
    Hence $C_u\cap V(G^{\texttt{side}})=\emptyset$.
    In particular, $v\in T_0\cup X\cup V_u$, so adding
    $uv$ does not merge distinct components of
    $G-(T_0\cup X)$.
    Thus $X$ is also a $\mathcal T$-\mwc\ in $G'-T_0$.
    Since $(T_0,\mathcal T)$ was arbitrary, the lemma follows.
    \end{proof}
    
    \begin{lemma}[Deleting edges]\label{lem:rr:delete}
        Let $(G,T)$ be a terminal network.
        Fix $u \in V(G)$ and
        let $C_u$ be some $(N_G(u) \setminus T, T \setminus \{u\})$-separator in $G- \{u\}$,
        such that for each $v \in C_u$, $uv \in E(G)$.
        If there exists $w \in N_G(u) \setminus (C_u \cup T)$, 
        then $(G'=G - \{uw\}, T)$ is a vertex multicut-mimicking network of $(G,T)$.
    \end{lemma}
    \begin{proof}
    Fix $T_0\subseteq T$ and a partition $\mathcal T$
    of $T\setminus T_0$.
    Since $G'$ is a subgraph of $G$, every
    $\mathcal T$-\mwc\ in $G-T_0$ is also one in $G'-T_0$.
    
    Conversely, let $X$ be a $\mathcal T$-\mwc\ in $G'-T_0$.
    If $u$ or $w$ belongs to $T_0\cup X$, adding $uw$
    does not change the remaining graph.
    Otherwise, let $V_u,V_w$ be the vertex sets of the
    components of $G'-(T_0\cup X)$ containing $u,w$.
    If $V_u=V_w$, adding $uw$ does not merge components.
    
    Suppose that $V_u\ne V_w$.
    Since $w\notin C_u$, every vertex of $C_u$ is still
    adjacent to $u$ in $G'$. Thus
    $C_u\subseteq V_u\cup T_0\cup X$, and
    $V_w\cap C_u=\emptyset$.
    If $V_w$ contained a terminal $t$, a $w$-$t$ path
    inside $G'[V_w]$ would avoid both $u$ and $C_u$.
    This contradicts the fact that $C_u$ separates
    $N_G(u)\setminus T$ from $T\setminus\{u\}$ in $G-\{u\}$.
    
    Therefore $V_w$ contains no terminal.
    Adding $uw$ only merges $V_u$ with this terminal-free
    component, so $X$ remains a $\mathcal T$-\mwc\
    in $G-T_0$.
    \end{proof}
    
    \begin{proof}[Proof of Lemma~\ref{lem:global-degree}]
    Repeatedly apply the following procedure as long as the current graph has a vertex of degree at least $2k+1$.
        Let $(G,T)$ be the current terminal network and let $u \in V(G)$ such that
        $|N_G(u)| \geq 2k+1$.
        Compute an arbitrary minimum $(N_G(u) \setminus T, T \setminus \{u\})$-separator in $G -\{u\}$, say $C_u$ in polynomial time. This set $C_u$ is fixed in all our remaining arguments.
        
        Use Lemma~\ref{lem:rr:add} with respect to this set $C_u$, iteratively until all edges between $u$ and the vertices of $C_u$ are added. 
        This takes at most $|C_u|$ calls to Lemma~\ref{lem:rr:add}.
        Note that in each call to Lemma~\ref{lem:rr:add},
        (the fixed set) $C_u$ satisfies the properties in the lemma for the new graph also. Indeed, $G_i-\{u\}=G-\{u\}$, and the new source set is $A':=A\cup(\{v\}\setminus T)$ with $v\in C_u$, so $A'\setminus C_u=A\setminus C_u$; hence $C_u$ is an $(A',B)$-separator, and it is minimum because every $(A',B)$-separator is an $(A,B)$-separator.
        Let the resulting terminal network be $(G_1,T)$.
        Note that the number of newly added edges in $G_1$ is at most $|C_u|\leq |T\setminus\{u\}|\leq k$. 
    
        Apply Lemma~\ref{lem:rr:delete} exhaustively starting with the terminal network $(G_1,T)$, while keeping $T,u,C_u$ fixed, until all edges incident on $u$ and a vertex in $N_G(u) \setminus (T \cup C_u)$ are deleted. 
        Again note that Lemma~\ref{lem:rr:delete} did not impose that $C_u$ should be a minimum separator between the said sets, therefore, during the exhaustive application of Lemma~\ref{lem:rr:delete} the fixed set $C_u$ still satisfies the properties of Lemma~\ref{lem:rr:delete}.
        
        Since each application of Lemma~\ref{lem:rr:delete} decreases the number of edges in the graph, Lemma~\ref{lem:rr:delete} is executed only a polynomial number of times.
        Also after the exhaustive application of Lemma~\ref{lem:rr:delete}, the number of deleted edges is at least $|N_G(u) \setminus T| - |C_u|$.
        Let $(G':=G_2,T)$ be the resulting terminal network.
        
        From the correctness of Lemmas~\ref{lem:rr:add} and~\ref{lem:rr:delete},
        $(G',T)$ is a vertex multicut-mimicking network of $(G,T)$.
        Note that the vertex set of $G'$ is the same as that of $G$. 
      After these operations, $N_{G'}(u)\subseteq T\cup C_u$.
    Hence
    \[
        \deg_{G'}(u)\leq k+|C_u|\le2k<\deg_G(u).
    \]
    Since only edges incident with $u$ were modified,
    \[
        |E(G')|
        =|E(G)|-\deg_G(u)+\deg_{G'}(u)
        <|E(G)|.
    \]
           This is the end of one application of this procedure. 
           Each application results in a vertex multicut-mimicking network, and keeps the vertex set unchanged,
           and strictly decreases the total number of edges.
           Consequently, there are at most $|E(G)|$ applications,
           and the procedure runs in polynomial time.
    \end{proof}

    \section{Proof of Theorem~\ref{thm:budgeted-vertex-mimicking}: formal components and outline} \label{sec:outline}
    
    We now explain the components of the formal proof of Theorem~\ref{thm:budgeted-vertex-mimicking}. 
    
    \thmbudgetedvertex*

    We fix the following constants, function, definitions and three auxiliary lemmas,
    that will be used throughout the proofs in Sections~\ref{sec:outline},~\ref{sec:no-small-tail-case} and~\ref{sec:small-tail-case}.

    \begin{enumerate}
        \item Let $c_{\texttt{cover}}\ge1$ be the constant such that 
    Lemma~\ref{lem:cut-covering}
    gives a set $M$ of size at most $c_{\texttt{cover}} |T|^3$.
    
    We refer to the algorithm of Lemma~\ref{lem:cut-covering} by $\cutcover(G,T)$.
    
    \item Fix $ \alpha:=
     160\max\left\{1,
          \left\lceil\log_2c_{\texttt{cover}}\right\rceil
          \right\}$.

        \item For integer $c \geq 1$, let $\bound(c): = c^{\lceil\alpha\log_2 c \rceil}$ and $\bound(0): = 0$.

        For integers $k,c \ge1$, let
     $\bound(k,c):=
       k^7 \bound(4c)$.
    Let $\bound(0,c):=0$ for every
    integer $c\ge1$.

    Here $\bound(|T|,c)$ is the target upper bound on the
    number of vertices in a connectivity-$c$ vertex
    multicut-mimicking network of $(G,T)$.

    \end{enumerate}

    For any $X \subseteq V(G)$, the {\em $T$-components} of $G-X$ are the connected components of $G-X$ that contain a vertex of $T$.

    \begin{definition}[\Bigslim\ terminal network]
    Let $(G,T)$ be a terminal network and put $k:=|T|$.
    We call $(G,T)$ \emph{\slim} if every vertex of $G$
    has degree at most $2k$ and every connected component
    of $G$ contains a terminal.
    In particular, a slim terminal network has at most
    $k$ connected components.
    We call $(G,T)$ \emph{\bg} if $|V(G)|>\bound(k)$,
    and \emph{\bigslim} if it is both \bg\ and \slim.
    \end{definition}
    
    \begin{definition}[$h$-recursively useful region]
        Let $(G,T)$ be a terminal network and let $0 \leq h < |T|$.
        A set $R \subseteq V(G)$ is called an {\em $h$-recursively useful region} of $(G,T)$ if 
        
        \begin{enumerate}
            \item $R \cap T = \emptyset$,
            \item $N_G[R] \neq V(G)$,
            \item $|N_G(R)| \leq h$, and
            \item $|R| > 2 \cdot \bound(|N_G(R)|)$.
        \end{enumerate}
        
    \end{definition}

    \begin{definition}[$h$-\lslf\ modulator]
        Let $(G,T)$ be a terminal network and let $0 \leq h < |T|$ be an integer.
        A set $M \subseteq V(G)$ is called an {\em $h$-\lslf\ modulator} of $(G,T)$,
        if for every connected component $D$ of $G-M$,
    
        $$|V(D)| > 3 \cdot \bound(h) \implies \flow_G(V(D),T) > h.$$
    \end{definition}
    
    \paragraph{The three supporting lemmas.}
    Here we present the three technical lemmas that are used in this section and in Section~\ref{sec:no-small-tail-case}.

    \begin{lemma}[Three separators isolate a vertex]
    \label{lem:three-binary-separators}
    Let $G$ be a graph and let 
    $\mathcal{Q}=\{Q_1,\ldots,Q_r\}$ be a partition of a set $Q \subseteq V(G)$ and let $Q\subseteq M\subseteq V(G)$.
    Suppose there is a $\mathcal{Q}$-\mwc\ of size at most $p$,
    and $M$ contains a minimum $(Q_I, Q\setminus Q_I)$-separator
    for every $I\subseteq[r]$, where $Q_I:=\bigcup_{i\in I}Q_i$.
    Let $v\in V(G) \setminus M$. 
    Further suppose that, 
    for each $i\in[r]$, 
    every minimum $(Q_i,Q\setminus Q_i)$-separator contained in $M$ separates
    $v$ from $Q_i$ in $G$.
    
    Then $M$ contains a $(\{v\},Q)$-separator in $G$ of size at most $3p$.
    \end{lemma}

    \begin{proof}
    If there is no $v$ to $Q$ path in $G$, then
    the empty set is the required separator, and therefore
    the lemma holds.
    Hence assume that there is a path from $v$ to $Q$ in $G$.
    
    Choose a pair $(I,B_I)$ minimizing $|I|$, where
    $I\subseteq[r]$ and $B_I\subseteq M$ is a minimum
    $(Q_I,Q\setminus Q_I)$-separator such that
    there is a $v$ to $Q_I\setminus B_I$ path in $G-B_I$.
    Such a pair exists: take $I=[r]$ and $B_I=\emptyset$.
    The empty set cannot be chosen as $I$, and
    the hypothesis of the lemma excludes singletons.
    Thus $|I|\ge2$.
    
    Arbitrarily partition $I$ into two nonempty sets
    $I_1$ and $I_2$.
    For each $a\in\{1,2\}$, choose a minimum
    $(Q_{I_a},Q\setminus Q_{I_a})$-separator $B_a\subseteq M$.
    Since $|I_a|<|I|$, the choice of $(I,B_I)$ implies
    that $B_a$ separates $v$ from $Q_{I_a}$ in $G$.
    
    By the choice of $B_I$, the component of $G-B_I$
    containing $v$ contains a vertex of $Q_I\setminus B_I$.
    Since $B_I$ separates $Q_I$ from $Q\setminus Q_I$,
    this component contains no vertex of $Q\setminus Q_I$.
    Thus $B_I$ separates $v$ from $Q\setminus Q_I$.
    
    Moreover, $B_1$ and $B_2$ separate $v$ from
    $Q_{I_1}$ and $Q_{I_2}$, respectively.
    Since $Q_I=Q_{I_1}\cup Q_{I_2}$, the set
    $B_I\cup B_1\cup B_2\subseteq M$ separates $v$ from $Q$.
    
    Every $\mathcal Q$-\mwc\ is a separator for each binary
    coarsening of $\mathcal Q$.
    Since there is a $\mathcal Q$-\mwc\ of size at most $p$,
    each of $B_I,B_1,B_2$ has size at most $p$.
    Hence $B_I\cup B_1\cup B_2\subseteq M$ is a
    $(\{v\},Q)$-separator in $G$ of size at most $3p$.
    \end{proof}

    \begin{lemma}[Relative large-size-large-flow property]
    \label{lem:relative-lslf}
    Let $(G,T)$ be a terminal network, 
    let $T\subseteq A\subseteq V(G)$ and $A^{\dagger}:=N_G[A]$, 
    and let $M:=\cutcover(G,A^{\dagger})$.
    For any $T_0\subseteq T$ and partition $\mathcal T$ of
    $T\setminus T_0$, choose a minimum $\mathcal{T}$-\mwc\ $X$ in
    $G-T_0$ minimizing $|X\setminus M|$.
    
    Then for every connected component $D$ of $G-M$ such that $V(D) \cap X \neq \emptyset$,
    we have that $$\flow_G(V(D),T)\le3|X\setminus A|.$$
    \end{lemma}
    
    \begin{proof}
    If $X \subseteq M$, then the lemma holds vacuously.
    Fix a connected component $D$ of $G-M$ such that $V(D) \cap X \neq \emptyset$.
    Fix a vertex $v \in V(D) \cap X$. Then, in particular, $v \in X \setminus M$.
    To prove the lemma we will show a stronger property:
    there exists $F \subseteq M$ of size at most $3|X \setminus A|$, such that $F$ is a $(\{v\}, A^{\dagger})$-separator in $G$ of size at most $3 |X \setminus A|$.
    Since $T \subseteq A \subseteq A^{\dagger}$, this implies that $\flow_G(\{v\},T) \leq 3 |X \setminus A|$. Further since $F \subseteq M$, this implies that $F$ is also a $(V(D),T)$-separator in $G$, proving the lemma.
    In the remaining we prove this. The idea is to build a scenario in order to use Lemma~\ref{lem:three-binary-separators}.

    Set $X^*:=X\cup T_0$,
     $k_A:=|X\setminus A|$ and $k_{A^\dagger}:=|X\setminus A^\dagger|$.
    Since $A\subseteq A^\dagger$, we have
    $k_{A^\dagger}\leq k_A$.
    Let $V_1,\ldots,V_r$ be the vertex sets of the connected
    components of $G-X^*$, and put $ A_0^\dagger:=X^*\cap A^\dagger$,
    $H:=G-A_0^\dagger$ and
    $X_{A^\dagger}:=X\setminus A^\dagger$.
    Since $T_0\subseteq A^\dagger$, we have
    $X^*=A_0^\dagger\cup X_{A^\dagger}$.
    For each $i\in[r]$, put $A_i^\dagger:=A^\dagger\cap V_i$.
    The nonempty sets among $A_1^\dagger,\ldots,A_r^\dagger$
    partition $A^\dagger\setminus A_0^\dagger$, and
    $X_{A^\dagger}$ is a \mwc\ for this partition in $H$.
    We first establish the following claim.
    
    \begin{claim}\label{claim:relative-lslf}
    Let $v \in X \setminus M$.
    For every $j\in[r]$ with $A_j^\dagger\ne\emptyset$,
    every minimum
    $(A_j^\dagger,A^\dagger\setminus(A_0^\dagger\cup A_j^\dagger))$-separator
    $B\subseteq M\setminus A_0^\dagger$ in $H$
    is a 
    $(\{v\}, A_j^\dagger)$-separator in $H$.
    \end{claim}
    
    \begin{proof}
    Fix $j$ and $B$ as in the statement, and set $R:=\reach_{H-B}(A_j^\dagger\setminus B)$.
    For the sake of contradiction, say $v\in R$, and let $X_R:=X_{A^\dagger}\cap R$.
    Since $v\in X\setminus M$ and $A^\dagger\subseteq M$,
    we have $v\in X_R$.
    
    The set $B':=(B\cap(V_j\cup X_{A^\dagger}))\cup X_R$
    is an
    $(A_j^\dagger,A^\dagger\setminus(A_0^\dagger\cup A_j^\dagger))$-separator
    in $H$.
    Indeed, any path from $A_j^\dagger$ to
    $A^\dagger\setminus(A_0^\dagger\cup A_j^\dagger)$ in $H$
    first leaves $V_j$ through a vertex $x\in X_{A^\dagger}$.
    If its prefix ending at $x$ intersects $B$, that prefix intersects
    $B\cap(V_j\cup X_{A^\dagger})$.
    Otherwise, it connects $A_j^\dagger\setminus B$ to $x$
    in $H-B$, and hence $x\in X_R \subseteq B'$.
    
    Since $B'$ is an $(A_j^\dagger,A^\dagger\setminus(A_0^\dagger\cup A_j^\dagger))$-separator
    in $H$, $B$ is a minimum such separator and 
     $X_R\cap B=\emptyset$, we get that $ |B|\leq |B'|
     =|B\cap(V_j\cup X_{A^\dagger})|+|X_R|$.
    Consequently, $|B\setminus(V_j\cup X_{A^\dagger})|\leq |X_R|$.
    
    Define $X':=(X\setminus X_R)\cup
           (B\setminus(V_j\cup X_{A^\dagger}))$.
    Then $|X'|
     \leq |X|-|X_R|
          +|B\setminus(V_j\cup X_{A^\dagger})|
     \leq |X|$.
    Moreover, $X'\cap T_0=\emptyset$, since both $X$ and $B$
    are disjoint from $T_0$.

    We now show that $X'$ is a $\mathcal T$-\mwc\ in $G-T_0$.
    Since $X_R\cap A^\dagger=\emptyset$, we have
    $A_0^\dagger\subseteq T_0\cup X'$.
    Moreover, $ N_H(V_j)\subseteq X_{A^\dagger}$,
    $ N_H(R)\subseteq B$ and 
    $X_{A^\dagger}\cap R=X_R$.
    It follows that
    \[
     N_H(V_j\cup R)
     \subseteq
     (X_{A^\dagger}\setminus X_R)
     \cup(B\setminus(V_j\cup X_{A^\dagger}))
     \subseteq X'.
    \]
    Every component of $G-(T_0\cup X')$ that intersects $X_R$
    is therefore contained in $V_j\cup R$.
    The separator property of $B$ gives
    $R\cap A^\dagger\subseteq A_j^\dagger$.
    Since $T\subseteq A^\dagger$, every terminal in
    $V_j\cup R$ belongs to $V_j$, and these terminals lie
    in at most one part of $\mathcal T$.
    
    Any component of $G-(T_0\cup X')$ that does not intersect $X_R$,
    is disjoint from $X^*$, because
    $X\setminus X_R\subseteq X'$.
    It is therefore contained in a component of $G-X^*$
    and also intersects at most one part of $\mathcal T$.
    Hence $X'$ is a $\mathcal T$-\mwc\ in $G-T_0$.
    
    By the optimality of $X$, the inequality $|X'|\leq |X|$
    implies that $X'$ is a minimum $\mathcal{T}$-\mwc\ in $G-T_0$.
    However, $(X' \setminus X) \subseteq M$ whereas
    $v\in X_R \setminus M$.
    Thus $ X'\setminus M=(X\setminus M)\setminus X_R
     \subsetneq X\setminus M$,
    contradicting the choice of $X$.
    This proves the claim.
    \end{proof}

    Since $M:= \cutcover(G,A^{\dagger})$,
    from Lemma~\ref{lem:cut-covering}
    we conclude that 
    $M\setminus A_0^\dagger$ contains a minimum separator
    for every binary coarsening of the partition formed
    by the nonempty sets $A_i^\dagger$ in $H$.
    In addition to this, Claim~\ref{claim:relative-lslf} implies that the
     instance $(H, A^{\dagger} \setminus A^{\dagger}_0, M \setminus A^{\dagger}_0)$ satisfies the properties of Lemma~\ref{lem:three-binary-separators} ($H$ is the graph, $A^{\dagger} \setminus A^{\dagger}_0$ corresponds to the terminal set $Q$ and $M \setminus A^{\dagger}_0$ is the marked set in the statement of Lemma~\ref{lem:three-binary-separators}).
     Since $X_{A^\dagger}$ is a multiway cut in $H$ for the partition
    consisting of the nonempty sets among
    $A_1^\dagger,\ldots,A_r^\dagger$,
     and $|X_{A^\dagger}|=k_{A^\dagger}$,
    Lemma~\ref{lem:three-binary-separators} applies with
    $p=k_{A^\dagger}$.
     Thus from Lemma~\ref{lem:three-binary-separators},
     there exists $F' \subseteq M \setminus A_0^{\dagger}$ of size at most $3k_{A^{\dagger}}$ such that $F'$ is a $(\{v\}, A^{\dagger} \setminus A^{\dagger}_0)$-separator in $H$.
     
     Then $F:= F' \cup (X \cap (A^{\dagger}    \setminus A))$ is a $(\{v\},A^{\dagger})$-separator in $G$.
     Suppose otherwise, and consider a path from $v$ to
    $A^\dagger$ avoiding $F$ in $G$.
    Let $z$ be its first vertex in $A^\dagger$.
    If $z\in A$, its predecessor belongs to
    $N_G[A]=A^\dagger$, a contradiction.
    Thus $z\in A^\dagger\setminus A$.
    Since the path avoids $X\cap(A^\dagger\setminus A)$, we have $z\notin X$.
    Also
    $z\notin T_0$, because $T_0\subseteq A$.
    Hence $z\notin X^*$, and therefore
    $z\in A^\dagger\setminus A_0^\dagger$.
    The prefix ending at $z$ avoids $A_0^\dagger\cup F'$
    and is therefore a path in $H-F'$ from $v$ to
    $z\in A^\dagger\setminus A_0^\dagger$.
    This contradicts that $F'$ is a
    $(\{v\},A^\dagger\setminus A_0^\dagger)$-separator in $H$.
    
    Observe that $F \subseteq M$.
    Finally we argue about the size of $F$.
    Since $A\subseteq A^\dagger$, we have
    $|X \cap (A^{\dagger}    \setminus A)|=k_A-k_{A^\dagger}$.
    Therefore, $|F|
     \le3k_{A^\dagger}+(k_A-k_{A^\dagger})
     =k_A+2k_{A^\dagger}
     \le3k_A$.
    \end{proof}

    \begin{lemma}[Region-finding procedure]
    \label{lem:exposed-region}
    Let $(G,T)$ be a terminal network, $ T \subseteq M \subseteq V(G)$, 
    and $0 \leq h < |T|$ be an integer.
    There is a deterministic polynomial-time algorithm that,
    on input $(G,T,M,h)$, 
    either 
    \begin{enumerate}
        \item finds an $h$-recursively useful region of $(G,T)$, 
    or 
        \item correctly concludes $M$ is an $h$-\lslf\ modulator of $(G,T)$.
    \end{enumerate}
    \end{lemma}
    
    \begin{proof}
    For every connected component $D$ of $G-M$ with
    $|V(D)|>3\cdot\bound(h)$, compute a minimum
    $(V(D),T)$-separator $F$ in $G$.
    If $|F|>h$, continue to the next component.
    Otherwise, set $R:=\reach_{G-F}(V(D)\setminus F)$.
    We show that $R$ is an $h$-recursively useful region.
    
    Since $F$ is a $(V(D),T)$-separator in $G$, we have
    $R\cap T=\emptyset$.
    Moreover, $N_G(R)\subseteq F$, and hence
    $|N_G(R)|\leq h$ and
    \[
     |R|
     \geq |V(D)|-|F|
     >3\cdot\bound(h)-h
     \ge2\cdot\bound(h)
     \ge2\cdot\bound(|N_G(R)|).
    \]
    Here we used that $\bound(h)\geq h$ and the monotonicity of
    $\bound$.
    
    Since $|F|\leq h<|T|$, there exists a terminal
    $t\in T\setminus F$.
    As $R\cap T=\emptyset$ and $N_G(R)\subseteq F$,
    we have $t\notin N_G[R]$.
    Thus $N_G[R]\ne V(G)$.
    This shows that $R$ is indeed an $h$-recursively useful region.
    Return $R$.
    
    If the procedure returns no such $R$, then every connected
    component $D$ of $G-M$ with
    $|V(D)|>3\cdot\bound(h)$ has a minimum
    $(V(D),T)$-separator of size greater than $h$.
    That is,
    \[
     |V(D)|>3\cdot\bound(h)
     \quad\Longrightarrow\quad
     \flow_G(V(D),T)>h.
    \]
    Consequently, $M$ is an $h$-\lslf\ modulator of $(G,T)$.
    
    The procedure performs at most $|V(G)|$
    minimum-separator computations, each of which takes
    deterministic polynomial time.
    Hence its total running time is deterministic polynomial.
    \end{proof}

    \paragraph{The main technical lemma.} 
    We now present the main technical lemma
    that underlies the proof of Theorem~\ref{thm:budgeted-vertex-mimicking}. 
    
    \begin{restatable}[Main technical lemma]{lemma}{thmessentialmarking}
    \label{lem:marking-essential-vertices}
    Let $(G,T)$ be a \slim\ terminal network with
    $k:=|T|\ge2$, and let $1\leq c\leq k$ be an integer.
    Suppose that $ |V(G)|>\bound(k)$ or 
    $ |V(G)|>\bound(k,c)$.
    There is a deterministic polynomial-time algorithm that,
    on input $(G,T,c)$, either returns
    \begin{enumerate}
    \item an $\frac{14}{15}k$-recursively useful region $R$; or
    \item a set $M\supseteq T$ of size at most $|V(G)|/4$
    such that, for every $T_0\subseteq T$ and every partition
    $\mathcal T$ of $T\setminus T_0$, if the minimum size
    of a $\mathcal T$-\mwc\ in $G-T_0$ is at most $c$,
    then $M$ contains every vertex essential for
    $(T_0,\mathcal T)$.
    \end{enumerate}
    \end{restatable}
    
    We now give a proof of Theorem~\ref{thm:budgeted-vertex-mimicking} assuming Lemma~\ref{lem:marking-essential-vertices}.

    \begin{proof}[Proof of Theorem~\ref{thm:budgeted-vertex-mimicking}]
    We give a recursive algorithm that, for every input
    $(G,T,c)$ with $k:=|T|$ and $1\leq c\leq k$,
    returns a terminal network $(H,T)$ satisfying:
    \begin{enumerate}
    \item $(H,T)$ is a connectivity-$c$ vertex
    multicut-mimicking network of $(G,T)$; and
    \item
    \[
     |V(H)|\le
     \begin{cases}
     \bound(k), & c=k,\\
     \bound(k,c), & c<k.
     \end{cases}
    \]
    \end{enumerate}
    We prove these guarantees by strong induction on $k$,
    simultaneously for all admissible budgets $c$.
    
    In each recursive call, write $k:=|T|$.
    If $k\le1$, return the graph consisting only of $T$.
    If $c=k$ and $|V(G)|\le\bound(k)$, or if
    $c<k$ and $|V(G)|\le\bound(k,c)$, return $(G,T)$.
    Otherwise, apply Lemma~\ref{lem:global-degree} and
    delete connected components containing no terminal.
    If the resulting graph satisfies the corresponding size
    bound, return it.
    Otherwise, apply
    Lemma~\ref{lem:marking-essential-vertices} to $(G,T,c)$.
    
    \textit{Case 1.}
    Suppose Lemma~\ref{lem:marking-essential-vertices} returns
    a set $M\supseteq T$, with $|M|\le|V(G)|/4$, containing
    every essential vertex for each pair $(T_0,\mathcal T)$
    satisfying $\opt_{G-T_0}(\mathcal T)\leq c$.
    Choose $v\in V(G)\setminus M$ and bypass it,
    that is, make $N_G(v)$ a clique and delete $v$.
    Let $G'$ be the resulting graph.
    For every pair $(T_0,\mathcal T)$ satisfying
    $\opt_{G-T_0}(\mathcal T)\leq c$, there is a minimum
    $\mathcal T$-\mwc\ in $G-T_0$ avoiding $v$,
    because $v$ was not essential.
    This remains a $\mathcal T$-\mwc\ in $G'-T_0$.
    Conversely, every $\mathcal T$-\mwc\ in $G'-T_0$
    is also a $\mathcal T$-\mwc\ in $G-T_0$, since a path
    through $v$ can be replaced by an edge between its neighbors.
    Hence $(G',T)$ is a connectivity-$c$ vertex
    multicut-mimicking network of $(G,T)$.
    Continue the current call on $(G',T,c)$.
    This decreases the vertex count by one.
    
    \textit{Case 2.}
    Suppose instead that the lemma returns a terminal-free
    set $R$.
    Put $B_R:=N_G(R)$, $b:=|B_R|$ and $G_R:=G[N_G[R]]$.
    Note that $b\le14k/15<k$.
    If $b\le1$, take $H_R$ to be the graph consisting only
    of $B_R$.
    Then $(H_R,B_R)$ is a vertex multicut-mimicking network
    of $(G_R,B_R)$ and has at most $\bound(b)$ vertices.
    
    Otherwise, apply the recursive algorithm to $(G_R,B_R,b)$.
    Since the budget equals $|B_R|=b$, the induction hypothesis
    gives a connectivity-$b$ vertex multicut-mimicking network
    $(H_R,B_R)$ on at most $\bound(b)$ vertices.
    For every $F\subseteq B_R$ and every partition $\mathcal Q$
    of $B_R\setminus F$, deleting all vertices of
    $B_R\setminus F$ is a $\mathcal Q$-\mwc\ in $G_R-F$.
    Consequently,
    $\opt_{G_R-F}(\mathcal Q)
     \leq |B_R\setminus F|\leq b$.
    Thus $(H_R,B_R)$ preserves every such optimum exactly
    and is a vertex multicut-mimicking network of $(G_R,B_R)$.
    
    Let $G'$ be obtained from the disjoint union of $G-R$
    and $H_R$ by identifying, for each $x\in B_R$,
    the terminal $x$ of $H_R$ with the vertex $x$ of $G-R$.
    
    We show that $(G',T)$ is a vertex multicut-mimicking
    network of $(G,T)$.
    Fix $T_0\subseteq T$, a partition $\mathcal T$ of
    $T\setminus T_0$, and a $\mathcal T$-\mwc\ $X$ in $G-T_0$.
    Put $B_0:=B_R\cap(T_0\cup X)$,
    and let $\mathcal P$ be the partition of
    $B_R\setminus B_0$ induced by the connected components
    of $G_R-(B_0\cup(X\cap R))$.
    Then $X\cap R$ is a $\mathcal P$-\mwc\ in $G_R-B_0$.
    Since $(H_R,B_R)$ is a vertex multicut-mimicking network
    of $(G_R,B_R)$, there exists a $\mathcal P$-\mwc\
    $Y$ in $H_R-B_0$ satisfying $|Y|\le|X\cap R|$.
    
    We claim that $X':=(X\setminus R)\cup Y$ is a
    $\mathcal T$-\mwc\ in $G'-T_0$.
    Suppose that $P$ is a path between terminals belonging
    to different parts of $\mathcal T$ in
    $G'-(T_0\cup X')$.
    Decompose $P$ into subpaths contained in $G-R$ or $H_R$.
    Every subpath contained in $G-R$ also lies in
    $G-(T_0\cup X)$.
    Every subpath contained in $H_R$ has endpoints in
    $B_R$ and avoids $B_0\cup Y$.
    Since $Y$ is a $\mathcal P$-\mwc\ in $H_R-B_0$,
    its endpoints belong to the same part of $\mathcal P$.
    By the definition of $\mathcal P$, this subpath can
    therefore be replaced by a path in
    $G_R-(B_0\cup(X\cap R))$.
    These replacements give a walk between the endpoints
    of $P$ in $G-(T_0\cup X)$, contradicting that $X$
    is a $\mathcal T$-\mwc.
    Furthermore,
    $|X'|\le|X\setminus R|+|Y|\le|X|$.
    
    Starting with a $\mathcal T$-\mwc\ in $G'-T_0$,
    apply the same argument with $G_R$ and $H_R$
    interchanged, keeping $G-R$ as the fixed exterior graph.
    Partition the surviving boundary according to the
    connected components of $H_R$ after deleting the
    boundary vertices and interior vertices of this cut.
    The mimicking guarantee then gives a
    $\mathcal T$-\mwc\ in $G-T_0$ of no larger size.
    Hence the size of a minimum $\mathcal T$-\mwc\ in $G-T_0$
    equals the size of a minimum $\mathcal T$-\mwc\ in
    $G'-T_0$, proving that $(G',T)$ is a vertex
    multicut-mimicking network of $(G,T)$.
    
    \textit{Recursive guarantee.}
    After every bypass or region replacement, continue the
    current call on the resulting terminal network with the
    same budget, including the degree reduction
    (Lemma~\ref{lem:global-degree})
    and deletion of terminal-free components.
    A bypass decreases the vertex count by one.
    For the replacement above, writing $r:=|R|$, we have
    \[
     d_R:=|V(G)|-|V(G')|
         =r+b-|V(H_R)|
         \geq r+b-\bound(b)>r/2+b.
    \]
    Thus every bypass or region replacement strictly
    decreases the vertex count.
    Every recursive call is made on a region boundary,
    whose terminal count is at most $14k/15<k$.
    By induction on the terminal count, every recursive
    replacement is correct.
    Each iteration preserves all optimum values after
    truncation at $c+1$ and strictly decreases the vertex count.
    The procedure therefore terminates with a connectivity-$c$
    vertex multicut-mimicking network on at most
    $\bound(k,c)$ vertices.
    When $c=k$, its size is at most $\bound(k)$ and all
    optimum values are preserved exactly.
    Here we use $\bound(k)\le\bound(k,k)$, so the stated
    budgeted size bound also holds when $c=k$.
    
    \textit{Running time.}
    We next bound the running time.
    Let $n_0,k_0,c_0$ be the original vertex count, terminal
    count and budget, respectively, and let $n_i$ be the
    number of vertices in the input graph of recursive call $i$.
    We may assume $k_0\ge2$, since the cases $k_0\le1$
    are handled immediately.
    The recursive call corresponding to a region $R$ has
    input vertex count $r+b<2d_R$.
    Since the decreases within call $i$ sum to at most $n_i$,
    its immediate children have total input vertex count
    at most $2n_i$.
    Along each branch, the terminal count decreases by a factor
    of at most $14/15$, so the recursion has depth $O(\log k_0)$.
    Consequently, the total input vertex count of all
    recursive calls is
    \[
     \sum_i n_i
     \leq n_0\,2^{O(\log k_0)}
     =n_0k_0^{O(1)}.
    \]
    Each call performs at most $n_i$ bypasses and region
    replacements.
    Each application of Lemma~\ref{lem:marking-essential-vertices},
    the degree preprocessing, and the graph modifications
    take deterministic polynomial time.
    The size thresholds can also be tested in polynomial time,
    and the working budget is at most the terminal count.
    Together with the bound on $\sum_i n_i$, this gives
    a deterministic polynomial-time algorithm.
    The output is a connectivity-$c_0$ vertex
    multicut-mimicking network on at most
    \[
     \bound(k_0,c_0)
     =k_0^7\bound(4c_0)
     =k_0^7(c_0+1)^{O(\log(c_0+1))}
    \]
    vertices.
    
    Finally, if the original budget satisfies $c_0>k_0$,
    apply the algorithm with budget $k_0$.
    Every original optimum is at most $k_0$, so the resulting
    network preserves all optimum values exactly.
    Its size is at most $\bound(k_0)\le\bound(k_0,c_0)$,
    as required.
    \end{proof}

    \subsection{Proving the main technical lemma using two distinguishing cases based on ``tails''}
    
    \begin{definition}[Partitions with small-tail solution]
    \label{def:small-tail}
    For $T_0\subseteq T$ and a partition $\mathcal{T}$ of
    $T\setminus T_0$, 
    we say that $(T_0,\mathcal{T})$ admits a {\em small tail solution} in $(G,T)$, if 
        there exists a minimum $\mathcal{T}$-\mwc\ in $G-T_0$, 
        say $X \subseteq V(G) \setminus T_0$, 
        such that if $V_1, \ldots, V_r$ are the $T$-components of $G-(X \cup T_0)$ 
        which are ordered in the non-increasing order of their cardinality, 
        then either $r \leq 3$, or  $\sum_{i=4}^r |V_i|\leq \frac{|V(G)|}{12|T|^3}$.
    \end{definition}
    
    Given a terminal network $(G,T)$,
    a {\em small-tail query} of $(G,T)$
    is a pair $(T_0,\mathcal{T})$ that admits a small-tail solution,
    where $T_0 \subseteq T$ and $\mathcal{T}$ is a partition of $T \setminus T_0$.
    Similarly, a {\em no-small-tail query} of $(G,T)$ is a pair $(T_0,\mathcal{T})$ that does not admit a small-tail solution.

    Towards proving Lemma~\ref{lem:marking-essential-vertices}
    we design two lemmas.
    Lemma~\ref{lem:nosmalltail} gives a marked set containing
    a minimum multiway cut for every no-small-tail query, and hence
    every essential vertex for each such query.
    Lemma~\ref{lem:smalltail} takes care of the other case, 
    and returns a set which contains every
    essential vertex for all small-tail queries.
    
    \begin{restatable}[No small-tail case]{lemma}{lemnosmalltail}
    \label{lem:nosmalltail}
    Let $(G,T)$ be a \bigslim\ terminal network and $|T| \geq 2$.
    There is a deterministic polynomial-time algorithm that on input $(G,T)$ either returns
    \begin{enumerate}
    \item an $\frac{14}{15}|T|$-recursively useful region $R$, or
    \item a set $M \supseteq T$ of size at most $|V(G)|/16$ such that
    for every no-small-tail query $(T_0,\mathcal{T})$ of $(G,T)$,
    $M$ contains a minimum $\mathcal{T}$-\mwc\ of $G-T_0$.
    \end{enumerate}
    \end{restatable}

    \begin{restatable}[Small-tail case]{lemma}{lemsmalltail}
    \label{lem:smalltail}
    Let $(G,T)$ be a \bg\ terminal network with $|T|\ge2$.
    There is a deterministic polynomial-time algorithm that returns
    a set $M\supseteq T$ of size at most $3|V(G)|/16$ containing every
    essential vertex for every small-tail query
    $(T_0,\mathcal T)$ of $(G,T)$.
    \end{restatable}
    
    In Section~\ref{sec:no-small-tail-case} we give the proof of Lemma~\ref{lem:nosmalltail} and
    in Section~\ref{sec:small-tail-case} we give the proof of Lemma~\ref{lem:smalltail}.
    The arguments in the small-tail case 
    are analogous to the arguments used in~\cite{DBLP:journals/talg/Wahlstrom22} for showing quasi-polynomial-sized (in the capacity of the terminals) edge-multicut-mimicking networks,
    whereas the complications raised by the no small-tail case are new to the current paper due to us addressing the vertex-deletion version of the problem.
    We now give the proof of Lemma~\ref{lem:marking-essential-vertices} assuming Lemmas~\ref{lem:nosmalltail} and~\ref{lem:smalltail}.

    \begin{proof}[Proof of Lemma~\ref{lem:marking-essential-vertices}]
    Put $n:=|V(G)|$ and $k:=|T|$.
    
    \textit{Case 1: $n>\bound(k)$.}
    Apply Lemma~\ref{lem:nosmalltail} to $(G,T)$.
    If it returns a region $R$, return $R$.
    Otherwise, let $M_{\mathrm{no-small}}$ be its marked set.
    Apply Lemma~\ref{lem:smalltail} to $(G,T)$, obtaining
    a marked set $M_{\mathrm{small}}$, and put
    \[
     M:=M_{\mathrm{no-small}}\cup M_{\mathrm{small}}.
    \]
    Both marked sets contain $T$ and $ |M|
     \leq |M_{\mathrm{no-small}}|+|M_{\mathrm{small}}|
     \leq n/16+3n/16=n/4$.
    
    Fix $T_0\subseteq T$ and a partition $\mathcal T$ of
    $T\setminus T_0$.
    If $(T_0,\mathcal T)$ is a small-tail query,
    Lemma~\ref{lem:smalltail} guarantees that
    $M_{\mathrm{small}}$ contains every essential vertex
    for this query.
    Otherwise, Lemma~\ref{lem:nosmalltail} guarantees that
    $M_{\mathrm{no-small}}$ contains a minimum
    $\mathcal T$-\mwc\ in $G-T_0$, and hence contains every
    essential vertex for this query.
    Thus $M$ satisfies the required guarantee.
    Return $M$.
    
    \textit{Case 2: $n\le\bound(k)$.}
    The hypothesis of the lemma gives $n>\bound(k,c)$.
    We first observe that $k>4c$.
    Indeed, if $k\le4c$, monotonicity would give
    $\bound(k)\le\bound(4c)\le\bound(k,c)<n$,
    contradicting $n\le\bound(k)$.
    
    Put $Q:=N_G[T]$ and compute
    $M_0:=\cutcover(G,Q)$.
    Then $T\subseteq Q\subseteq M_0$.
    Since $(G,T)$ is \slim, we have
    $ |Q|\le(2k+1)k\le3k^2$, and $ |M_0|\le27c_{\texttt{cover}}k^6$.
    Apply Lemma~\ref{lem:exposed-region} to
    $(G,T,M_0,3c)$.
    This is permitted because $3c<k$.
    If it returns a region $R$, return $R$.
    Such a region is also an $\frac{14}{15}k$-recursively
    useful region, since $3c<3k/4<14k/15$.
    
    Otherwise, $M_0$ is a $3c$-\lslf\ modulator.
    Let $\mathcal D$ consist of all connected components
    of $G-M_0$ having at most $3\bound(3c)$ vertices,
    and put
    \[
     M:=M_0\cup\bigcup_{D\in\mathcal D}V(D).
    \]
    In particular, $T\subseteq M$.
    
    Fix $T_0\subseteq T$ and a partition $\mathcal T$ of
    $T\setminus T_0$ with
    $\opt_{G-T_0}(\mathcal T)\leq c$.
    Choose a minimum $\mathcal T$-\mwc\ $X$ in $G-T_0$
    minimizing $|X\setminus M_0|$.
    By Lemma~\ref{lem:relative-lslf}, applied with $A=T$,
    every component $D$ of $G-M_0$ meeting $X$ satisfies
    \[
     \flow_G(V(D),T)\le3|X\setminus T|\le3c.
    \]
    Since $M_0$ is a $3c$-\lslf\ modulator, every such
    component has at most $3\bound(3c)$ vertices.
    Consequently, $X\subseteq M$.
    Hence $M$ contains every vertex essential for
    $(T_0,\mathcal T)$.
    
    It remains to bound $|M|$.
    Write $m:=|M_0|\geq k$.
    Since $(G,T)$ is \slim, the graph $G$ has at most
    $k$ connected components and maximum degree at most $2k$.
    Deleting $M_0$ therefore leaves at most $k+2km$
    connected components.
    Using $k\ge2$, $m\geq k$ and $\bound(3c)\ge1$, we obtain
    \[
     \begin{aligned}
     |M|
     &\leq m+3(k+2km)\bound(3c)\\
     &\le8km\bound(3c)\\
     &\le216c_{\texttt{cover}}k^7\bound(3c)\\
     &<\frac{1}{4}k^7\bound(4c)\\
     &=\frac{\bound(k,c)}4
     <\frac n4.
     \end{aligned}
    \]
    The strict comparison involving $\bound(4c)$ follows
    from Lemma~\ref{lem:budgeted-size-ratio}.
    Return $M$.
    
    All procedures used in both cases are deterministic
    polynomial-time algorithms.
    \end{proof}

    \section{The no-small-tail case}\label{sec:no-small-tail-case}

    In this section we prove Lemma~\ref{lem:nosmalltail}.
    
    \lemnosmalltail*

    \begin{definition}[\Bch]
        Let $(G,T)$ be a terminal network and $|T|\geq 1$.
        A set $S \subseteq V(G) \setminus T$ is called a {\em \bch}
        if for every $Y \subseteq V(G)$ of size at most $2|T|$,
        and for every union $U$ of connected components of $G-Y$,
    
        $$|U \setminus T| \geq \frac{|V(G)|}{24|T|^4} \implies (S \cap U) \neq \emptyset.$$
    \end{definition}

    \subsection{Additional supporting lemmas for the no-small-tail case}
    
    \begin{lemma}[Hitting set of large components]
     \label{lem:sample-flow-support}
    Let $(G,T)$ be a \slim\ terminal network and $|T| \geq 1$. 
    There is a deterministic polynomial-time algorithm 
    that on input $(G,T)$,
    outputs a \bch\ of $(G,T)$ of size at most $120|T|^6$.
    \end{lemma}
    
    \begin{proof}
    Set $n:=|V(G)|$, $k:=|T|\ge1$, $r:=4k^2+k$ and $ L:=\left\lceil\frac{n}{24k^4r}\right\rceil$.

    Let $\pi:V(G)\to[n]$ be a depth-first search (DFS) traversal ordering
    of $G$, where $\pi(v)$ is the position at which $v$
    is first visited in the fixed DFS traversal.
    Let $v_1,\ldots,v_{n-k}$ be the vertices of
    $V(G)\setminus T$, indexed so that $ \pi(v_1)<\pi(v_2)<\cdots<\pi(v_{n-k})$.
    
    Define
    \[
     S:=\{v_{jL}:1\leq j\le\lfloor(n-k)/L\rfloor\}.
    \]
    By construction, $S\subseteq V(G)\setminus T$, and $|S|
     \leq (n-k)/L
     \le24rk^4
     =96k^6+24k^5
     \le120k^6$.

     We claim that $S$ is a \bch\ of $(G,T)$.
     Fix $Y\subseteq V(G)$ with $|Y|\le2k$, and let $U$
    be a union of connected components of $G-Y$.
    
    \begin{claim}\label{claim:intervals-in-ordering}
        $U \setminus T$ occupies at most $r$ intervals in the ordering $v_1, \ldots, v_{n-k}$.
    \end{claim}
    
    \begin{proof}
    Let $F$ be the search forest of the DFS defining $\pi$,
    and let $\delta_F(U)$ denote the set of edges of $F$
    with exactly one endpoint in $U$.
    Since $U$ is a union of connected components of $G-Y$,
    every edge of $G$ with exactly one endpoint in $U$
    has its other endpoint in $Y$.
    Since $(G,T)$ is \slim, the  degree of each vertex of $G$ is at most $2k$, therefore,
    \[
     |\delta_F(U)|
     \leq \sum_{y\in Y}\deg_G(y)
     \leq 2k|Y|
     \leq 4k^2.
    \]
    
    For each tree of $F$, consider its full DFS traversal
    walk, starting at the root and traversing every tree
    edge once in each direction.
    Every maximal consecutive block of visits to vertices
    of $U$ starts either at the initial root or immediately
    after traversing an edge from outside $U$ into $U$.
    Each edge of $\delta_F(U)$ is traversed into $U$
    exactly once.
    Since $(G,T)$ is \slim, $G$ has at most $k$ connected components and therefore $F$ has at most $k$ roots.
    So the total number
    of such blocks over all traversal walks is at most $ k+|\delta_F(U)|\leq k+4k^2=r$.
    
    Concatenating these walks, retaining only the first
    occurrence of each vertex, and then removing terminals
    gives the ordering $v_1,\ldots,v_{n-k}$.
    None of these operations increases the number of
    blocks of vertices belonging to $U$.
    Consequently, $U\setminus T$ occupies at most $r$
    intervals in this ordering.
    \end{proof}

    Assume that $|U\setminus T|\geq n/(24k^4)$.
    We show that $S\cap U\ne\emptyset$.
    Suppose, for a contradiction, that $S\cap U=\emptyset$.
    By Claim~\ref{claim:intervals-in-ordering},
    $U\setminus T$ occupies at most $r$ intervals in
    the ordering $v_1,\ldots,v_{n-k}$.
    Each such interval contains at most $L-1$ vertices,
    since every $L$ consecutive positions contain a
    vertex of $S$.
    Consequently,
    \[
     |U\setminus T|
     \leq r(L-1)
     <\frac{n}{24k^4}.
    \]
    This contradicts the assumed size bound on $U \setminus T$.
    
    Since a depth-first search traversal ordering of a graph can be found in polynomial time,  the set $S$ can be computed
    in deterministic polynomial time.
    This completes the proof.
    \end{proof}

    \begin{lemma}[\Lslf]\label{lem:lslf}
        Let $(G,T)$ be a slim terminal network and let $S$ be a \bch\ of $(G,T)$.
        Let $1 \leq h <|T|$ be an integer and let $T^*:= T \cup S$.
        Let $M:=\cutcover(G,T^*)$ such that $M$ is an $h$-\lslf\ modulator of $(G,T)$.

        Then for every $X \subseteq V(G)$ of size at most $|T|$ and every connected component $C$ of $G-X$,
    
        $$|C| > \frac{|V(G)|}{24|T|^4} + 12 \cdot |T| \cdot |M| \cdot \bound(h) \implies \flow_G(V(C) \cap T^*, T) >h.$$
    \end{lemma}

    \begin{proof}
    Put $n:=|V(G)|$ and $k:=|T|$.
    Fix $X\subseteq V(G)$ with $|X|\leq k$ and a connected
    component $C$ of $G-X$.
    Set $P:=V(C)\cap T^*$.
    We prove the implication by contraposition.
    
    Suppose that $\flow_G(P,T)\leq h$.
    Since $P,T\subseteq T^*$ and $M=\cutcover(G,T^*)$,
    Lemma~\ref{lem:cut-covering} gives a minimum
    $(P,T)$-separator $F\subseteq M$ with $|F|\leq h$.
    
    Let $\mathcal D$ be the set of connected components $D$
    of $G-M$ with $\flow_G(V(D),T)\leq h$, and put
    \[
     Z:=M\cup\bigcup_{D\in\mathcal D}V(D).
    \]
    Since $M$ is an $h$-\lslf\ modulator, every component
    in $\mathcal D$ has at most $3\bound(h)$ vertices.
    Moreover, slimness implies that $G-M$ has at most
    $k+2k|M|$ connected components. Hence
    \[
     |Z|
     \leq |M|+3(k+2k|M|)\bound(h)
     \leq 12k|M|\bound(h),
    \]
    where we use $|M|\geq k\ge2$ and $\bound(h)\ge1$.
    
    Let $W$ be the union of the vertex sets of all connected
    components of $G-(X\cup F)$ disjoint from $T^*$.
    Since $|X\cup F|\leq k+h\le2k$ and $W$ misses both
    $T$ and $S$, the definition of a \bch\ gives
    $|W|<n/(24k^4)$.
    
    We claim that $V(C)\subseteq Z\cup W$.
    Suppose otherwise, and let $v\in V(C)\setminus(Z\cup W)$.
    Since $F\subseteq M\subseteq Z$, the vertex $v$ belongs
    to a connected component of $G-(X\cup F)$.
    As $v\notin W$, this component meets $T^*$.
    It is contained in $C$, so it meets $P$.
    Thus $v$ is connected to $P\setminus F$ in $G-F$.
    
    Let $D$ be the connected component of $G-M$ containing $v$.
    Since $F\subseteq M$, every vertex of $D$ is connected
    to $P\setminus F$ in $G-F$.
    As $F$ separates $P$ from $T$, it also separates $V(D)$
    from $T$. Consequently,
    $ \flow_G(V(D),T)\leq |F|\leq h$.
    Thus $D\in\mathcal D$ and $v\in Z$, a contradiction.
    
    It follows that
    \[
     |V(C)|
     \leq |Z|+|W|
     <12k|M|\bound(h)+\frac{n}{24k^4},
    \]
    which proves the required implication.
    \end{proof}

    \begin{lemma}[Capturing most of an optimum \mwc\ under flow-condition]
    \label{lem:binary-capture}
    Let $(G,T)$ be a terminal network and
    let $S\subseteq V(G)\setminus T$.
    Let $T_0,\mathcal T,X$ be such that
    $T_0\subseteq T$, $\mathcal T$ is a partition of
    $T\setminus T_0$, and $X$ is a minimum
    $\mathcal T$-\mwc\ of $G-T_0$.
    Set $X^*:=T_0\cup X$ and $T^*:=T\cup S$.
    
    Suppose that $G-X^*$ has at least $4$ connected components.
    Let $V_1,\ldots,V_4$ be the vertex sets of some $4$
    connected components of $G-X^*$.
    For each $i\in[4]$,
    \begin{itemize}
        \item set $P_i:=T^*\cap V_i$, and
        \item assume that
        $\flow_G(P_i,T)\ge\frac{14}{15}|T|$.
    \end{itemize}
    Further set
    $P_0:=T^*\cap X^*$,
    $A:=P_1\cup P_2$ and $B:=P_3\cup P_4$.
    If $C$ is any minimum $(A,B)$-separator in $G-P_0$,
    then
    \[
     |X\setminus(C\cup P_0)|\le\frac{4}{15}|T|.
    \]
    \end{lemma}
    
    \begin{proof}
    Put $k:=|T|$, $C^*:=C\cup P_0$,
    $d:=|X^*|$, and $d_{C^*}:=|X^*\cap C^*|$.
    Since the set $T\setminus T_0$ is a $\mathcal T$-\mwc\
    in $G-T_0$,
    we have that $|X|\le|T\setminus T_0|=k-|T_0|$.
    Since $X\cap T_0=\emptyset$, it follows that
    $d=|X^*|=|T_0|+|X|\leq k$.
    
    The set $X^*\setminus P_0$ is an $(A,B)$-separator
    in $G-P_0$, since $V_1,\ldots,V_4$ are the vertex
    sets of distinct connected components of $G-X^*$.
    Since $C$ is a minimum $(A,B)$-separator in $G-P_0$,
    we get
    \[
     |C^*|=|C|+|P_0|
     \leq |X^*\setminus P_0|+|P_0|
     =d.
    \]
    
    For each $i\in[4]$, set
    $T_i:=T\cap V_i$ and
    \[
     X_i^{\mathrm{reach}}
     :=\reach_{G-C^*}(P_i\setminus C^*)\cap X^*.
    \]
    We claim that
    \[
     F_i:=(C^*\cap(V_i\cup X^*))
           \cup X_i^{\mathrm{reach}}\cup T_i
    \]
    is a $(P_i,T)$-separator in $G$.
    Indeed, every path from $P_i$ to a terminal in $V_i$
    intersects $T_i$.
    Any path from $P_i$ to a terminal outside $V_i$
    first leaves $V_i$ through a vertex $x\in X^*$.
    Its prefix ending at $x$ lies in $V_i\cup X^*$.
    If this prefix intersects $C^*$, it intersects
    $C^*\cap(V_i\cup X^*)\subseteq F_i$.
    Otherwise, it is a path in $G-C^*$ from
    $P_i\setminus C^*$ to $x$, so
    $x\in X_i^{\mathrm{reach}}\subseteq F_i$.
    Thus every path from $P_i$ to $T$ intersects $F_i$.
    
    By the definition of $\flow_G$ and the assumption
    in the lemma statement, for every $i\in[4]$ we have
    \begin{equation}\label{eq:capture-component-flow}
     \frac{14}{15}k
     \leq \flow_G(P_i,T)
     \leq |F_i|
     \leq |C^*\cap V_i|+d_{C^*}
          +|X_i^{\mathrm{reach}}|+|T_i|.
    \end{equation}
    Here we used $V_i\cap X^*=\emptyset$.
    We next bound the sums of the terms on the right-hand
    side of \eqref{eq:capture-component-flow} over $i\in[4]$.
    
    Since $C$ is an $(A,B)$-separator in $G-P_0$,
    the set $C^*$ is an $(A,B)$-separator in $G$.
    Consequently,
    \begin{equation}\label{eq:disjoint}
     \left(\bigcup_{i=1}^2X_i^{\mathrm{reach}}\right)
     \cap
     \left(\bigcup_{i=3}^4X_i^{\mathrm{reach}}\right)
     =\emptyset.
    \end{equation}
    Indeed, a vertex in this intersection would be
    reachable from both $A\setminus C^*$ and
    $B\setminus C^*$ in $G-C^*$, contradicting
    that $C^*$ is an $(A,B)$-separator in $G$.
    
    Every $X_i^{\mathrm{reach}}$ is contained in
    $X^*\setminus C^*$.
    By the disjointness of the two unions in
    \eqref{eq:disjoint}, each vertex of $X^*\setminus C^*$
    belongs to at most two of the sets
    $X_1^{\mathrm{reach}},\ldots,X_4^{\mathrm{reach}}$.
    Therefore,
    \begin{equation}\label{eq:sum-x-reach}
     \sum_{i=1}^4|X_i^{\mathrm{reach}}|
     \le2|X^*\setminus C^*|
     =2(d-d_{C^*}).
    \end{equation}
    Moreover, the sets $V_i$ are pairwise disjoint
    and do not intersect $X^*$, so
    \begin{equation}\label{eq:sum-comp}
     \sum_{i=1}^4|C^*\cap V_i|
     \leq |C^*|-d_{C^*}.
    \end{equation}
    
    Put $t:=\sum_{i=1}^4|T_i|$.
    Summing \eqref{eq:capture-component-flow} over $i\in[4]$
    and applying \eqref{eq:sum-x-reach} and
    \eqref{eq:sum-comp}, we get
    \begin{align}
     \frac{56k}{15}-t
     &\leq \sum_{i=1}^4|C^*\cap V_i|
           +4d_{C^*}
           +\sum_{i=1}^4|X_i^{\mathrm{reach}}|
           \notag\\
     &\leq (|C^*|-d_{C^*})
           +4d_{C^*}+2(d-d_{C^*})
           \notag\\
     &=|C^*|+2d+d_{C^*}
           \notag\\
     &\le3d+d_{C^*}.
     \label{eq:capture-summed-flow}
    \end{align}
    
    For each $i\in[4]$, the terminals in $T_i$
    belong to at most one part of $\mathcal T$,
    because $X$ is a $\mathcal T$-\mwc\ in $G-T_0$.
    Thus $T\setminus(T_0\cup T_i)$ is also a
    $\mathcal T$-\mwc\ in $G-T_0$.
    By the optimality of $X$,
    $|X|\leq k-|T_0|-|T_i|$,
    and therefore $d+|T_i|\leq k$.
    Summing over $i\in[4]$ gives
    \begin{equation}\label{eq:sum-four}
     4d+t
     =\sum_{i=1}^4(d+|T_i|)
     \le4k.
    \end{equation}
    
    Combining \eqref{eq:capture-summed-flow}
    and \eqref{eq:sum-four}, we obtain
    \[
     d-d_{C^*}
     \le4d+t-\frac{56k}{15}
     \le\frac{4k}{15}.
    \]
    Finally, $T_0\subseteq P_0\subseteq C^*$, so
    \[
     |X\setminus(C\cup P_0)|
     =|X^*\setminus C^*|
     =d-d_{C^*}
     \le\frac{4k}{15}.
    \]
    This proves the lemma.
    \end{proof}

    \subsection{The two main lemmas}

    We will see that Lemmas~\ref{lem:exposed-region}, \ref{lem:sample-flow-support}, \ref{lem:lslf} and~\ref{lem:binary-capture}, together prove Lemma~\ref{lem:marking-a-large-fraction}.
    
    \begin{restatable}[Marking a large fraction of the vertices of every optimum
    \mwc]{lemma}{markingalargefraction}\label{lem:marking-a-large-fraction}
    Let $(G,T)$ be a \bigslim\ terminal network.
    There is a deterministic polynomial-time algorithm that on input $(G,T)$,
    either returns
    \begin{enumerate}
        \item a $\frac{14}{15}|T|$-recursively useful region $R$; or
        \item a set $M_1\supseteq T$ of size at most $c_{\texttt{cover}} (|T|+120|T|^6)^3$
        such that 
        for every no-small-tail query $(T_0,\mathcal{T})$ of $(G,T)$,
        and every minimum $\mathcal{T}$-\mwc\ $X$ in $G-T_0$,
        $|X\setminus M_1|\leq 4|T|/15$.
    \end{enumerate}
    \end{restatable}
    
    \begin{proof}
    Put $n:=|V(G)|$ and $k:=|T|$.
    Using Lemma~\ref{lem:sample-flow-support} on the input $(G,T)$, compute a \bch\
    $S\subseteq V(G)\setminus T$ of size at most $120k^6$.
    Set $T^*:=T\cup S$, $M_1:=\cutcover(G,T^*)$ and
    $h:=\left\lfloor\frac{14k}{15}\right\rfloor$.
    By Lemma~\ref{lem:cut-covering},
    $T^*\subseteq M_1$ and
    $|M_1|\leq c_{\texttt{cover}}(k+120 k^6)^3$.

    Apply Lemma~\ref{lem:exposed-region} to $(G,T,M_1,h)$.
    If it returns an $h$-recursively useful region $R$,
    return $R$.
    Since $h\le14k/15$, this region satisfies the first
    outcome of the lemma.
    Otherwise, $M_1$ is an $h$-\lslf\ modulator.
    Return $M_1$.
    
    It remains to show that, when $M_1$ is returned,
    every minimum $\mathcal T$-\mwc\ $X$ in $G-T_0$
    satisfies $|X\setminus M_1|\le4k/15$,
    for every no-small-tail query $(T_0,\mathcal T)$.
    Fix a no-small-tail query $(T_0,\mathcal T)$ and any
    minimum $\mathcal T$-\mwc\ $X$ in $G-T_0$.
    Set $X^*:=T_0\cup X$.
    Since $T\setminus T_0$ is a $\mathcal{T}$-\mwc\ in $G-T_0$,
    $ |X^*|=|T_0|+|X|\leq k$.

    Let $V_1,\ldots,V_r$ be the vertex sets of the
    $T$-components of $G-X^*$, ordered
    by non-increasing cardinality.
    Since the query admits no small-tail solution,
    $r\ge4$ and $\sum_{i=4}^r|V_i|>\frac{n}{12k^3}$.
    
    Each of these components  $V_1, \ldots, V_r$ 
    contains a distinct terminal,
    so $r\leq k$ and, in particular, $k\ge4$.
    Therefore,
    \[
     \frac{n}{12k^3}
     <\sum_{i=4}^r|V_i|
     \leq k|V_4|,
    \]
    and consequently
    \begin{equation}\label{eq:vi-bound}
         |V_i|\ge|V_4|>\frac{n}{12k^4}
     \qquad\text{for every }i\in[4].
    \end{equation}

    Since $(G,T)$ is \bigslim, we have $n>\bound(k)$.
    As $k\ge4$, the bound on $|M_1|$ and
    \eqref{eq:exposure-slack} in
    Appendix~\ref{app:constructive-size-estimates} give
    \begin{equation}\label{eq:size-bound-one}
         \frac{n}{24k^4}
     +12k|M_1|\bound(h)
     <\frac{n}{12k^4}.
    \end{equation}
    
    For each $i\in[4]$, set $P_i:=T^*\cap V_i$.
    Since $S$ is a \bch, $M_1$ is an $h$-\lslf\
    modulator, and $|X^*|\leq k$, we may apply
    Lemma~\ref{lem:lslf} with deletion set $X^*$
    and component vertex set $V_i$.
    Together with \eqref{eq:vi-bound} and
    \eqref{eq:size-bound-one}, this gives
    $\flow_G(P_i,T)>h$.
    Since $\flow_G(P_i,T)$ is an integer,
    \[
     \flow_G(P_i,T)
     \geq h+1
     >\frac{14k}{15}
     \qquad\text{for every }i\in[4].
    \]
    
    Set $ P_0:=T^*\cap X^*$,
    $ A:=P_1\cup P_2$ and
    $ B:=P_3\cup P_4$.
    By Lemma~\ref{lem:cut-covering},
    $M_1\setminus P_0$ contains a minimum
    $(A,B)$-separator $C$ in $G-P_0$.
    Lemma~\ref{lem:binary-capture} therefore gives $|X\setminus(C\cup P_0)|\le4k/15$.
    Since $(C\cup P_0) \subseteq M_1$, we conclude that $ |X\setminus M_1|\le4k/15$.
    
    Note that the query $(T_0,\mathcal{T})$ and the minimum $\mathcal{T}$-\mwc\ $X$ in $G-T_0$,
    were arbitrary.
    Since Lemmas~\ref{lem:sample-flow-support}, \ref{lem:cut-covering} and~\ref{lem:exposed-region} run in deterministic polynomial time,
    the algorithm of this lemma also runs in deterministic polynomial time.
    \end{proof}

    We will see that Lemmas~\ref{lem:relative-lslf} and~\ref{lem:exposed-region}, will imply Lemma~\ref{lem:second-cut-marking}.

    \begin{restatable}[Confining the unmarked vertices of an optimum
    to small components]{lemma}{secondcutmarking}
    \label{lem:second-cut-marking}
    Let $(G,T)$ be a terminal network.
    Let $T\subseteq M_1\subseteq V(G)$ be such that,
     for every no-small-tail query $(T_0,\mathcal{T})$ of $(G,T)$,
        and every minimum $\mathcal{T}$-\mwc\ $X$ in $G-T_0$,
        $|X\setminus M_1|\leq 4|T|/15$.
    
    There is a polynomial-time algorithm that either 
    \begin{enumerate}
        \item returns a $\frac{4}{5}|T|$-recursively useful region $R$, or
       \item the set $M_2=\cutcover(G,N_G[M_1])$ is such that,
    for every no-small-tail query $(T_0,\mathcal T)$,
    there exists a minimum $\mathcal T$-\mwc\ $X$ in $G-T_0$
    such that every vertex of $X\setminus M_2$ lies in a
    connected component of $G-M_2$ of size at most
    $3\cdot\bound(\lfloor4|T|/5\rfloor)$.
    \end{enumerate}
    \end{restatable}
    
    \begin{proof}
    Set $k:=|T|$ and compute
    $M_2:=\cutcover(G,N_G[M_1])$ using Lemma~\ref{lem:cut-covering}.
    
    By Lemma~\ref{lem:cut-covering},
    $T\subseteq M_1\subseteq N_G[M_1]\subseteq M_2$.
    If $k=0$, return $M_2$.
    In this case every minimum terminal multiway cut is
    empty, so the required conclusion holds.
    
    Hence assume that $k\ge1$ and set
    $h:=\lfloor4k/5\rfloor$.
    Apply Lemma~\ref{lem:exposed-region} to $(G,T,M_2,h)$.
    If it returns an $h$-recursively useful region $R$,
    return $R$.
    Since $h\le4k/5$, this region satisfies the first
    outcome of the lemma.
    Otherwise, $M_2$ is an $h$-\lslf\ modulator.
    Return $M_2$.
    
    It remains to prove the required conclusion when
    $M_2$ is returned.
    Fix a no-small-tail query $(T_0,\mathcal T)$.
    Among all minimum $\mathcal T$-\mwc{} s in $G-T_0$,
    choose $X$ minimizing $|X\setminus M_2|$.
    By the hypothesis of the lemma,
    $|X\setminus M_1|\le4k/15$.
    
    Apply Lemma~\ref{lem:relative-lslf} with
    $A:=M_1$ and marked set $M:=M_2$.
    Every connected component $D$ of $G-M_2$ with
    $V(D)\cap X\ne\emptyset$ satisfies
    \[
     \flow_G(V(D),T)
     \le3|X\setminus M_1|
     \le\frac{4k}{5}.
    \]
    Since $\flow_G(V(D),T)$ is an integer,
    $\flow_G(V(D),T)\le\lfloor4k/5\rfloor=h$.
    As $M_2$ is an $h$-\lslf\ modulator, it follows that
    $ |V(D)|\le3\cdot\bound(h)$.
    Thus every vertex of $X\setminus M_2$ lies in a
    connected component of $G-M_2$ of size at most
    $3\cdot\bound(\lfloor4k/5\rfloor)$.
    The query was arbitrary, so this conclusion holds
    for every no-small-tail query.
    
    The algorithm consists of using the algorithms of Lemma~\ref{lem:cut-covering} and~\ref{lem:exposed-region}, both of which run in deterministic polynomial time.
    \end{proof}

    \lemnosmalltail*

    \begin{proof}
    Set $n:=|V(G)|$ and $k:=|T|\ge2$.
    
    If $k\le3$, return $M:=T$.
    Every query $(T_0,\mathcal{T})$ admits a small-tail solution, since there are at
    most $k\le3$ $T$-components after deletion of any solution.
    Thus the lemma holds vacuously
    and because $(G,T)$ is \bg,
    $|M| =|T| =k<n/16$.
    
    Hence assume $k\ge4$.
    Apply Lemma~\ref{lem:marking-a-large-fraction} to $(G,T)$.
    If it returns an $\frac{14}{15}k$-recursively useful region,
    return that region.
    Otherwise, let $M_1$ be its returned marked set.
    Then $T\subseteq M_1$,
    $ |M_1|\leq c_{\texttt{cover}}(k+120k^6)^3$,
    and, for every no-small-tail query $(T_0,\mathcal T)$ and
    every minimum $\mathcal T$-\mwc\ $X$ in $G-T_0$,
    $|X\setminus M_1|\le4k/15$.
    
    Apply Lemma~\ref{lem:second-cut-marking} with this $M_1$.
    If it returns a $\frac{4}{5}k$-recursively useful region $R$,
    return $R$.
    Since $4k/5\le14k/15$, this region also satisfies the first
    outcome of the present lemma.
    
    Otherwise, let
    $M_2:=\cutcover(G,N_G[M_1])$, and 
    $ h:=\lfloor4k/5\rfloor$.
    Let $\mathcal{D}$ consist of all connected components $D$
    of $G-M_2$ satisfying
    $|V(D)|\le3\cdot\bound(h)$.
    Return
    \[
     M:=M_2\cup\bigcup_{D\in\mathcal D}V(D).
    \]
    From Lemma~\ref{lem:cut-covering},
    we have
    $T\subseteq M_1\subseteq N_G[M_1]\subseteq M_2\subseteq M$.
    
    We now observe that $M$ satisfies the guarantee of the lemma.
    Fix any no-small-tail query $(T_0,\mathcal T)$.
    By Lemma~\ref{lem:second-cut-marking}, there is a minimum
    $\mathcal T$-\mwc\ $X$ in $G-T_0$ such that every vertex of
    $X\setminus M_2$ belongs to a component of $G-M_2$ with at
    most $3\cdot\bound(h)$ vertices.
    All these components belong to $\mathcal D$.
    Consequently, $X\subseteq M$.
    Since the query $(T_0,\mathcal{T})$ was an arbitrary no-small-tail query, $M$ contains a minimum cut
    for every no-small-tail query, simultaneously.
    
    It remains to bound $|M|$.
    From Lemma~\ref{lem:marking-a-large-fraction}, $|M_1| \leq c_{\texttt{cover}} (k + 120k^6)^3$.
    Since $(G,T)$ is slim, its maximum degree is at most $2k$.
    Therefore
    $|N_G[M_1]|\le(2k+1)|M_1|\le3k|M_1|$.
    From Lemma~\ref{lem:cut-covering},
    \[
     |M_2| \leq  c_{\texttt{cover}} |N[M_1]|^3 
     \leq c_{\texttt{cover}}(3k|M_1|)^3
    \]

    Since $G$ is \slim,
    $G$ has at most $k$ connected components and
    maximum degree of $G$ is at most $2k$.
    Hence $G-M_2$ has at most $k+2k|M_2|$ connected components.
    It follows that
    \begin{align*}
     |M|
     &\leq |M_2|+3(k+2k|M_2|)\bound(h)\\
     &\le7k|M_2|\bound(h)\\
     &<n/16.
    \end{align*}
    For the second inequality, we used $|M_2|\geq k\ge4$ and
    $\bound(h)\ge1$, which imply
    $|M_2|+3k\bound(h)\leq k|M_2|\bound(h)$.
    The final inequality is Lemma~\ref{lem:relative-slack}.
    
    Since both Lemmas~\ref{lem:marking-a-large-fraction} and~\ref{lem:second-cut-marking} have deterministic polynomial-time
    algorithms,
    computing the components of $G-M_2$ and forming
    $M$ also take polynomial time.
    Thus the entire procedure is deterministic polynomial-time.
    \end{proof}

    \section{The small-tail case}
    \label{sec:small-tail-case}
    
    We prove Lemma~\ref{lem:smalltail} using representative families.
    Recall that a \emph{small-tail query} is a pair $(T_0, \mathcal{T})$
    such that $G-T_0$ admits a minimum $\mathcal{T}$-\mwc~$X$
    with the following property. Let $V_1, \ldots, V_r$ be the
    $T$-components of $G-(T_0 \cup X)$ in order of decreasing cardinality.
    Then the number of vertices in components $V_4$ through $V_r$, if
    these components exist, is in total at most $n/(12k^3)$.

    \lemsmalltail*
    
    The algorithm returns a marked set containing every vertex
    essential for any small-tail query.
    The construction follows the representative-family marking
    arguments of~\cite{DBLP:journals/jacm/KratschW20,
    DBLP:journals/talg/Wahlstrom22}.
    The original algorithms use randomized gammoid
    representations. 
    For determinism, we use almost multilinear representations of~\cite{DBLP:journals/corr/abs-2609-29141}.
    The required unequal-rank representative-family bound is proved
    in Appendix~\ref{app:deterministic-representative-families}.

    We proceed with some preliminaries below.
    \subsection{Definitions, matroids and representative families}
    
    \paragraph{Matroids and representative families.}
    A \emph{matroid} is a pair $\mathcal M=(E,\cI)$ where
    $\cI\subseteq 2^E$ is a set of \emph{independent sets} in $\mathcal M$ such that (i) $\emptyset \in \cI$,
    (ii) if $A \in \cI$ and $B \subset A$ then $B \in \cI$, 
    and (iii) if $A, B \in \cI$ such that $|A|<|B|$ then there
    exists $x \in B \setminus A$ such that $A \cup \{x\} \in \cI$.
    For a matroid $\mathcal M=(E,\mathcal I)$ and $F\subseteq E$,
    the \emph{rank} $r_{\mathcal M}(F)$ is the maximum size of an independent
    subset of $F$. The rank of $\mathcal M$ is $r_{\mathcal M}(E)$.
    For more on matroids, see Oxley~\cite{OxleyBook2};
    for more on their applications in kernelization, see~\cite{Book_kernelization_FLSZ19}.
    We follow the matroid-based framework of previous work~\cite{DBLP:journals/jacm/KratschW20,DBLP:journals/talg/Wahlstrom22} and here include only the most essential definitions. 
    
    For $0\leq R\le|E|$, the {\em uniform matroid} of rank $R$
    on $E$ has independent sets
    $\{F\subseteq E:|F|\leq R\}$.
    For a directed graph $D$ and sets $S,E\subseteq V(D)$,
    the {\em gammoid} $\Gamma(D,S,E)$ has ground set $E$.
    A set $F\subseteq E$ is {\em independent} if there are $|F|$
    pairwise vertex-disjoint directed paths from $S$ to $F$.
    Paths of length zero are allowed.
    
    The {\em direct sum of matroids} on pairwise disjoint ground
    sets is the matroid in which a set is independent precisely
    when its intersection with each ground set is independent.
    We refer to the summands as layers.
    
    Let $A$ and $B$ be independent sets of a matroid.
    We say that {\em $A$ extends $B$} if $A\cap B=\emptyset$
    and $A\cup B$ is independent.
    A subfamily $\widehat{\mathcal Y}\subseteq\mathcal Y$
    {\em represents} a family $\mathcal Y$ of independent sets if,
    for every independent set $B$, whenever some member of
    $\mathcal Y$ extends $B$, some member of
    $\widehat{\mathcal Y}$ also extends $B$.
    
    The following bound retains the product of the layer ranks.
    Its proof uses the deterministic almost multilinear machinery of Koana and
    Kumabe~\cite{DBLP:journals/corr/abs-2609-29141}.
    
    \begin{lemma}[Deterministic representative families]
    \label{lem:repset-product}
    Let $d\ge2$ be fixed.
    Let $N$ be a gammoid of rank $\rho\ge1$, specified by
    its directed graph, source set and ground set, and let
    $L$ be a uniform matroid of rank $R\ge1$ on an explicitly
    given ground set.
    Let $\mathcal M$ be the direct sum of $d-1$ copies of $N$
    and one copy of $L$, on pairwise disjoint ground sets.
    
    Given an explicit family $\mathcal Y$ of independent sets
    of $\mathcal M$, each containing precisely one element from
    each layer, one can deterministically compute a representative
    family $\widehat{\mathcal Y}\subseteq\mathcal Y$ of size
    at most $2\rho^{d-1}R$ in polynomial time.
    \end{lemma}
    
    The proof of Lemma~\ref{lem:repset-product} is given in
    Appendix~\ref{app:deterministic-representative-families}.
    
    \paragraph*{Properties of gammoids and multiway cuts.}
    We now summarize the properties of gammoids which make them a useful tool for the kernelization of multiway cut problems. 
    
    \begin{definition}[Closest separator]
    A minimum $(A,B)$-separator $X$ is
    {\em closest} to $B$ if
    $\reach_{G-X}(A\setminus X)$ is inclusion-wise
    maximum among all minimum $(A,B)$-separators.
    The closest minimum $(A,B)$-separator is unique
    (cf.~\cite{DBLP:journals/jacm/KratschW20})
    and we denote it by $\mathcal C_G(A,B)$,
    or simply by $\mathcal C(A,B)$ when the graph is
    clear from context.
    \end{definition}
    
    We use the following characterization of closest
    minimum separators by linkages to sink copies
    (cf.~\cite[Proposition~3.5]{DBLP:journals/jacm/KratschW20}).
    To add a \emph{sink-only copy} (or sink copy) of a vertex $v$ in a
    digraph $D$ refers to creating a new vertex $v'$ whose in-neighbors
    are the in-neighbors of $v$ in $D$ and with no outgoing arcs.

    \begin{proposition}
    \label{prop:closest-sinkcopy}
    Let $H$ be a graph.
    Construct a directed graph $D_H$ by replacing every edge
    $uv$ of $H$ by the arcs $uv$ and $vu$ and adding a 
    sink-only copy $x'$ of every vertex $x\in V(H)$, i.e.\ adding the
    vertex $x'$ and the arc $ux'$ for every $u\in N_H(x)$,
    with no outgoing arcs.
    
    Let $S,X\subseteq V(H)$, and suppose that there are
    $|X|$ pairwise vertex-disjoint paths from $S$ to $X$ in $H$.
    For every $x\in X\setminus S$, the set $X\cup\{x'\}$
    is independent in $\Gamma(D_H,S,V(D_H))$ if and only if
    $x\in\mathcal C_H(X,S)$.
    \end{proposition}
    
    \begin{definition}[Central vertices]
    Let $(G,T)$ be a terminal network, $T_0 \subseteq T$ and $\mathcal{T}=\{T_1, \ldots, T_r\}$ be a partition of $T \setminus T_0$.
    Let $X$ be a minimum $\mathcal T$-\mwc
    in $G-T_0$.
    We say that $v\in V(G)\setminus T$ is {\em central to
    $X$} with respect to $(T_0,\mathcal T)$ if,
    for every $i\in[r]$, $v$ belongs to the
    minimum $(X,T\setminus(T_0\cup T_i))$-separator
    in $G-T_0$ closest to
    $T\setminus(T_0\cup T_i)$.
    \end{definition}

    The following proposition establishes a relation
    between essential and central vertices.
    \par
    \begin{restatable}[Lemma~5.11,~\cite{DBLP:journals/jacm/KratschW20}]{proposition}{propessentialcentral}
    \label{prop:relation-essential-central}
    Let $(G,T)$ be a terminal network.
    Let $T_0\subseteq T$ and let
    $\mathcal T=\{T_1,\ldots,T_r\}$ be a partition of
    $T\setminus T_0$.
    Let $X$ be a minimum $\mathcal T$-\mwc
    in $G-T_0$.
    The following holds:
    \begin{enumerate}
    \item for every $i\in[r]$,
    $X_i:=\mathcal C_{G-T_0}(X,T\setminus(T_0\cup T_i))$
    is a $\mathcal T$-\mwc in $G-T_0$ and
    $|X_i|\leq |X|$, and
    \item if $\mathcal T\ne\emptyset$, then for every
    $v\in V(G)\setminus T$, if $v$ is central
    for $X$, then $v\in X$.
    \end{enumerate}
    Therefore, if $v\in V(G)\setminus T$ is essential
    for $(T_0,\mathcal T)$, then for any minimum
    $\mathcal T$-\mwc $X$ in $G-T_0$,
    $v$ is central for $X$.
    \end{restatable}

    \subsection{Proof of Lemma~\ref{lem:smalltail}}
    
    \begin{proof}
    Set $n:=|V(G)|$ and $k:=|T|\ge2$.
    Since the terminal network $(G,T)$ is \bg, $n>\bound(k)$.
    
    Construct a directed graph $D$ from $G$ as in Proposition~\ref{prop:closest-sinkcopy}
    by replacing every edge
    $uv$ of $G$ by the arcs $uv$ and $vu$, adding a sink-only
    copy $v'$ of every vertex $v\in V(G)$, and adding the
    arcs $uv'$ and $vu'$ for every edge $uv$ of $G$.
    Thus a path can enter $v'$ through a neighbor of $v$,
    but cannot continue through $v'$.
    Set
    \[
     R:=\min\{n,\lfloor n/(12k^3)\rfloor+k\}.
    \]
    Let $\mathcal M$ be the direct sum of four matroids
    on pairwise disjoint ground sets:
    \begin{itemize}
    \item for $i\in[3]$, $\mathcal M_i$ is a copy of
    $\Gamma(D,T,V(D))$, whose rank is $k$;
    \item $\mathcal M_4$ is a uniform matroid of rank $R$
    on a copy of $V(G)$.
    \end{itemize}

    Let $W$ be the set of non-terminals belonging to
    $T$-components of $G$.
    For every $v\in W$, let $Y(v)$  consist of the copy of $v'$ in
    each of the first three layers and the copy of $v$
    in the uniform layer.
    Let $\mathcal Y$ be the family of these tuples.
    
    Every member of $\mathcal Y$ is independent:
    $v'$ is reachable from $T$ in $D$, and the
    uniform-layer element is independent since $R\ge1$.
    The family $\mathcal Y$ can be constructed by a
    reachability computation.
    
    Apply Lemma~\ref{lem:repset-product} with $d=4$
    and $\rho=k$ to obtain a representative family
    $\widehat{\mathcal Y}\subseteq\mathcal Y$ of size
    at most $2k^3R$.
    Set
    \[
    U:=\{v\in W:Y(v)\in\widehat{\mathcal Y}\}.
    \]
    and return $ M:=T\cup U$.
    
    We prove that $M$ contains every vertex essential for
    any small-tail query.
    Fix a small-tail query $(T_0,\mathcal T)$ and let $X$ be a minimum
    $\mathcal T$-\mwc\ in $G-T_0$
    such that if $V_1,\ldots,V_r$ are the $T$-components of $G-(X \cup T_0)$,
    ordered by non-increasing size
    then
    \[
     \sum_{i=4}^r|V_i|
     \le\left\lfloor\frac{n}{12k^3}\right\rfloor,
    \]
    where an empty sum is zero.
    
    Every terminal is marked.
    By Proposition~\ref{prop:relation-essential-central},
    every non-terminal essential for $(T_0,\mathcal T)$
    is central to $X$.
    It therefore suffices to show that every central
    vertex $v\in X\setminus T$ belongs to $U$.
    Fix such a vertex $v$ and write
    $\mathcal T=\{T_1,\ldots,T_q\}$.
    Since $X$ is minimum, it contains no vertex of an
    original terminal-free component: restoring such a
    vertex cannot change terminal connectivity.
    In particular, $v\in W$, so $Y(v)$ is defined.
    
    For each $i\le\min\{r,3\}$, choose $j_i\in[q]$
    such that $T\cap V_i\subseteq T_{j_i}$.
    For each remaining index $i\in[3]$, choose any
    $j_i\in[q]$.
    
    Construct a set $B$ in $\mathcal M$ as follows.
    In each gammoid layer $\mathcal M_i$, put the copies
    of the original vertices in $X \cup T_0\cup T_{j_i}$.
    In the uniform layer, put the copies of
    \[
     B_4:=(X\setminus\{v\})\cup\bigcup_{i=4}^rV_i.
    \]
    We show that $B$ is independent and that $Y(v)$
    is the unique member of $\mathcal Y$ extending $B$.
    
    Fix $i\in[3]$.
    By Proposition~\ref{prop:relation-essential-central}
    and the minimum cardinality of $X$, the minimum size
    of an
    $(X,T\setminus(T_0\cup T_{j_i}))$-separator in $G-T_0$
    is $|X|$.
    Hence there is a linkage from
    $T\setminus(T_0\cup T_{j_i})$ to $X$ in $G-T_0$.
    Centrality of $v$ and
    Proposition~\ref{prop:closest-sinkcopy} give a linkage
    from this source set to $X\cup\{v'\}$ in $D$ avoiding $T_0$.
    
    All vertices of $X$ are occupied targets of the linkage,
    so no path uses another vertex of $X$ internally.
    Since $X$ is a $\mathcal T$-\mwc\ in $G-T_0$,
    these paths also avoid $T_{j_i}\setminus X$.
    Replace the paths ending in $X\cap T_{j_i}$
    by zero-length paths at those vertices, and add
    zero-length paths at $(T_{j_i}\setminus X)\cup T_0$.
    This produces a linkage from $T$ to
    \[
     X \cup T_0\cup T_{j_i}\cup\{v'\}.
    \]
    Thus the copy of $v'$ extends the part of $B$
    in layer $\mathcal M_i$.
    
    In the uniform layer, $v\notin B_4$ and
    \[
     |B_4|
     \leq |X|-1+\left\lfloor\frac{n}{12k^3}\right\rfloor
     \leq k-1+\left\lfloor\frac{n}{12k^3}\right\rfloor.
    \]
    Also, $|B_4|\leq n-1$.
    Consequently $|B_4|<R$, so the copy of $v$
    extends the part of $B$ in the uniform layer.
    It follows that $B$ is independent and $Y(v)$
    extends $B$. In particular, $Y(v)\in\mathcal Y$.
    
    Now consider any other candidate $u\ne v$.
    If $u\in X\setminus\{v\}$ or
    $u\in\bigcup_{i=4}^rV_i$, its uniform-layer copy
    belongs to $B$, so $Y(u)$ does not extend $B$.
    
    Otherwise, $u$ lies in one of the first three
    $T$-components or in a terminal-free component of $G-(X \cup T_0)$.
    In the former case, choose the corresponding layer $i$;
    in the latter case, choose any gammoid layer.
    Every terminal of the component containing $u$
    belongs to $T_{j_i}$.
    
    Suppose that the copy of $u'$ extends the copies
    of $X \cup T_0\cup T_{j_i}$ in this layer.
    Take a corresponding linkage from $T$ to
    $X \cup T_0 \cup T_{j_i}\cup\{u'\}$, and replace every path
    ending at an original terminal by the zero-length
    path at that terminal.
    The path ending at $u'$ must then start in
    $T\setminus(X \cup T_0 \cup T_{j_i})$ and avoid $X \cup T_0$.
    
    Every original neighbor of $u$ belongs to its component
    of $G-(X \cup T_0)$ or to $X \cup T_0$.
    Since the path avoids $X \cup T_0$, its last original vertex
    belongs to the component containing $u$.
    Its entire original portion, including its terminal
    source, therefore lies in this component.
    This contradicts the fact that all terminals of this
    component belong to $T_{j_i}$.
    
    Thus $Y(v)$ is the unique member of $\mathcal Y$
    extending $B$.
    Representativeness forces
    $Y(v)\in\widehat{\mathcal Y}$, and hence $v\in U$.
    Therefore $M$ contains every vertex essential for
    $(T_0,\mathcal T)$.
    Since the query was arbitrary, this holds
    simultaneously for all small-tail queries.
    
    Finally, distinct vertices define distinct tuples, so
    \[
     |M|\leq k+|\widehat{\mathcal Y}|
          \leq k+2k^3R<\frac{3n}{16},
    \]
    where the last inequality is
    Lemma~\ref{lem:deterministic-smalltail-size}.
    The reachability computation and the representative-family
    computation are deterministic and take polynomial time.
    \end{proof}

\section{Kernelization applications}\label{sec:kernelization}
    
Recall the definitions of {\sc Vertex Multiway Cut}, {\sc Vertex Multicut} and {\sc Group Feedback Vertex Set} from Section~\ref{sec:intro}, which are restated below for convenience.

\defparproblem{\textsc{Vertex Multiway Cut}}{An undirected graph $G$, a set $T\subseteq V(G)$ of terminals and a non-negative integer $k$}
    {$k$}
    {Does there exist a set $X\subseteq V(G)\setminus T$
    of size at most $k$ such that every connected component
    of $G-X$ contains at most one terminal?}

    \defparproblem{\textsc{Vertex Multicut}}
    {An undirected graph $G$, a set
    $\mathcal T\subseteq\binom{V(G)}2$ of size $p$,
    and a non-negative integer $k$}
    {$k+p$}
    {Does there exist a set
    $X\subseteq V(G)\setminus
     \bigcup_{\{s,t\}\in\mathcal T}\{s,t\}$
    of size at most $k$ such that in $G-X$ there is no path from $s$ to $t$ for any $(s,t) \in \mathcal{T}$?}

  \defparproblem{\textsc{Group Feedback Vertex Set}}
{A finite group $(\Gamma,\otimes)$ specified by its
multiplication table, a directed graph $D$, an arc labelling
$\phi:E(D)\to\Gamma$, and a non-negative integer $k$}
{$k+|\Gamma|$}
{Does there exist $X\subseteq V(D)$ of size at most $k$
such that $D-X$ has a consistent labelling, that is,
a function $\pi:V(D)\setminus X\to\Gamma$ satisfying
$\pi(u)\otimes\phi((u,v))=\pi(v)$
for every $(u,v)\in E(D-X)$?}

    \thmkernelmwc*
    
    \begin{proof}
        Let $(G,T,k)$ be an instance of {\sc Vertex Multiway Cut}.
        We use the result of Guillemot~\cite{DBLP:journals/disopt/Guillemot11a}, that takes an instance $(G,T,k)$ and runs in polynomial-time,
        and  either reports that it is a no-instance,
        or returns an equivalent instance $(G',T,k)$ of {\sc Vertex Multiway Cut} such that
        $G'$ is a contraction of $G$ and 
        $|N_{G'}(T)| \leq 2k$.
        In the former case, return a trivial constant-sized no-instance.
        In the latter case,
        it is enough to solve the problem on $(G',T,k)$ where $|N_{G'}(T)| \leq 2k$.

        If there exists $t,t' \in T$ such that $tt' \in E(G')$,
        then report a no-instance.
        Thus, we can assume, going forward, that $T$ is an independent set in $G'$.

        Let $ W:=\{v\in V(G')\setminus T:|N_{G'}(v)\cap T|\ge2\}$.
        Observe that every solution of $(G',T,k)$ contains $W$, since each vertex
        of $W$ lies on a path of length two between distinct terminals.
        If $|W|>k$, return a constant-sized no-instance.
        Otherwise, 
        let $G'':=G'-(T\cup W)$,
        $T^*:=N_{G'}(T)\setminus W$ and $k':=k-|W|$.
       
        For each $t\in T$, let $B_t:=N_{G'}(t)\setminus W$ and let $ \mathcal B:=\{B_t:t\in T,\ B_t\ne\emptyset\}$. By the definition of $W$, the nonempty sets $B_t$ are
    pairwise disjoint. Thus $\mathcal B$ is a partition of $T^*$
    and $|T^*|\le2k$.
    If $T^*=\emptyset$, return a constant-sized yes-instance.
    
    For every $X\subseteq V(G'')$, we claim that $X\cup W$
    is a multiway cut for $T$ in $G'$ if and only if $X$
    is a $\mathcal B$-\mwc\ in $G''$.
    Indeed, a path in $G''-X$ between $B_s\setminus X$ and
    $B_t\setminus X$, for distinct $s,t\in T$, extends to an
    $s$--$t$ path in $G'-(X\cup W)$.
    Conversely, if $X$ is a $\mathcal B$-\mwc, every component
    of $G''-X$ meets at most one part of $\mathcal B$.
    Adding the terminals of $T$ therefore joins only components
    associated with the same terminal, so distinct terminals
    remain disconnected.
    Since every multiway cut for $T$ in $G'$ contains $W$, it follows that
    $(G',T,k)$ is a yes-instance if and only if $G''$ has a
    $\mathcal B$-\mwc\ of size at most $k'$.
    
    Apply Corollary~\ref{thm:full-vertex-mimicking}, that runs in deterministic polynomial time, to the terminal
    network $(G'',T^*)$ (where the terminals in $T^*$ are deletable).
    Let $(G^\dagger,T^*)$ be the resulting vertex
    multicut-mimicking network.
    Then
    $|V(G^\dagger)|\leq |T^*|^{O(\log |T^*|)}$,
    and the minimum size of a $\mathcal B$-\mwc\ is the same
    in $G''$ and $G^\dagger$.

    For each $B\in\mathcal B$, add a new vertex $z_B$ to
    $G^\dagger$, adjacent precisely to the vertices of $B$.
    Denote the resulting graph by $H$ and let $Z:=\{z_B:B\in\mathcal B\}$.
    Output the instance $(H,Z,k')$.
    Since $|T^*|\le2k$ and
    $|Z|=|\mathcal B|\le|T^*|$, the graph $H$ has
    $k^{O(\log k)}$ vertices.

    For every $Y\subseteq V(G^\dagger)$, the set $Y$ is a
    multiway cut for $Z$ in $H$ if and only if it is a
    $\mathcal B$-\mwc\ in $G^\dagger$.
    Indeed, a path between distinct parts of $\mathcal B$
    extends to a path between their corresponding terminals
    in $Z$.
    Conversely, if the parts are separated in $G^\dagger-Y$,
    adding $z_B$ joins only components meeting the same part $B$,
    so no two terminals in $Z$ become connected.
    Consequently,
    \[
    \begin{aligned}
     (G,T,k)\text{ is a yes-instance}
     &\iff (G',T,k)\text{ is a yes-instance}\\
     &\iff G''\text{ has a }\mathcal B\text{-}\mwc
                             \text{ of size at most }k'\\
     &\iff G^\dagger\text{ has a }\mathcal B\text{-}\mwc
                             \text{ of size at most }k'\\
     &\iff (H,Z,k')\text{ is a yes-instance}.
    \end{aligned}
    \]
    
    \end{proof}

    \thmkernelmc*

    \begin{proof}
    Let $(G,\mathcal T,k)$ be an instance of
    \textsc{Vertex Multicut}, where $|\mathcal T|=p$, and put
    $T:=\bigcup_{\{s,t\}\in\mathcal T}\{s,t\}$.
    Thus $|T|\le2p$, and the vertices of $T$ are undeletable.
    Without loss of generality, $k,p \geq 2$.
    
    For each $t\in T$, replace $t$ by an independent set $D_t$
    of $k+1$ new vertices.
    For each edge $tv\in E(G)$ with $t\in T$ and
    $v\in V(G)\setminus T$, join $v$ to every vertex of $D_t$.
    For each edge $st\in E(G)$ with $s,t\in T$, add all edges
    between $D_s$ and $D_t$. Retain all edges of $G-T$.
    Denote the resulting graph by $G^+$, and put
    $T^+:=\bigcup_{t\in T}D_t$ and $q:=|T^+|=|T|(k+1)\le2p(k+1)$.
    Regard $(G^+,T^+)$ as a terminal network with deletable
    terminals. 
    For a partition $\mathcal P$ of $T$, let
    $\mathcal P^+:=
     \left\{\bigcup_{t\in P}D_t:P\in\mathcal P\right\}$.
    
    For every $X\subseteq V(G)\setminus T$, the set $X$ is a
    $\mathcal P$-\mwc\ in $G$ if and only if it is a
    $\mathcal P^+$-\mwc\ in $G^+$.
    Indeed, paths between terminals in $G-X$ lift to paths
    between their corresponding classes in $G^+-X$.
    Conversely, identifying each class $D_t$ with $t$ maps
    any such path to a walk in $G-X$.
    
    Furthermore, suppose that $X$ is a $\mathcal P^+$-\mwc\
    in $G^+$ with $|X|\leq k$.
    Each $D_t$ contains a vertex outside $X$.
    Since the vertices of $D_t$ have identical neighborhoods,
    restoring the vertices of $D_t\cap X$ cannot connect
    terminals belonging to different parts of $\mathcal P^+$.
    Consequently, $X\setminus T^+$ is also a
    $\mathcal P^+$-\mwc\ in $G^+$.
    
    Apply Theorem~\ref{thm:budgeted-vertex-mimicking} to
    $(G^+,T^+)$ with budget $k$, and let $(H,T^+)$ be
    the resulting graph. Then
    $|V(H)|=q^{7}k^{O(\log k)}$.

    We claim that, for every partition $\mathcal P$ of $T$,
    every minimum $\mathcal P^+$-\mwc\ in $H$ of size at most
    $k$ is disjoint from $T^+$.
    To prove the claim, let $X$ be a minimum
    $\mathcal P^+$-\mwc\ in $H$ with $|X|\leq k$.
    Put $a:=|X|$ and $F:=X\cap T^+$.
    Let $\mathcal P^+_F$ be the partition obtained by removing
    $F$ from the parts of $\mathcal P^+$ and omitting empty
    parts.
    Then $X\setminus F$ is a $\mathcal P^+_F$-\mwc\
    in $H-F$ of size $a-|F|\leq k$.
    By Theorem~\ref{thm:budgeted-vertex-mimicking}, the minimum
    sizes of a $\mathcal P^+_F$-\mwc\ of size at most $k$, in $H-F$ and in
    $G^+-F$ agree.
    Since the former is at most $a-|F|\leq k$, there is a
    $\mathcal P^+_F$-\mwc\ $Y$ in $G^+-F$ with
    $|Y|\leq a-|F|$.
    Thus $F\cup Y$ is a $\mathcal P^+$-\mwc\ in $G^+$
    of size at most $a\leq k$.
    By the preceding observation, $(F\cup Y)\setminus T^+$
    is also such a multiway cut, and its size is at most
    $a-|F|$.
    The minimum size of a $\mathcal P^+$-\mwc\ in $G^+$
    is $a$, by the mimicking guarantee.
    Therefore $|F|=0$, proving the claim.
    
    Construct $H'$ from $H$ by identifying, for each $t\in T$,
    all vertices of $D_t$ into one vertex named $t$.
    Remove loops and duplicate edges.
    Output $(H',\mathcal T,k)$.
    The endpoints of its requests are precisely the vertices
    of $T$, so every solution is required to be disjoint from
    $T$ by the definition of \textsc{Vertex Multicut}.
    
    Suppose first that $(G,\mathcal T,k)$ is a yes-instance,
    and let $X\subseteq V(G)\setminus T$ be a
    $\mathcal T$-multicut of size at most $k$.
    Let $\mathcal P$ be the partition of $T$ induced by the
    connected components of $G-X$.
    The endpoints of every request belong to different parts
    of $\mathcal P$.
    By the first observation, $X$ is a $\mathcal P^+$-\mwc\
    in $G^+$.
    Hence $H$ has a minimum $\mathcal P^+$-\mwc\ $Y$
    of size at most $k$, and the claim gives
    $Y\cap T^+=\emptyset$.
    Identifying the vertices of $D_t$ joins only components
    whose terminals belong to the same part of $\mathcal P^+$.
    Consequently, $Y$ is a $\mathcal P$-\mwc\ in $H'$,
    disjoint from $T$.
    It therefore separates every pair in $\mathcal T$,
    so $(H',\mathcal T,k)$ is a yes-instance.
    
    Conversely, suppose that $Y\subseteq V(H')\setminus T$
    is a $\mathcal T$-multicut in $H'$ of size at most $k$.
    Let $\mathcal P$ be the partition of $T$ induced by the
    connected components of $H'-Y$.
    Since identification can only increase connectivity, $Y$ is
    a $\mathcal P^+$-\mwc\ in $H$.
    The mimicking guarantee gives a $\mathcal P^+$-\mwc\
    $X$ in $G^+$ of size at most $|Y|\leq k$.
    By the second observation, $X\setminus T^+$ is also
    such a multiway cut.
    By the first observation, it is a $\mathcal P$-\mwc\
    in $G$, disjoint from $T$.
    As every requested pair has its endpoints in different
    parts of $\mathcal P$, this is a $\mathcal T$-multicut
    of size at most $k$ in $G$.
    This proves the equivalence of the two instances.
    
    Since $q\le2p(k+1)$ and identification does not increase
    the vertex count, we have
    $ |V(H')|\le|V(H)|
     =p^{O(1)} k^{O(\log k)}$.
    The total encoding length of the output, including the
    requests and the budget, is polynomial in $|V(H')|+p+k$,
    and hence is also
    $p^{O(1)} k^{O(\log k)}$.
    \end{proof}

    \thmgroupfvs*
    
    \begin{proof}
     Let $(\Gamma,\otimes)$ be a finite group with identity
$1_\Gamma$, and enumerate its elements as
$\Gamma=\{\alpha_1,\ldots,\alpha_s\}$, where
$s:=|\Gamma|$ and $\alpha_1=1_\Gamma$.
    Let $(D,\phi,\Gamma,k)$ be an instance of
\textsc{Group Feedback Vertex Set}.
    Without loss of generality,
    $s, k \geq 2$ and $k < |V(D)|$.
    
    By the approximation result of Kratsch and
    Wahlstr\"om~\cite[Section~7.3]{DBLP:journals/jacm/KratschW20},
    we can, in deterministic polynomial time, either correctly
    conclude that the instance $(D,\phi,\Gamma,k)$ 
    is a no-instance or find a set
    $Z\subseteq V(D)$ with $|Z|=O(s^2k^2)$ such that $D-Z$
    has a consistent labelling.
    This procedure remains polynomial when $\Gamma$ is supplied
by its multiplication table: its auxiliary graph has
$s|V(D)|$ vertices and $O(s^2|V(D)|)$ terminal pairs,
and group operations are evaluated using the table.
 If $Z=\emptyset$, return a yes-instance consisting of an
empty directed graph, the same group $\Gamma$, and budget $k$.

    Let $\sigma:V(D)\setminus Z\to\Gamma$ be a consistent
    labelling of $D-Z$, and extend it to $V(D)$ by setting
    $\sigma(z):=1_\Gamma$ for $z\in Z$.
    Define a new arc labelling by
    \[
     \phi_0((u,v)):=
     \sigma(u)\otimes\phi((u,v))\otimes\sigma(v)^{-1}.
    \]
    For every $X\subseteq V(D)$, the map
    $\pi\mapsto\pi_0$, where
    $\pi_0(v):=\pi(v)\otimes\sigma(v)^{-1}$, is a bijection
    between the consistent labellings of $(D-X,\phi)$ and
    $(D-X,\phi_0)$.
    Thus the two instances have the same solutions.
    Moreover, every arc of $D-Z$ has label $1_\Gamma$
    under $\phi_0$.
    Henceforth write $\phi$ for $\phi_0$.
    
    Construct an undirected graph $G$ from the underlying graph
    of $D-Z$ by adding a vertex $z(\alpha)$ for each
    $z\in Z$ and $\alpha\in\Gamma$.
    For each arc $(z,v)$ with $z\in Z$ and $v\notin Z$,
    add the edge $z(\phi((z,v)))v$.
    For each arc $(v,z)$ with $z\in Z$ and $v\notin Z$,
    add the edge $z(\phi((v,z))^{-1})v$.
    Put
    $T:=\{z(\alpha):z\in Z,\ \alpha\in\Gamma\}$ and $q:=|T|=s|Z|=O(s^3k^2)$,
    and regard $(G,T)$ as a terminal network with deletable
    terminals.
    
    For $W\subseteq Z$, let
    $ F(W):=\{z(\alpha):z\in W,\ \alpha\in\Gamma\}$.
    For a consistent labelling
    $\eta:Z\setminus W\to\Gamma$ of $D[Z\setminus W]$,
    let $\mathcal P_\eta$ be the partition of $T\setminus F(W)$
    in which $z(\alpha)$ and $z'(\alpha')$ belong to the same
    part if and only if
    $\eta(z)\otimes\alpha=\eta(z')\otimes\alpha'$.
    
    We claim that $(D,\phi,\Gamma,k)$ is a yes-instance if and only if
    there exist $W\subseteq Z$, a consistent labelling $\eta$
    of $D[Z\setminus W]$, and a $\mathcal P_\eta$-\mwc\
    $Y$ in $G-F(W)$ such that
    \begin{equation}\label{eq:gfvs-budgeted-characterization}
     |W|+|Y|\leq k.
    \end{equation}
    Here $Y$ may intersect $T\setminus F(W)$.
    
    Suppose first that $X$ is a solution of size at most $k$,
    and let $\pi$ be a consistent labelling of $D-X$.
    Set $W:=X\cap Z$, $Y:=X\setminus Z$, and
    $\eta:=\pi|_{Z\setminus W}$.
    Assign the value $\eta(z)\otimes\alpha$ to each terminal
    $z(\alpha)\in T\setminus F(W)$, and retain the values of
    $\pi$ on the vertices outside $Z\cup X$.
    The endpoints of every edge of $G-F(W)-Y$ receive the
    same value: this follows from the identity labels outside
    $Z$ and the labelling equations on arcs between $Z$ and
    $V(D)\setminus Z$.
    Hence the assignment is constant on each connected
    component, so $Y$ is a $\mathcal P_\eta$-\mwc.
    Moreover, $|W|+|Y|=|X|\leq k$.
    
    Conversely, suppose that $W,\eta,Y$ satisfy
    \eqref{eq:gfvs-budgeted-characterization}, and put
    $ Y_0:=Y\setminus T$, and 
    $W':=W\cup\{z\in Z:z(\alpha)\in Y
                             \text{ for some }\alpha\in\Gamma\}$.
    Then $|W'|+|Y_0|\le|W|+|Y|$.
    Since $G-F(W')-Y_0$ is a subgraph of $G-F(W)-Y$,
    all terminals in each of its connected components have
    the same prescribed value $\eta(z)\otimes\alpha$.
    Assign this value to every non-terminal in that component,
    or assign $1_\Gamma$ if the component contains no terminal.
    Together with $\eta|_{Z\setminus W'}$, this defines a
    consistent labelling of $D-(W'\cup Y_0)$.
    Indeed, the equations outside $Z$ follow from the identity
    labels, those inside $Z$ follow from $\eta$, and those
    between $Z$ and $V(D)\setminus Z$ follow from the edges
    incident with the terminals $z(\alpha)$.
    Thus $W'\cup Y_0$ is a solution of size at most $k$.
    This proves the claim.
    
    Apply Theorem~\ref{thm:budgeted-vertex-mimicking} to $(G,T)$
    with budget $k$, and let $(H,T)$ be the resulting graph.
    Then $N:=|V(H)|=O(q^7\,k^{O(\log k)})$.
    For every $W\subseteq Z$ and every consistent labelling
    $\eta$ of $D[Z\setminus W]$, the minimum sizes of a
    $\mathcal P_\eta$-\mwc\ in $G-F(W)$ and in $H-F(W)$
    agree after truncation at $k+1$.
    Consequently, the existence of $W,\eta,Y$ satisfying
    \eqref{eq:gfvs-budgeted-characterization} is unchanged
    when $G$ is replaced by $H$.
    
   We next construct an equivalent instance of
\textsc{Group Feedback Vertex Set} from $H$.
    Put $R:=V(H)\setminus T$, renaming its vertices if
    necessary so that $R\cap Z=\emptyset$, and put $U:=Z\cup R$.
    For each $v\in V(H)$, define
    \[
     \nu(v):=
     \begin{cases}
     v,&v\in R,\\
     z,&v=z(\alpha)\in T,
     \end{cases}
     \qquad
     \delta(v):=
     \begin{cases}
     1_\Gamma,&v\in R,\\
     \alpha,&v=z(\alpha)\in T.
     \end{cases}
    \]
    Form a list $\mathcal L$ of triples $(u,v,\beta)$ as follows:
    \begin{enumerate}
    \item For every arc $(u,v)$ of $D[Z]$, include
          $(u,v,\phi((u,v)))$.
    \item For every edge $ab$ of $H$, choose an ordering of its
          endpoints and include
    $\bigl(\nu(a),\nu(b), \delta(a)\otimes\delta(b)^{-1}\bigr)$.
    \end{enumerate}
    The triple $(u,v,\beta)$ represents the equation
    $\lambda(u)\otimes\beta=\lambda(v)$.
    Construct a directed graph $D'$ with vertex set $U$ and
    one additional vertex $x_\ell$ for each occurrence
    $\ell=(u,v,\beta)$ in $\mathcal L$.
    For this occurrence, add the arcs
    \[
     u\longrightarrow x_\ell
           \text{ with label }\beta,
     \qquad
     x_\ell\longrightarrow v
           \text{ with label }1_\Gamma.
    \]
    Let $\phi'$ be the resulting arc labelling, and output
$(D',\phi',\Gamma,k)$.
    
    We show that this last conversion preserves the answer.
    For $Q\subseteq U$ and a map $\lambda:U\setminus Q\to\Gamma$,
    define
    $\widehat\lambda(a):=\lambda(\nu(a))\otimes\delta(a)$
    whenever $\nu(a)\notin Q$.
    For every edge $ab$ of $H$ with
    $\nu(a),\nu(b)\notin Q$, its equation in $\mathcal L$
    is equivalent to
    $\widehat\lambda(a)=\widehat\lambda(b)$.
    For a triple $\ell=(u,v,\beta)$ with $u,v\notin Q$,
    the values $\lambda(u)$ and $\lambda(v)$ extend to
    $x_\ell$ precisely when
    $\lambda(u)\otimes\beta=\lambda(v)$.
    If one or both endpoints belong to $Q$, a value at
    $x_\ell$ can always be chosen to satisfy the remaining arcs.
    
    Suppose that $W,\eta,Y$ satisfy
    \eqref{eq:gfvs-budgeted-characterization} with $H$ in
    place of $G$.
    Define $W'$ and $Y_0$ as in the proof of the claim.
    Assign values to $R\setminus Y_0$ using the connected
    components of $H-F(W')-Y_0$: use the common prescribed
    terminal value for a component meeting $T$, and
    $1_\Gamma$ for a component disjoint from $T$.
    On $Z\setminus W'$, use $\eta|_{Z\setminus W'}$.
    These assignments satisfy every equation of $\mathcal L$
    whose endpoints lie outside $W'\cup Y_0$.
    They therefore extend to a consistent labelling of
    $D'-(W'\cup Y_0)$.
    Since $|W'|+|Y_0|\leq k$, the output is a yes-instance.
    
    Conversely, let $S$ be a solution in $D'$ of size at most $k$,
    and put
    \[
     S_0:=(S\cap U)\cup
           \{u:x_\ell\in S,\ \ell=(u,v,\beta)\in\mathcal L\}.
    \]
    Then $S_0\subseteq U$ and $|S_0|\le|S|$.
    We claim that $D'-S_0$ also has a consistent labelling.
    Start with a consistent labelling of $D'-S$, and retain
    its values on $U\setminus S_0$ and on the vertices
    $x_\ell\notin S$.
    For each $x_\ell\in S$, the first endpoint $u$ belongs
    to $S_0$. Thus $x_\ell$ has at most one incident arc
    in $D'-S_0$, and its value can be chosen to satisfy
    that arc. This proves the claim.
    
    Let $\lambda$ be a consistent labelling of $D'-S_0$,
    put $W:=S_0\cap Z$ and $Y:=S_0\cap R$, and let
    $\eta:=\lambda|_{Z\setminus W}$.
    The triples corresponding to arcs of $D[Z]$ imply that
    $\eta$ is consistent on $D[Z\setminus W]$.
    The triples corresponding to edges of $H$ imply that
    $\widehat\lambda$ is constant on every connected component
    of $H-F(W)-Y$.
    For each remaining terminal $z(\alpha)$, its value is
    $\widehat\lambda(z(\alpha))=\eta(z)\otimes\alpha$.
    Hence $Y$ is a $\mathcal P_\eta$-\mwc, and
    $|W|+|Y|=|S_0|\leq k$.
    The first claim and the budgeted mimicking guarantee
    therefore imply that $(D,\phi,\Gamma,k)$ is a yes-instance.
    This proves the equivalence of the input and output.
    
    Finally, $|\mathcal L|\leq |Z|^2+\binom N2$,
    so $D'$ has $O(|Z|^2+N^2)$ vertices and arcs.
    Since $|Z|=O(s^2k^2)$, $q=O(s^3k^2)$, and
    $N=O(q^7\,k^{O(\log k)})$, the total encoding length
    of the output is $ s^{O(1)}k^{O(\log k)}$.
 The output retains the multiplication table of $\Gamma$,
whose encoding requires $O(s^2\log(s+1))$ bits.
This is absorbed in the stated encoding bound.
Thus the construction gives a kernel parameterized by $k+s$.
    The construction does not enumerate the sets $W$ or
    the labellings $\eta$.
    All steps take deterministic polynomial time, and
    the output uses the same group and has budget $k$.
    \end{proof}

    \section{Multicut-mimicking hypergraph networks}\label{sec:hypergraphs}

    In this section we prove Theorem~\ref{thm:hypergraph}.
    
    \thmhypergraph*
    
\begin{proof}
Put $k:=|T|$ and $m:=|E(H)|$.
If $m=0$, return the hypergraph with vertex set $T$
and no edges.
Without loss of generality $c\leq m$.

Construct a graph $G$ with a vertex $v_e$
for every $e\in E(H)$.
For distinct $e,f\in E(H)$, add the edge $v_ev_f$
if $e\cap f\ne\emptyset$.
For each $t\in T$, introduce an independent set $D_t$
of $c+1$ new vertices, each adjacent precisely to the
vertices $v_e$ for which $t\in e$.
Put
\[
 T^+:=\bigcup_{t\in T}D_t,
 \qquad q:=|T^+|=k(c+1).
\]
For a partition $\mathcal T$ of $T$, let
\[
 \mathcal T^+:=
 \left\{\bigcup_{t\in P}D_t:P\in\mathcal T\right\}.
\]

For every $X\subseteq E(H)$, the set $X$ is a
$\mathcal T$-edge-\mwc\ in $H$ if and only if
$\{v_e:e\in X\}$ is a $\mathcal T^+$-\mwc\ in $G$.
Indeed, a path between terminals in $H-X$ gives a path
between their corresponding terminal copies in
$G-\{v_e:e\in X\}$.
Conversely, a path between terminal copies in this graph
gives a sequence of intersecting edges connecting the
corresponding terminals in $H-X$.

Furthermore, let $X$ be a $\mathcal T^+$-\mwc\ in $G$
with $|X|\leq c$.
Each class $D_t$ contains a vertex outside $X$.
Its vertices have identical neighborhoods and belong
to the same part of $\mathcal T^+$.
Removing the vertices of $D_t\cap X$ from the deletion
set therefore cannot connect terminals in different parts:
each such vertex joins only the component containing
a surviving copy, or is isolated if no neighbor survives.
Consequently, $X\setminus T^+$ is also a
$\mathcal T^+$-\mwc\ in $G$.

Recall that $\opt_H^{\mathrm{edge}}(\mathcal T)$ denotes
the minimum size of a $\mathcal T$-edge-\mwc\ in $H$,
whereas $\opt_G(\mathcal T^+)$ denotes the minimum size
of a vertex $\mathcal T^+$-\mwc\ in $G$.
It follows that, for every partition $\mathcal T$ of $T$,
\[
 \min\{\opt_H^{\mathrm{edge}}(\mathcal T),c+1\}
 =
 \min\{\opt_G(\mathcal T^+),c+1\}.
\]

Apply Theorem~\ref{thm:budgeted-vertex-mimicking}
to $(G,T^+)$ with budget $c$.
Let $(G',T^+)$ be the resulting connectivity-$c$ vertex
multicut-mimicking network, and put $n:=|V(G')|$.
Then
\[
 n\leq q^7(c+1)^{O(\log(c+1))}
   =k^7(c+1)^{O(\log(c+1))}.
\]

\paragraph{Identifying the terminal copies.}
We claim that, for every partition $\mathcal T$ of $T$,
every minimum $\mathcal T^+$-\mwc\ in $G'$ of size
at most $c$ is disjoint from $T^+$.
Let $X$ be such a minimum multiway cut, put $a:=|X|$,
and let $F:=X\cap T^+$.
Since $a\leq c$, the mimicking guarantee gives
$\opt_G(\mathcal T^+)=a$.

Let $\mathcal T_F^+$ be the partition obtained by removing
$F$ from the parts of $\mathcal T^+$ and omitting empty parts.
The set $X\setminus F$ is a $\mathcal T_F^+$-\mwc\
in $G'-F$, so
$ \opt_{G'-F}(\mathcal T_F^+)\leq a-|F|\leq c$.

From
Theorem~\ref{thm:budgeted-vertex-mimicking}, there is a
$\mathcal T_F^+$-\mwc\ $Y$ in $G-F$ satisfying
$|Y|\leq a-|F|$.
Thus $F\cup Y$ is a $\mathcal T^+$-\mwc\ in $G$
of size at most $a\leq c$.
By the preceding observation, $(F\cup Y)\setminus T^+$
is also a $\mathcal T^+$-\mwc\ in $G$,
of size at most $a-|F|$.
Since $\opt_G(\mathcal T^+)=a$, this implies $F=\emptyset$,
proving the claim.

Construct $G''$ from $G'$ by identifying, for each $t\in T$,
all vertices of $D_t$ into one vertex named $t$.
Remove loops and duplicate edges.
We consider multiway cuts in $G''$ that are disjoint from $T$.
For every partition $\mathcal T$, such a cut of size at most
$c$ exists if and only if a $\mathcal T^+$-\mwc\ of size
at most $c$ exists in $G'$, and their minimum sizes agree
whenever either is at most $c$.
Indeed, the claim gives a minimum multiway cut in $G'$
avoiding all terminal copies.
Identifying a class $D_t$ only joins components whose
terminals belong to the same part of $\mathcal T^+$,
so this multiway cut remains feasible in $G''$.
Conversely, a multiway cut in $G''$ disjoint from $T$
is also a $\mathcal T^+$-\mwc\ in $G'$, since identification
can only add connectivity.

\paragraph{Construction of the output hypergraph.}
Put $U:=V(G'')\setminus T$.
For every $u\in U$, introduce a new vertex $p_u$
and an edge $E_u$ initially containing only $p_u$.
For every edge $e=\{u,v\}$ of $G''[U]$, introduce
a new vertex $p_e$ and put it in both $E_u$ and $E_v$.
For every edge $ut$ of $G''$ with $u\in U$ and $t\in T$,
put $t$ in $E_u$.

For every edge $st$ of $G''[T]$, introduce $c+1$
new vertices $x_{st,1},\ldots,x_{st,c+1}$ and the edges $\{s,x_{st,i}\}$ and $\{x_{st,i},t\}$ for every $i\in[c+1]$.
All introduced vertices are distinct unless explicitly identified.
Let $H'$ have these vertices together with $T$, and all
the edges just constructed.
The vertices $p_u$ ensure that the edges $E_u$ are distinct.

Fix a partition $\mathcal T$ of $T$.
If $Z\subseteq U$ is a $\mathcal T$-\mwc\ in $G''$,
then $\{E_u:u\in Z\}$ is a $\mathcal T$-edge-\mwc\
in $H'$ of the same size.
Indeed, connections through an edge $E_u$ correspond
to connections through the vertex $u$ in $G''$.
The vertices $p_e$ represent edges between non-terminals,
and each path through $x_{st,i}$ represents the edge $st$.
Consequently, a connection between terminals in $H'$
after these edge deletions would give a connection
between the same terminals in $G''-Z$.

Conversely, let $\mathcal X\subseteq E(H')$ be a
$\mathcal T$-edge-\mwc\ with $|\mathcal X|\leq c$,
and put
$Z:=\{u\in U:E_u\in\mathcal X\}$.
Then $|Z|\le|\mathcal X|$.
For every edge $st$ of $G''[T]$, at least one of the
$c+1$ two-edge $s$--$t$ paths remains intact in
$H'-\mathcal X$, since these paths are pairwise edge-disjoint.
Every path between terminals in $G''-Z$ therefore gives
a connection between the same terminals in $H'-\mathcal X$:
surviving non-terminal vertices correspond to surviving
edges $E_u$, and every edge $st$ of $G''[T]$ can be replaced by
a path $s,x_{st,i},t$ whose two edges lie outside $\mathcal X$.
Hence $Z$ is a $\mathcal T$-\mwc\ in $G''$ disjoint from $T$.

Combining the preceding equivalences gives
\[
 \min\{\opt_H^{\mathrm{edge}}(\mathcal T),c+1\}
 =
 \min\{\opt_{H'}^{\mathrm{edge}}(\mathcal T),c+1\}.
\]
The partition $\mathcal T$ was arbitrary.
Thus $(H',T)$ is a connectivity-$c$ edge
multicut-mimicking network of $(H,T)$.

\paragraph{Size and running time.}
Identification does not increase the vertex count, so
$|V(G'')|\leq n$.
The output hypergraph has at most
\[
 n+2(c+1)\binom k2
\]
edges, and
\[
 |V(H')|+|E(H')|+\sum_{e\in E(H')}|e|
 =O\bigl(n^2+kn+(c+1)k^2+k\bigr).
\]
Using the bound on $n$, this is
$k^{O(1)} c^{O(\log c)}$.

Since $c\leq m$, the graph $G$ and all sets of terminal
copies have polynomial size in the input.
Theorem~\ref{thm:budgeted-vertex-mimicking} runs in
deterministic polynomial time, so $n$ is polynomially
bounded in the original input size.
The identification and the construction of $H'$ also
take deterministic polynomial time.
This proves the theorem.
\end{proof}

\section{Concluding remarks}\label{sec:conclusion}

We gave a deterministic polynomial-time construction
of connectivity-$c$ vertex multicut-mimicking networks
with $|T|^7 \cdot c^{O(\log c)}$ vertices.
This yields exact vertex multicut-mimicking networks
of quasi-polynomial size and deterministic quasi-polynomial
kernels for \textsc{Vertex Multiway Cut},
\textsc{Vertex Multicut}, and
\textsc{Group Feedback Vertex Set}.
The latter two bounds have only polynomial dependence
on the number of requests and the group order,
respectively.

The construction also gives hypergraph edge
multicut-mimicking networks whose total encoding length
is bounded independently of the hypergraph rank.

Whether exact multicut-mimicking networks of polynomial
size exist remains open for both edge and vertex cuts.
Even from the viewpoint of just existence,
the approaches discussed here
give bounds of the form $k^{O(\log k)}$ and seems to stall at the exponent of $O(\log k)$.
New approaches seems to be discovered to get even a bound of $k^{O(\sqrt{\log k})}$, even for edge multicut-mimicking networks.
Furthermore, polynomial-time constructions of polynomial-sized
networks would yield polynomial kernels for
\textsc{Edge Multiway Cut} and
\textsc{Vertex Multiway Cut}, respectively.

\bibliographystyle{siam}
\bibliography{references}
    
\appendix
    
\section{Deterministic representative families}
    \label{app:deterministic-representative-families}
    
    We prove Lemma~\ref{lem:repset-product} using the
    deterministic representation and selection machinery
    of Koana and Kumabe~\cite{DBLP:journals/corr/abs-2609-29141}.
    The additional observation is that their selection
    algorithm admits a bound in terms of the product
    of the individual layer ranks.
    
    We use $\varepsilon$-almost multilinear representations
    in the sense of~\cite[Definition~4.1]
    {DBLP:journals/corr/abs-2609-29141}.
    Such a representation has a block of $t$ columns
    for each matroid element, where $t\ge1$ is its scale.
    If $S(F)$ denotes the span of the blocks indexed by $F$,
    then
    \[
     \left|\frac{\dim S(F)}{t}-r_N(F)\right|
     \le\varepsilon
     \qquad\text{for every }F\subseteq E(N).
    \]
    
    \subsection{An unequal-rank selection lemma}
    
    \begin{lemma}
    \label{lem:almost-repset-product}
    Let $d\ge2$ be fixed, and let $N_1,\ldots,N_d$ be
    matroids on pairwise disjoint ground sets $E_1,\ldots,E_d$,
    with positive integer rank upper bounds $r_1,\ldots,r_d$.
    Put
    \[
     a:=\max_i r_i,
     \qquad P:=\prod_{i=1}^d r_i,
     \qquad
     \varepsilon:=\frac{1}{8d(a+1)^{d-1}}.
    \]
    Suppose that the matroids are supplied with
    $\varepsilon$-almost multilinear representations
    over a common prime field $\mathbb F_p$
    and at a common scale $t$.
    
    Given an explicit family $\mathcal Y$ of independent
    sets in their direct sum, each containing one element
    from each layer, one can deterministically compute
    a representative family of size at most $2P$
    in polynomial time in the input size.
    \end{lemma}
    
    \begin{proof}
    If $\mathcal Y=\emptyset$, return the empty family.
    Otherwise, extend each matroid by loops to the common
    ground set $V:=\bigcup_i E_i$, appending zero blocks
    to its representation.
    Put $r:=\min\{a,|V|\}$.
    The extended matroids have rank at most $r$, and
    \[
     \varepsilon
     =\frac{1}{8d(a+1)^{d-1}}
     \le\frac{1}{4d(r^{d-1}+1)}.
    \]
    We may therefore apply the tensor-space selection
    algorithm of~\cite[Theorem~1.5 and Section~5]
    {DBLP:journals/corr/abs-2609-29141}.
    Its correctness gives a representative family,
    and its running time is polynomial for fixed $d$.
    
    We refine only the size analysis.
    Let $\mathcal V_i$ be the span of all columns in
    the representation of layer $i$.
    Then
    \[
     \dim\mathcal V_i\leq t(r_i+\varepsilon).
    \]
    The algorithm retains a candidate only when its
    tensor subspace increases the current span dimension
    by at least $t^d/2$.
    The ambient tensor space has dimension at most
    \[
     \prod_i\dim\mathcal V_i
     \leq t^d\prod_i(r_i+\varepsilon).
    \]
    Since $\varepsilon\le1$, telescoping the product gives
    \[
     \prod_i(r_i+\varepsilon)-P
     \leq d\varepsilon(a+1)^{d-1}
     =\frac18.
    \]
    Consequently,
    \[
     |\widehat{\mathcal Y}|
     \le2\prod_i(r_i+\varepsilon)
     \le2P+\frac14.
    \]
    Integrality gives $|\widehat{\mathcal Y}|\le2P$.
    \end{proof}
    
    \subsection{Proof of the representative-family bound}
    
    \begin{proof}[Proof of Lemma~\ref{lem:repset-product}]
    Write $E_L$ for the ground set of the uniform matroid $L$,
    and put
    \[
     a:=\max\{\rho,R\},
     \qquad
     \varepsilon:=\frac{1}{8d(a+1)^{d-1}}.
    \]
    
    Before constructing a representation, pad the directed
    graph specifying $N$, if necessary, with isolated vertices
    outside its source and ground sets until it has at least
    $|E_L|$ vertices.
    This does not change $N$.
    
    By~\cite[Corollary~1.4]{DBLP:journals/corr/abs-2609-29141},
    we can deterministically construct an
    $\varepsilon$-almost multilinear representation of $N$
    over a prime field $\mathbb F_p$ in time polynomial
    in the input size and $1/\varepsilon$.
    Let its scale be $t$.
    Use copies of this representation in the first $d-1$
    layers.
    
    The prime chosen by the construction
    in~\cite[Section~4]{DBLP:journals/corr/abs-2609-29141}
    exceeds the number of vertices of the padded graph.
    Hence we can assign distinct labels
    $\lambda_e\in\mathbb F_p$ to the elements of $E_L$.
    For each $e\in E_L$, put
    \[
     u_e:=(1,\lambda_e,\ldots,\lambda_e^{R-1})^{\mathsf T},
    \]
    and assign to $e$ the block $u_e\otimes I_t$.
    By the Vandermonde rank property, the blocks indexed
    by any $F\subseteq E_L$ have rank
    $t\min\{|F|,R\}$.
    They therefore give an exact multilinear representation
    of $L$ over $\mathbb F_p$ at scale $t$.
    
    All layers now have representations over the same field
    and at the same scale.
    Lemma~\ref{lem:almost-repset-product}, applied with rank
    bounds $\rho,\ldots,\rho,R$, gives a representative
    family of size at most $2\rho^{d-1}R$.
    
    Since $d$ is fixed and
    $1/\varepsilon=8d(a+1)^{d-1}$ is polynomially bounded
    in the input size, the entire construction takes
    deterministic polynomial time.
    \end{proof}
    
    \subsection{The marking-size budget}
    \begin{lemma}
    \label{lem:deterministic-smalltail-size}
    Let $k\ge2$, $\alpha\ge14$, and
    $n>\bound(k)=k^{\lceil\alpha\log_2k\rceil}$.
    Put
    \[
     R:=\min\{n,\lfloor n/(12k^3)\rfloor+k\}.
    \]
    Then $k+2k^3R<3n/16$ and $k<n/16$.
    \end{lemma}
    
    \begin{proof}
    Since $k\ge2$ and $\alpha\ge14$,
    \[
     48(2k^4+k)
     \le144k^4
     \leq k^{14}
     \le\bound(k)
     <n.
    \]
    Hence $2k^4+k<n/48$, and therefore
    \[
     k+2k^3R
     \le\frac n6+2k^4+k
     <\frac n6+\frac n{48}
     =\frac{3n}{16}.
    \]
    The same estimate gives $k<n/48<n/16$.
    \end{proof}

    \section{Strong vertex multicut-mimicking in hypergraphs}
    \label{app:strong-vertex-deletion}
    
    We show that, unless $\mathrm{FPT}=\mathrm{W}[2]$, there is
    no deterministic polynomial-time algorithm that constructs
    a strong vertex multicut-mimicking network whose number
    of vertices is bounded by a computable function of the
    number of terminals and the cut budget.
    The obstruction concerns hypergraphs of unrestricted rank
    and already holds for a single binary terminal partition,
    without free terminal deletions.
    
    \begin{proposition}
    Unless $\mathrm{FPT}=\mathrm{W}[2]$, there is no deterministic
    polynomial-time algorithm that, given a hypergraph terminal
    network $(H,T)$ and an integer $c\ge1$, outputs an explicitly
    represented hypergraph terminal network $(H',T)$ satisfying
    the following properties:
    \begin{enumerate}
     \item $|V(H')|\leq f(|T|,c)$ for some computable function $f$;
     \item for every partition $\mathcal T$ of $T$ into $2$ parts, a strong
           $\mathcal T$-multiway cut (so essentially a minimum separator) of size at most $c$ exists
           in $H$ if and only if one exists in $H'$.
    \end{enumerate}
    \end{proposition}
    
    \begin{proof}
    We give a reduction from \textsc{Set Cover} parameterized
    by the solution size.
    Let $U$ be its universe, let $S_1,\ldots,S_m\subseteq U$
    be the given sets, and let $c$ be the budget.
    We may assume $1\leq c\leq m$, since the remaining cases
    can be decided in polynomial time.
    
    Construct a hypergraph $H$ with an ordinary vertex $x_\ell$
    for each set $S_\ell$, and two disjoint terminal sets
    \[
     A:=\{a_1,\ldots,a_{c+1}\},
     \qquad
     B:=\{b_1,\ldots,b_{c+1}\}.
    \]
    Put $T:=A\cup B$.
    For every $u\in U$ and $i,j\in[c+1]$, introduce the edge
   $ e_{u,i,j}:=
     \{a_i,b_j\}\cup\{x_\ell:u\in S_\ell\}$.
    Identical edges may be merged, since their multiplicity
    does not affect connectivity under strong vertex deletion.
    The construction takes polynomial time.
    Consider the partition $\mathcal T:=\{A,B\}$ with budget $c$.
    
    Suppose first that $I\subseteq[m]$ satisfies $|I|\leq c$
    and $\bigcup_{\ell\in I}S_\ell=U$.
    Every edge of $H$ meets
    $X:=\{x_\ell:\ell\in I\}$.
    Strong deletion of $X$ therefore removes every edge,
    so $X$ is a strong $\mathcal T$-multiway cut of size
    at most $c$.
    
    Conversely, let $X\subseteq V(H)$ be a strong
    $\mathcal T$-multiway cut with $|X|\leq c$.
    Since $|A|=|B|=c+1$, there are terminals
    $a_i\in A\setminus X$ and $b_j\in B\setminus X$.
    Put
    $ I:=\{\ell\in[m]:x_\ell\in X\}$.
    If the sets indexed by $I$ fail to cover some $u\in U$,
    then $e_{u,i,j}\cap X=\emptyset$.
    This edge survives strong deletion and connects $a_i$
    to $b_j$, contradicting the choice of $X$.
    Hence the sets indexed by $I$ cover $U$, and $|I|\leq c$.
    
    Thus the \textsc{Set Cover} instance is a yes-instance
    if and only if $H$ has a strong $\mathcal T$-multiway cut
    of size at most $c$.
    Moreover, $|T|=2(c+1)$.

    Suppose now that the stated algorithm exists.
    Apply it to $(H,T)$ with budget $c$.
    Its output $H'$ has at most $f(2c+2,c)$ vertices.
    Enumerate all subsets of $V(H')$ of size at most $c$,
    and test whether strongly deleting one of them separates
    the surviving terminals of $A$ from those of $B$.
    Each test takes polynomial time in the explicit encoding
    length of $H'$.
    This encoding length is polynomial in the original input
    length, because the construction runs in polynomial time.
    
    By the preservation guarantee, this procedure decides
    the original \textsc{Set Cover} instance.
    Its running time is $g(c)N^{O(1)}$ for some computable
    function $g$, where $N$ is the input length.
    Since \textsc{Set Cover} parameterized by the solution size
    is $\mathrm{W}[2]$-complete, this implies
    $\mathrm{FPT}=\mathrm{W}[2]$.
    \end{proof}
    
    The proposition concerns polynomial-time constructions.
    It does not exclude a purely existential bound on the
    size of a strong vertex multicut-mimicking network.

\section{Size estimates}
\label{app:constructive-size-estimates}

Recall that
\[
 q:=\max\{1,\lceil\log_2 c_{\texttt{cover}}\rceil\},
 \qquad \alpha:=160q,
 \qquad
 \bound(t):=t^{\lceil\alpha\log_2t\rceil}
 \quad(t\ge1).
\]

\begin{lemma}[Size estimates]
\label{lem:relative-slack}
Let $k\ge4$, $n>\bound(k)$, and suppose that
\[
 |M_1|\le c_{\texttt{cover}}(k+120k^6)^3,
 \qquad
 |M_2|\le
 c_{\texttt{cover}}\bigl((2k+1)|M_1|\bigr)^3.
\]
Put $h_0:=\lfloor14k/15\rfloor$ and
$h_1:=\lfloor4k/5\rfloor$. Then
\begin{equation}\label{eq:exposure-slack}
 \frac{n}{24k^4}+12k|M_1|\bound(h_0)
 <\frac{n}{12k^4},
\end{equation}
and
\[
 7k|M_2|\bound(h_1)<\frac n{16}.
\]
\end{lemma}

\begin{proof}
Write $C:=c_{\texttt{cover}}\le2^q$ and $x:=\log_2k\ge2$.
Since $k+120k^6\le128k^6$ and $2k+1\le3k$, we have
\[
 |M_1|\le\mu:=2^{21}Ck^{18},
 \qquad
 |M_2|\le\nu:=27\cdot2^{63}C^4k^{57}.
\]
Put $A_0:=288k^5\mu$ and $A_1:=128k\nu$.
It suffices to prove
\[
 \bound(k)>A_0\bound(h_0),
 \qquad
 \bound(k)>A_1\bound(h_1).
\]

Using $y\le\lceil y\rceil<y+1$, for every integer
$1\le u<k$ we obtain
\begin{equation}\label{eq:size-ratio}
 \log_2\frac{\bound(k)}{\bound(u)}
 \ge160q\bigl(x^2-(\log_2u)^2\bigr)-\log_2u.
\end{equation}
Moreover,
\[
 \log_2 A_0<59/2+q+23x,
 \qquad
 \log_2 A_1<75+4q+58x.
\]
Using $\log_2(15/14)>2/21$ and
$\log_2(5/4)>8/25$ in \eqref{eq:size-ratio} gives
\begin{align*}
 \log_2\frac{\bound(k)}{\bound(h_0)}-\log_2 A_0
 &>(640q/21-24)x-(1+640/441)q-59/2,\\
 \log_2\frac{\bound(k)}{\bound(h_1)}-\log_2 A_1
 &>(512q/5-59)x-2548q/125-75.
\end{align*}
Both right-hand sides increase with $q\ge1$ and $x\ge2$.
The first is positive when $q\ge2$ or $k\ge32$,
as its values at $(q,x)=(2,2)$ and $(1,5)$ are positive.
The second is positive when $k\ge5$, since
$\log_2 5>11/5$ and its value at $(q,x)=(1,11/5)$
is $12/125$.

For the first inequality, it remains to consider
$q=1$ and $4\le k\le31$.
The function
\[
 D_a(t):=(\log_2t)^2-(\log_2(t-a))^2
\]
decreases for $t\ge a+3$.
Using $D_1(14)>4/5$ and $D_2(30)>145/153$,
together with \eqref{eq:size-ratio}, gives
\[
\begin{array}{c|c|c}
 k
 & \log_2\bigl(\bound(k)/\bound(h_0)\bigr)
 & \log_2 A_0\\
 \hline
 4\le k\le14 & >124 & <245/2\\
 16\le k\le31 & >291/2 & <291/2
\end{array}
\]
Indeed, $h_0=k-1$ in the first range and $h_0=k-2$
for $16\le k\le30$; also $h_0(31)=h_0(30)=28$.
The second row uses $160(145/153)-5>291/2$.

For $k=15$, the exact exponents give
$\bound(15)=15^{626}$ and $\bound(14)=14^{610}$.
Hence
\[
\begin{aligned}
 \log_2\frac{\bound(15)}{\bound(14)}-\log_2 A_0
 &>603\log_2(15/14)-7\log_2 14-61/2\\
 &>53/62>0,
\end{aligned}
\]
using $\log_2(15/14)>3/31$ and $\log_2 14<27/7$.
This proves the first inequality.

For the second inequality, the remaining case is $k=4$.
Here $h_1=3$, $\bound(4)=4^{320q}$ and
$\bound(3)\le3^{256q}$.
Since $\log_2 3<8/5$,
\[
 \log_2\frac{\bound(4)}{\bound(3)}
 >\frac{1152q}{5}
 >191+4q
 >\log_2 A_1.
\]
This proves the second inequality.

Finally, since $n>\bound(k)$, these inequalities imply
\[
 12k|M_1|\bound(h_0)<\frac{n}{24k^4},
 \qquad
 7k|M_2|\bound(h_1)<\frac{7n}{128}<\frac n{16},
\]
as required.
\end{proof}

\begin{lemma}
\label{lem:budgeted-size-ratio}
For every integer $c\ge1$,
\[
 \bound(4c)>864c_{\texttt{cover}}\bound(3c).
\]
\end{lemma}

\begin{proof}
Put $x:=\log_2(3c)>1$ and
$\delta:=\log_2(4/3)>2/5$.
Using $y\le\lceil y\rceil<y+1$, we obtain
\[
\begin{aligned}
 \log_2\frac{\bound(4c)}{\bound(3c)}
 &\ge160q\bigl((x+\delta)^2-x^2\bigr)-x\\
 &>(128q-1)x+\frac{128q}{5}\\
 &>q+10\\
 &>\log_2(864c_{\texttt{cover}}),
\end{aligned}
\]
where the last two inequalities use $q\ge1$, $x>1$,
and $c_{\texttt{cover}}\le2^q$.
Exponentiating proves the assertion.
\end{proof}

\end{document}